\documentstyle[sprocl]{article}
\input psfig.sty
\def\Journal#1#2#3#4{{#1} {\bf #2}, #3 (#4)}
\def\NPB{{\em Nucl. Phys.} B}
\def\PLB{{\em Phys. Lett.}  B}
\def\PRL{\em Phys. Rev. Lett.} 
\def\PRD{{\em Phys. Rev.} D}
\def\ZPC{{\em Z. Phys.} C}

\def\be{\begin{equation}}
\def\ee{\end{equation}}
\def\bea{\begin{eqnarray}}
\def\eea{\end{eqnarray}}

\def\what{\widehat}

\def\mstop{m_{\tilde t}}
\def\twoloop{two-loop/RGE-improved}
\def\Twoloop{Two-loop/RGE-improved}

\def\delmhsm{\Delta\mhsm}

\def\DM{D$^-$}
\def\DP{D$^+$}
\def\NSM{NS$^-$}
\def\NSP{NS$^+$}
\def\HSM{HS$^-$}
\def\HSP{HS$^+$}
\def\etc{{\it etc.}}
\def\leff{L_{\rm eff}}
\def\sign{{\rm sign}}

\def\anti{\overline}

\def\hi{\h_1}

\def\mhi{m_{\hi}}

\def\etc{{\em etc.}}
\def\chisq{\chi^2}

\def\thdm{2HDM}

\def\gamhsm{\Gamma_{\hsm}^{\rm tot}}
\def\gamhl{\Gamma_{\hl}^{\rm tot}}
\def\tauptaum{\tau^+\tau^-}
\def\gamgam{\gam\gam}

\def\wstar{W^{\star}}

\def\zstar{Z^{\star}}

\def\br{BR}

\def\rts{\sqrt s}
\def\eps{\epsilon}
\def\h{h}
\def\a{a}
\def\mh{m_{\h}}
\def\ma{m_{\a}}

\def\lam{\lambda}

\def\eg{{\it e.g.}}
\def\etal{{\it et al.}}
\def\epem{e^+e^-}
\def\mupmum{\mu^+\mu^-}

\def\ee{e^+e^-}
\def\lplm{\ell^+\ell^-}
\def\taup{\tau^+}
\def\taum{\tau^-}

\def\lsim{\mathrel{\raise.3ex\hbox{$<$\kern-.75em\lower1ex\hbox{$\sim$}}}}
\def\gsim{\mathrel{\raise.3ex\hbox{$>$\kern-.75em\lower1ex\hbox{$\sim$}}}}
\def\@versim#1#2{\vcenter{\offinterlineskip
        \ialign{$\m@th#1\hfil##\hfil$\crcr#2\crcr\sim\crcr } }}
\def\zstar{Z^\star}
\def\wstar{W^\star}
\def\slash#1{#1\hskip-8pt/\hskip4pt}

\def\ie{{\it i.e.}}

\def\gam{\gamma}

\def\anti{\overline}
\def\pbi{~{\rm pb}^{-1}}
\def\fbi{~{\rm fb}^{-1}}
\def\fb{~{\rm fb}}
\def\pb{~{\rm pb}}

\def\mev{\,{\rm MeV}}
\def\gev{\,{\rm GeV}}
\def\tev{\,{\rm TeV}}

\def\wt{\widetilde}

\def\rta{\rightarrow}
\def\mhalf{m_{1/2}}

\def\stop{\wt t}
\def\stopone{\wt t_1}
\def\stoptwo{\wt t_2}
\def\mstop{m_{\stop}}

\def\sq{\wt q}

\def\slep{\wt \ell}

\def\sbot{\wt b}

\def\hsm{h_{\rm SM}}
\def\mhsm{m_{\hsm}}
\def\hl{h^0}
\def\hh{H^0}
\def\ha{A^0}
\def\hp{H^+}
\def\hm{H^-}
\def\hpm{H^{\pm}}
\def\mhl{m_{\hl}}
\def\mhh{m_{\hh}}
\def\mha{m_{\ha}}

\def\mhpm{m_{\hpm}}
\def\tanb{\tan\beta}

\def\mt{m_t}
\def\mb{m_b}
\def\mz{m_Z}
\def\mw{m_W}

\def\wp{W^+}

\def\chitil{\wt\chi}
\def\cnone{\wt\chi^0_1}

\def\snu{\wt\nu}

\def\h{h}
\def\mh{m_{\h}}
\def\cpone{\wt \chi^+_1}
\def\cmone{\wt \chi^-_1}
\def\cpmone{\wt \chi^{\pm}_1}

\def\mcpmone{m_{\cpmone}}

\begin{document}
{\hspace*{\fill}{\vbox{\begin{flushright} \bf UCD-97-13 \\ May, 1997
\end{flushright}}}\\}
\vskip -.35in
\title{DETECTING AND STUDYING HIGGS BOSONS~\footnote{To appear in
{\it Perspectives on Higgs Physics}, ed. G. Kane, 2nd edition (World
Scientific Publishing).}}
\author{ J.F. GUNION }
\address{Department of Physics, University of California, Davis, CA 95616}
\maketitle\abstracts{We review the ability of the LHC (large hadron collider), 
NLC (next linear lepton collider) and FMC (first muon collider) to detect
and study Higgs bosons, with emphasis on the
Higgs bosons of extended Higgs sectors, especially
those of the minimal supersymmetric standard model (MSSM).
Particular attention is given to means for distinguishing the lightest 
neutral CP-even
Higgs boson of the MSSM from the single Higgs boson of the minimal 
Standard Model (SM).}

\section{Introduction}

Indirect evidence from precision electroweak analyses is increasingly
suggestive that there is a relatively light Higgs boson
with SM-like properties.~\cite{blondel} Since there
is also a complete absence to date of any signals for new physics, the SM
with a single neutral Higgs boson ($\hsm$) remains a very viable model.
However, there are simple 
extensions of the one-Higgs-doublet~\footnote{`Doublet'
refers to the transformation properties under weak isospin.}
SM Higgs sector that are equally
consistent with all known theoretical and phenomenological
constraints.~\cite{hhg}
Models containing extra Higgs doublets and/or singlets are the most attractive.
Such models automatically preserve the custodial SU(2)
symmetry tree-level prediction of $\rho\equiv \mw^2/[\mz^2\cos^2\theta_W]=1$.
\begin{itemize}
\item
In a general two-Higgs-doublet model (2HDM), there 
are three neutral CP-mixed Higgs eigenstates ($\h_{1,2,3}$) and
a charged Higgs pair ($\hpm$). In the CP-conserving limit, the neutral sector
of the 2HDM
divides into two CP-even states ($\hl$ and $\hh$) and one CP-odd state ($\ha$).
The Higgs sector of the MSSM is a highly constrained CP-conserving 2HDM.
\item
If a single complex Higgs singlet field is added to the 2HDM,
then there will be a pair of charged Higgs bosons
and, in the CP-conserving limit, three CP-even and two CP-odd
neutral states. In the CP-violating case,
there would simply be five CP-mixed neutral states.
The Higgs sector of the next-to-minimal
supersymmetric model (NMSSM) is a constrained two-doublet, one-singlet model
that is often taken to be CP-conserving, but could 
also be explicitly or spontaneously CP-violating.
\end{itemize}
Models containing triplet Higgs fields (in addition to at least one
doublet, as required to give masses to quarks and leptons) 
are also frequently discussed.~\cite{hhg} 
However, lack of space precludes discussing such models here.
A brief review of some recent phenomenology is given
in the Snowmass96 Higgs boson summary report~\cite{snowmasssummary}
(hereafter referred to as Higgs96).

In most extended Higgs sector models, it is very natural to
be in a decoupling regime in which the lightest Higgs boson ($\hl$)
is neutral and CP-even and has very SM-like properties. Other Higgs
bosons would be heavier; depending on the model, the
additional neutral scalars could be of pure or mixed CP nature.

Supersymmetric models provide a particularly natural
framework for light scalars, and are attractive in many ways, including
the fact that they resolve
the naturalness and hierarchy problems associated with the SM Higgs sector.
The minimal Higgs sector required in order to give both up and
down quarks masses and to avoid anomalies consists of two doublets.
Two doublets also imply quite accurate gauge coupling unification
if the supersymmetric partners of the SM particles have masses
below 1 to 10 TeV. More doublets would destroy this unification.
In the CP-conserving MSSM 2HDM Higgs sector with
physical eigenstates $\hl,\hh,\ha,\hpm$,
the Higgs boson masses and couplings are entirely specified by just
two parameters at tree-level. These
are normally taken to be $\mha$ and $\tanb$ (the ratio of
vacuum expectations values for the neutral members of the two doublets).
When $\mha\gg\mz$ (as is natural in a GUT context), 
the $\hl$ remains light ($\mhl\lsim 130\gev$~\cite{haber,pokorski})
and is very SM-like while
the other scalars have $\mhh\sim\mhpm\sim\mha$ and decouple from $ZZ,WW$.

However, there is no guarantee that the supersymmetric model Higgs
sector will consist of just two doublets.
In the next-to-minimal supersymmetric extension of the SM
(NMSSM),~\cite{hhg,eghrz}
a complex singlet superfield ($\what N$) is added to the two doublet
superfields ($\what H_1$ and $\what H_2$) of the MSSM.  The scalar
component of the singlet superfield joins with the scalar components
of the two doublet superfields to form a two-doublet, one-singlet
Higgs sector. One or more
such singlet superfields are a common feature of the low energy
supersymmetric effective field theories that emerge from the typical super
string compactification. Singlets leave intact the compelling
attractive theoretical successes of the MSSM.  In particular,
gauge coupling unification is unmodified.
The NMSSM has the added virtue of allowing
for a natural source for a $\sim\tev$-scale $\mu$ parameter
through a $\what H_1\what H_2 \what N$ term in the superpotential when
the scalar  component of $\what N$ acquires a non-zero vacuum
expectation value. However, adding one or more
singlet fields could have a substantial
impact on the ease with which the Higgs bosons
of the model can be discovered and studied. In the GUT context,
a decoupling limit in which one of the CP-even Higgs bosons remains
light ($\mhl\lsim 150\gev$ assuming that the coupling
$\lam$ characterizing the strength of
the cubic singlet superfield term, $\lam\what N^3$,
in the superpotential remains perturbative during
evolution to the GUT scale) and has SM-like couplings,
the other Higgs bosons being heavy, remains a possibility.
The $\hl$ would then be easily discovered,
while the discovery of the heavier Higgs bosons would not be guaranteed.
But, it is also very possible
that there will be two or even three relatively
light CP-even Higgs scalars that share the $ZZ,WW$ coupling strength.
In this case, the strength of the signals for any one of these scalars
is reduced relative to a SM Higgs boson of the same mass. An important
question is whether discovery of at least one of the Higgs bosons
of the model can then be guaranteed at future, if not present,
accelerators.

Given the possibility (or in the MSSM, the probability) that the extended
Higgs sector will have a light SM-like $\hl$, 
it is clear that discriminating between it and
the minimal SM Higgs sector will require
either the detection of differences between the $\hl$ properties
and those predicted for the $\hsm$ 
(at a precision level) or direct observation of the heavier scalar
eigenstates. It is the prospects for success in these two tasks 
upon which much of this review focuses.
Both tasks become increasingly difficult as the mass scale of the heavier
Higgs boson eigenstates increases.

The outline of the rest of this review is the following.
In Section~\ref{ssm}, we explore the precisions expected for measurements
of the properties of a SM-like Higgs boson at various accelerators.
Our primary focus will be on the next generation of machines:
the Large Hadron Collider (LHC), a next linear $\epem$
collider (NLC), and a possible future $\mupmum$ collider (FMC).
Expectations for LEP2 and Tev33 will also be noted.
In Section~\ref{snonsm}, we examine search strategies
for and the measurement of the properties of Higgs bosons
with very non-SM-like properties, in particular the heavier Higgs bosons
of the MSSM. Section~\ref{sconcl} presents conclusions.
This review is designed to give an overview of the current expectations
and strategies, and consequently
does not include many details. Familiarity with the basics
presented in the Higgs Hunters Guide~\cite{hhg} is presumed.
The review is designed to be read as a guide to
the recent DPF95~\cite{dpfreport} and Higgs96~\cite{snowmasssummary}
reviews, the NLC Physics report,~\cite{nlcreport} and the muon-collider
$s$-channel Higgs physics study.~\cite{bbgh}
The LEP2~\cite{lep2study} and European $\epem$ collider~\cite{eurostudy}
study group reports are also valuable references.

\section{Discovery and Precision Measurements of a SM-like Higgs}\label{ssm}

There is no question that a SM-like Higgs boson with 
mass below 1 TeV or so can be discovered
at the LHC assuming that the latter reaches its
design goal of $L=100\fbi$ per year per detector. Indeed,
for much of the mass range in which the Higgs sector is
weakly-coupled  ($\mhsm\lsim 600\gev$) significantly 
less integrated luminosity is required. The discovery reach
of the NLC~\footnote{An FMC operated in the same
manner as the NLC will have the same discovery reach.}
will be entirely determined by the available energy, $\rts$. 
A $\rts=500\gev$ NLC would discover a SM-like Higgs boson
with $\mhsm\leq 350\gev$, assuming $L=50\fbi$ is accumulated.
Much less luminosity is required for masses in the range
$\lsim 150\gev$ predicted for the SM-like supersymmetric model Higgs
boson in the MSSM or NMSSM. (Summary figures can
be found in DPF95.~\cite{dpfreport}) Finally, a FMC run
at $\rts\simeq\mhsm$ would produce a SM-like Higgs boson at a very high rate,
provided $\mhsm<2\mw$. The real issue is the precision with
which the properties of the SM-like Higgs boson
can be measured at the three machines. The most
extensive exploration of this matter appears in the Higgs96 
review.~\cite{snowmasssummary}
There, errors are estimated assuming that multi-year running
will achieve the following accumulated luminosities at the different
accelerator facilities: $L=300\fbi$ for
both ATLAS and CMS at the LHC; $L=200\fbi$ in $\rts=500\gev$
operation at the NLC; $L=50\fbi$ for $\gam\gam$ collisions at the NLC
operating in the photon-collider mode with
$E_{\epem}\sim \mhsm/0.8$; and $L=200\fbi$ at the FMC running
at $\rts\simeq\mhsm$ for a scan of the Higgs mass peak.
Depending upon the mass of the Higgs, data from LEP2 and/or the Tevatron could
also have an important impact. We assume detector-summed integrated
luminosity of $L=1000\pbi$ at LEP2 ($\rts=192\gev$) and of $L=60\fbi$
at TeV33.

\subsection{LHC, including Tevatron and LEP2 data}\label{sslhc}

At the LHC, it is useful to divide the discussion into five mass regions.
\begin{description}
\item {M1:} $\mhsm\lsim 95\gev-100\gev$. Detection of the $\hsm$
should be possible at all three machines: LEP2, the Tevatron,
and the LHC.
\item {M2:} $95-100\gev \lsim\mhsm\lsim 130\gev$. Detection should be possible
at the Tevatron and the LHC, but not at LEP2. Note
that we are adopting the optimistic conclusion~\cite{mrenna,kky,wmyao}
that the mass range for which detection at TeV33 will 
be viable in the $W\hsm$, $\hsm\to b\anti b$ mode includes the region 
between 120 and 130 GeV, and that up to 130 GeV 
some information can also be extracted
at TeV33 from the $Z\hsm$ mode. 
At the LHC, modes involving $\hsm\to b\anti b$ are 
currently regarded as being quite problematic above 120 GeV.
Of course, $\hsm\to Z\zstar$ and $W\wstar$
decay modes will not yet be significant; the Higgs remains very
narrow. 
\item {M3:} $130\gev\lsim\mhsm\lsim 150-155\gev$. Detection is only
possible at the LHC, $Z\zstar$ and $W\wstar$ decay modes
emerge and become highly viable. Still, the Higgs remains narrow.
\item {M4:} $155\lsim\mhsm\lsim 2\mz$.  The real $WW$ mode turns on,
$Z\zstar$ reaches a minimum at $\mhsm\sim 170\gev$. The inclusive $\gam\gam$
mode is definitely out of the picture. The Higgs starts to get
broad, but $\gamhsm\lsim 1\gev$.
\item {M5:} $\mhsm\gsim 2\mz$. Detection will only be possible
at the LHC, $ZZ$ and $WW$ modes are dominant,
and the Higgs becomes broad enough that a {\it direct}
determination of its width becomes conceivable
by reconstructing the $ZZ\to 4\ell$ final state mass
(probable resolution being of order $1\%\times\mhsm$ at CMS
and $1.5\%\times\mhsm$ at ATLAS).
\end{description}
The possible modes of potential use for determining
the properties of the $\hsm$ at each of
the three machines are listed in Table~\ref{modes}.  Even very
marginal modes are included when potentially crucial
to measuring an otherwise inaccessible Higgs property.
For $\mhsm\gsim 2\mw,2\mz$, we ignore $b\anti b$ decays of the $\hsm$
as having much too small a branching ratio, and 
$t\anti t$ decays are not relevant for $\mhsm\lsim 2\mt$.
Our focus here will be on masses in the $\lsim 400\gev$ range for
which the Higgs is clearly weakly coupled.  We recall that $\mhl< 2\mw$
is expected in supersymmetric models.~\cite{haber,pokorski}

\begin{table}[h]\caption{Modes for $\hsm$ production and observation at
LEP2, Tevatron and the LHC.}
\begin{center}
\begin{tabular}{ll}
\underline{LEP2} & \ \\
{LP1:} $\epem\to \zstar\to Z\hsm\to Z b\anti b$ &
{LP2:} $\epem\to \zstar\to Z\hsm\to Z \tauptaum$ \\
{LP3:} $\epem\to \zstar\to Z\hsm\to Z X$ &
\phantom{{LP3:} $\epem\to \zstar\to Z\hsm\to Z X$} \\
\underline{Tevatron/TeV33}&\ \\
{T1:} $\wstar\to W\hsm\to Wb\anti b$ &
{T2:} $\wstar\to W\hsm\to W\tauptaum$ \\
{T3:} $\zstar\to Z\hsm\to Zb\anti b$ &
{T4:} $\zstar\to Z\hsm\to Z\tauptaum$ \\
\underline{LHC: $\mhsm\lsim 2\mw,2\mz$}& \ \\
{L1:} $gg\to\hsm\to\gamgam$ &
{L2:} $gg\to\hsm\to Z\zstar$ \\
{L3:} $gg\to\hsm\to W\wstar$ & 
{L4:} $WW\to\hsm\to\gamgam$ \\
{L5:} $WW\to\hsm\to Z\zstar$ &
{L6:} $WW\to\hsm\to W\wstar$ \\
{L7:} $\wstar\to W\hsm\to W\gam\gam$ &
{L8:} $\wstar\to W\hsm\to W b\anti b$ \\
{L9:} $\wstar\to W\hsm\to W\tauptaum$ &
{L10:} $\wstar\to W\hsm \to WZ\zstar$ \\
{L11:} $\wstar\to W\hsm \to WW\wstar$ &
{L12:} $t\anti t \hsm\to t\anti t \gam\gam$ \\
{L13:} $t\anti t \hsm\to t\anti t b\anti b$   &
{L14:} $t\anti t \hsm \to t\anti t \tauptaum$  \\
{L15:} $t\anti t \hsm \to t\anti t Z\zstar$ &
{L16:} $t\anti t \hsm \to t\anti t W\wstar$ \\
\underline{LHC: $\mhsm\gsim 2\mw,2\mz$} & \ \\
{H1:} $gg\to\hsm\to ZZ$&
{H2:} $gg\to\hsm\to WW$\\
{H3:} $WW\to\hsm\to ZZ$&
{H4:} $WW\to\hsm\to WW$\\
{H5:} $\wstar\to W\hsm\to WWW$&
{H6:} $\wstar\to W\hsm\to W ZZ$\\
\end{tabular}
\end{center}
\label{modes}
\end{table}

Of the listed modes, the reactions that clearly allow $\hsm$ discovery and
that have proven or likely potential for measuring $\hsm$ properties 
in the M1, M2, M3, M4 and M5 mass regions are the following.
\begin{description}
\item{M1:} LP1, LP2, LP3, T1, T2, T3, T4, L1, L7, L8, L12, L13.
\item{M2:} T1, T2, T3, T4, L1, L7, L8, L12, L13.
\item{M3:} L1, L2, L3, L7.
\item{M4:} L2, L3.
\item{M5:} H1, H2.
\end{description}
It may be that techniques for employing some of 
the other reactions listed earlier will eventually
be developed, but we do not assume so here.

\noindent\underline{M1}

Rates for reactions LP1, LP3, T1, T3, L1, L7, L8, L12, L13
will be well measured.  Our ability
to observe reactions LP2, T2, T4, L4, L9
and determine with some reasonable accuracy the ratio
of the rates for these reactions to the better measured
reactions and to each other is less certain.
We consider only the well-measured rates to begin with.
\begin{itemize}
\item The rate for LP3 (\ie\ $Z\hsm\to Z X$
with $Z\to \epem,\mupmum$)
determines the $ZZ\hsm$ coupling (squared). 
\item LP1/LP3 gives $\br(\hsm\to b\anti b)$, which can
be checked against the SM prediction, but on
its own does not allow a model-independent determination
of the $\hsm\to b\anti b$ coupling.
\item The ratio T1/LP1 yields the  $(WW\hsm)^2/(ZZ\hsm)^2$ coupling-squared
ratio, and multiplying by the LP3 determination of $(ZZ\hsm)^2$
we get an absolute magnitude for $(WW\hsm)^2$.
\item The ratio T1/T3 gives a second determination of
$(WW\hsm)^2 / (ZZ\hsm)^2$.
\item The ratio T1/$\br(b\anti b)$ gives $(WW\hsm)^2$ and T3/$\br(b\anti b)$
gives $(ZZ\hsm)^2$. 
\item The ratios L7/L8 and L12/L13 yield two independent
determinations of  $\br(\gam\gam)/\br(b\anti b)$. 
Combining with $\br(b\anti b)$ from LEP2 yields
$\br(\gam\gam)$.
\item L1/$\br(\gamgam)$ yields the magnitude of the $(gg\hsm)^2$
coupling-squared, which is primarily sensitive to the $t\anti t\hsm$
coupling. The errors on the L1 rate are different for ATLAS
and CMS; a tabulation appears in Table~\ref{2gamerrors}.

\begin{table}[hbt]
\caption[fake]{We tabulate the approximate error in the determination
of $\sigma(gg\to\hsm)\br(\hsm\to \gam\gam)$ as a function
of $\mhsm$ (in GeV) assuming $L=300\fbi$ (each) for the CMS 
and ATLAS detectors at the LHC.}
\begin{center}
\begin{tabular}{|c|c|c|c|c|}
\hline
 Mass & 90 & 110 & 130 & 150 \\
\hline
CMS Error & $\pm 9\%$ & $\pm 6\%$ & $\pm 5\%$ & $\pm 8\%$ \\
ATLAS Error & $\pm 23\%$ & $\pm 7\%$ & $\pm 7\%$ & $\pm 10\%$ \\
\hline
Combined Error & $\pm 8.5\%$ & $\pm 4.5\%$ & $\pm 4.0\%$ & $\pm 6.2\%$ \\
\hline
\end{tabular}
\end{center}
\label{2gamerrors}
\end{table}

\item L12/L7 and L13/L8 yield independent results
for $(t\anti t\hsm)^2 / (WW\hsm)^2$.
By multiplying by the previously determined value
of $(WW\hsm)^2$ we get an absolute magnitude for the $(t\anti t\hsm)^2$
coupling-squared which can be checked against the $gg\hsm$ result.
\end{itemize}
Error expectations are tabulated in Table~\ref{m1errors} for $\mhsm\sim \mz$.

\begin{table}[hbt]
\caption[fake]{Summary of approximate
errors for branching ratios and couplings-squared
at $\mhsm\sim \mz$ in the M1 mass region.
Where appropriate, estimated systematic errors are included.
Quantities not listed cannot be determined in a model-independent manner.
Directly measured products of couplings-squared times
branching ratios can often be determined with better accuracy.}
\begin{center}
\begin{tabular}{|c|c|}
\hline
 Quantity & Error \\
\hline
 $\br(b\anti b)$ & $\pm 26\%$ \\
\hline
 $(WW\hsm)^2/(ZZ\hsm)^2$ & $\pm 14\%$ \\
\hline
 $(WW\hsm)^2$ & $\pm 20\%$ \\
\hline
 $(ZZ\hsm)^2$ & $\pm 22\%$ \\
\hline
 $(\gam\gam\hsm)^2/(b\anti b\hsm)^2$ & $\pm 17\%$ \\
\hline
 $\br(\gam\gam)$ & $\pm 31\%$ \\
\hline
 $(gg\hsm)^2$ & $\pm 31\%$ \\
\hline
 $(t\anti t\hsm)^2/(WW\hsm)^2$ & $\pm 21\%$ \\
\hline
 $(t\anti t\hsm)^2$ & $\pm 30\%$ \\
\hline
\end{tabular}
\end{center}
\label{m1errors}
\end{table}

What is missing from the list is any determination
of the $(b\anti b \hsm)$, $(\tau\tau\hsm)$ and $(\gamgam\hsm)$ couplings,
any check that fermion couplings are proportional to the
fermion mass (other than the $(t\anti t\hsm)$ coupling magnitude),
and the Higgs total width. Given the $(WW\hsm)$ and $(t\anti t\hsm)$
couplings we could compute the expected value for the $(\gamgam\hsm)$
coupling, and combine the $\Gamma(\hsm\to\gam\gam)$
computed therefrom with $\br(\gamgam)$ to get
a value for $\gamhsm$.  $\br(b\anti b)\gamhsm$ then yields
$(b\anti b\hsm)^2$ and we could thereby indirectly check
that $b\anti b\hsm/t\anti t\hsm=\mb/\mt$. Some systematic
uncertainty in the correct values of $\mb$ and $\mt$
would enter into this check, but the propagation of 
the already rather significant statistical
errors would be the dominant uncertainty.

\noindent\underline{M2}

Rates for reactions T1, T3, L1, L7, L8, L12, L13
will be well measured.  Reactions T2, T4
are less robust. Relative to mass region M1, we suffer
the crucial loss of a measurement of the $(ZZ\hsm)^2$
squared coupling constant.
Considering the well-measured rates, we should be able to
determine the following quantities.
\begin{itemize}
\item 
The ratio T1/T3 gives a determination of
$(WW\hsm)^2/(ZZ\hsm)^2$. 
\item 
The ratios L7/L8 and L12/L13 yield two independent
determinations of  $\br(\gam\gam)/\br(b\anti b)$.  At the moment we
can only estimate the accuracy of the L12/L13 determination
of $\br(\gam\gam)/\br(b\anti b)$.
\item 
\begin{sloppypar}
L12/L7 and L13/L8 yield independent determinations of
$(t\anti t\hsm)^2 / (WW\hsm)^2$.
However, L8 is dubious, so only results for L12/L7 are reliable.
\end{sloppypar}
\end{itemize}
Thus, we will have ways of determining the
$(WW\hsm):(ZZ\hsm):(t\anti t\hsm)$  coupling ratios, but no absolute
coupling magnitudes are directly determined, and there is no test
of the fermion-Higgs coupling being proportional to fermion
mass. 

To proceed further, requires more model input.  Given that
we know (in the SM) how to compute $\br(\gamgam)$
from the $WW\hsm$ and $t\anti t\hsm$ couplings, and given
that we know the ratio of the latter, $\br(\gamgam)/\br(b\anti b)$
would yield a result for $(t\anti t\hsm)/(b\anti b\hsm)$
which could then be checked against the predicted $\mt/\mb$.

We summarize as a function of $\mhsm$ in Table~\ref{m2errors} 
the errors for the few coupling-squared ratios that
can be determined in the M2 mass region.

\begin{table}[hbt]
\caption[fake]{Summary of approximate errors for coupling-squared ratios
at $\mhsm=100,110,120,130\gev$ in the M2 mass region.
As discussed in the text, directly measured products of couplings-squared times
branching ratios can often be determined with better accuracy.}
\begin{center}
\begin{tabular}{|c|c|c|c|c|}
\hline
 Quantity & \multicolumn{4}{c|}{Errors} \\
\hline
\hline
 Mass (GeV) & 100 & 110 & 120 & 130 \\
\hline
 $(WW\hsm)^2/(ZZ\hsm)^2$ & $\pm 23\%$ & $\pm 26\%$ & $\pm 34\%$ & $-$ \\
\hline
 $(\gam\gam\hsm)^2/(b\anti b\hsm)^2$ & $\pm 17\%$ & $\pm 19\%$ & $\pm 22\%$ &
 $\pm 25\%$ \\
\hline
 $(t\anti t\hsm)^2/(WW\hsm)^2$ & $\pm 21\%$ & $\pm 21\%$ & $\pm 21\%$ &
 $\pm 21\%$  \\
\hline
\end{tabular}
\end{center}
\label{m2errors}
\end{table}

\noindent\underline{M3}

Of the potential channels listed under M3, only L1 and L2
are thoroughly studied and certain to be measurable over
this mass interval. L1 should be viable for $\mhsm\lsim
150\gev$. L2 (the $gg\to \hsm\to
Z\zstar$ reaction) should be good for $\mhsm\gsim 130\gev$.
Errors are tabulated in Table~\ref{4lerrors}.
With these two modes alone, we discover the Higgs, 
and for $130\lsim \mhsm\lsim 150\gev$ we can determine
$\br(\gamgam)/\br(Z\zstar)$.  

The errors for $(\gam\gam\hsm)^2/(ZZ\hsm)^2$
deriving from the L1/L2 ratio appear in the summary Table~\ref{m3errors}.
This ratio is interesting, but cannot be unambiguously 
interpreted.

\begin{table}[hbt]
\caption[fake]{We tabulate the error in the determination
of $\sigma(gg\to\hsm)\br(\hsm\to 4\ell)$ as a function
of $\mhsm$ (in GeV) assuming $L=600\fbi$ at the LHC.}
\begin{center}
\begin{tabular}{|c|c|c|c|c|c|}
\hline
 Mass & 120 & 130 & 150 & 170 & 180 \\
 Error & $\pm 25\%$ & $\pm 9.5\%$ & $\pm 5.3\%$ & $\pm 11\%$ & $\pm 6.1\%$ \\
\hline
 Mass & 200 & 220 & 240 & 260 & 280 \\
 Error & $\pm 7.8\%$ & $\pm 6.9\%$ & $\pm 6.2\%$ & $\pm6.2\%$ & $\pm6.2\%$ \\
\hline
 Mass & 300 & 320 & 340 & 360 & 380 \\
 Error & $\pm 6.2\%$ & $\pm6.2\%$ & $\pm 6.1\%$ & $\pm 6.0\%$  & $\pm 6.4\%$ \\
\hline
 Mass & 400 & 500 & 600 & 700 & 800 \\
 Error & $\pm 6.7\%$ & $\pm 9.4\%$ & $\pm 14\%$ & $\pm 20\%$ & $\pm28\%$ \\
\hline
\end{tabular}
\end{center}
\label{4lerrors}
\end{table}

The L3 mode was first examined in detail~\cite{gloveretal,hanetal} 
some time ago. It was 
found that with some cuts it might be possible to dig out 
a signal in the $\ell\nu\ell\nu$ decay mode of the $W\wstar$ final state.
A more recent study~\cite{ditdr} 
focusing on the $\mhsm\gsim 155\gev$ mass region
finds that additional cuts are necessary in the context of a more
complete simulation, but that very promising $S/\sqrt B$ can
be obtained. For the M3 mass region we employ
a rough extrapolation into the $130-150\gev$ mass region
of these results by simply using the mass dependence of $\br(\hsm\to W\wstar)$.
Expected errors for L3 appear in Table~\ref{ggwwstarerrors}.

\begin{sloppypar}
The resulting statistical $(WW\hsm)^2/(ZZ\hsm)^2$ 
errors are tabulated in Table~\ref{m3errors}. Apparently
L3/L2 will provide a decent measurement of the $(WW\hsm)^2 / (ZZ\hsm)^2$ 
coupling-squared ratio, thereby allowing a check
that custodial SU(2) is operating, so long as the systematic error
is $\lsim 10\%$.
\end{sloppypar}

\begin{table}[hbt]
\caption[fake]{We tabulate the statistical error in the determination
of $\sigma(gg\to\hsm\to W\wstar)$ as a function
of $\mhsm$ (in GeV) assuming $L=600\fbi$ at the LHC. For $\mhsm\leq150\gev$,
the errors are based on extrapolation from $\mhsm\geq 155\gev$ results.}
\begin{center}
\begin{tabular}{|c|c|c|c|c|c|}
\hline
 Mass & 120 & 130 & 140 & 150 & $155-180$ \\
 Error & $\pm 12\%$ &  $\pm 6\%$ &  $\pm 3\%$ &  $\pm 3\%$ & $\pm 2\%$ \\
\hline
\end{tabular}
\end{center}
\label{ggwwstarerrors}
\end{table}

\begin{table}[hbt]
\caption[fake]{We tabulate the statistical errors at $\mhsm=120,130,150\gev$
in the determinations
of $(\gam\gam\hsm)^2/(ZZ\hsm)^2$ and $(WW\hsm)^2/(ZZ\hsm)^2$,
assuming $L=600\fbi$ at the LHC.}
\begin{center}
\begin{tabular}{|c|c|c|c|}
\hline
 Quantity & \multicolumn{3}{c|}{Errors} \\
\hline
\hline
Mass (GeV) & 120 & 130 & 150 \\
\hline
 $(\gam\gam\hsm)^2/(ZZ\hsm)^2$
& $\pm 25\%$ & $\pm 11\%$ & $\pm 10\%$ \\
\hline
 $(WW\hsm)^2/(ZZ\hsm)^2$
 & $\pm 27\%$ & $\pm 11\%$ & $\pm 6\%$ \\
\hline
\end{tabular}
\end{center}
\label{m3errors}
\end{table}
                                     
\noindent\underline{M4}

Let us now turn to the $155\lsim\mhsm\lsim 2\mz$ mass region.
The most significant variation in this region arises due to
the fact that as $\hsm\to WW$ becomes kinematically
allowed at $\mhsm\sim 160\gev$, the $\hsm\to Z\zstar$ branching ratio
dips, the dip being almost a factor of 4 at $\mhsm=170\gev$.
L2 can still be regarded as iron-clad throughout this region
provided adequate $L$ is accumulated.
For $L=600\fbi$, an accurate measurement of $(gg\hsm)^2\br(\hsm\to Z\zstar)$
is clearly possible; results were already tabulated in Table~\ref{4lerrors}.

L3 is now an on-shell $WW$ final state, and,
according to the results summarized in Table~\ref{ggwwstarerrors},
can be measured with good statistical accuracy in the $\ell\nu\ell\nu$
final state of the $\hsm\to WW$ Higgs decay.
The statistical accuracy for $(WW\hsm)^2/(ZZ\hsm)^2$ deriving
from L3/L2 is tabulated in Table~\ref{m4errors}.
The error on the L3/L2 determination of $(WW\hsm)^2/(ZZ\hsm)^2$
in the M4 mass region is dominated by that for the $4\ell$
channel (tabulated in Table~\ref{4lerrors}).

\begin{table}[hbt]
\caption[fake]{We tabulate the statistical errors at $\mhsm=155,170,180\gev$
in the determination of $(WW\hsm)^2/(ZZ\hsm)^2$ from L3/L2,
assuming $L=600\fbi$ at the LHC.}
\begin{center}
\begin{tabular}{|c|c|c|c|}
\hline
 Quantity & \multicolumn{3}{c|}{Errors} \\
\hline
\hline
Mass (GeV) & 155 & 170 & 180 \\
\hline
 $(WW\hsm)^2/(ZZ\hsm)^2$ & $\pm6\%$ & $\pm 11\%$ & $\pm 7\%$ \\
\hline
\end{tabular}
\end{center}
\label{m4errors}
\end{table}

\noindent\underline{M5} 

Finally we consider $\mhsm\gsim 2\mz$.
The first important remark is that $\gamhsm$ becomes
measurable in the $4\ell$ channel once $\gamhsm\gsim (1\%-1.5\%)\times\mhsm$,
which occurs starting at $\mhsm\sim 200\gev$ where $\gamhsm\sim 2\gev$.
At $\mhsm=210$, $250$, $300$, and $400\gev$,
rough percentage error expectations (assuming $L=600\fbi$ for ATLAS+CMS)
for $\gamhsm$ are $\pm 21\%$, $\pm 7\%$, $\pm 4\%$ and $\pm 3\%$,
respectively. Additional discussion is given later.

Only H1 is gold-plated, and of course
it alone provides very limited information about the actual
Higgs properties. As described for the M4 mass region,
the mode H2 has been studied for $\mhsm$ in the vicinity of $2\mz$
in the $\ell\nu\ell\nu$ final state.~\cite{gloveretal,hanetal,ditdr}
These results indicate that reasonable to good accuracy for the H2/H1 ratio,
implying a reasonably accurate implicit
determination of $(WW\hsm)^2/(ZZ\hsm)^2$,
might be possible for Higgs masses not too far above $2\mz$.
One could also ask if it would be possible to separate out the $WW$
final state in the $\ell\nu jj$ mode where a mass peak could be
reconstructed (subject to the usual two-fold ambiguity procedures).
Event rates would be quite significant, and a Monte Carlo 
study should be performed. 

Processes H3 and H4 would have to be separated from H1 and H2
using spectator jet tagging to isolate the former $WW$
fusion reactions. If this were possible, then
H3/H1 and H4/H2 would both yield a
determination of $(t\anti t\hsm)^2/(WW\hsm)^2$ under
the assumption that the $t$-loop dominates the $(gg\hsm)$ coupling.
However, the mass range for which separation of H3 and H4 from H1 and H2
would be possible
is far from certain.~\footnote{A recent study \cite{wwpoggioli} has shown
that forward jet tagging allows isolation of H4 in the $\ell\nu jj$ final state
for $\mhsm\gsim 600\gev$ (\ie\ beyond
the mass range being explicitly considered here), but suggests that
the $W$+jets background is difficult to surmount for lower masses.
However, strategies in the mass range down near $2\mz$ could be quite
different given the much larger signal rates.}

\subsubsection{Impact of LEP2, Tevatron and LHC measurements
for {\boldmath $\hl$} vs. {\boldmath $\hsm$} discrimination}
\label{lhcimpactss}

LEP2, the Tevatron and the LHC will certainly allow
detection of a SM-like $\hl$ of a supersymmetric model or a
light $\hsm$ of the SM. However, the errors listed for important
coupling ratios in Tables~\ref{m1errors}, \ref{m2errors}, and \ref{m3errors}
(those tables relevant for the $\mhl< 150\gev$ light supersymmetric
Higgs mass range) are large.  As discussed
in more detail in the following section,
discrimination between the $\hl$ and $\hsm$ will require rather
high accuracy for coupling
ratio measurements unless the parameters of the Higgs sector are 
far from the decoupling limit.
Thus, for this group of accelerators,
direct detection at the LHC of the heavier Higgs bosons (which,
as reviewed later, might be possible but is certainly not guaranteed)
could be the only means of establishing that nature has
chosen an extended (\eg\ supersymmetric) Higgs sector.

\subsection{NLC and $s$-channel FMC data}\label{ssnlcfmc}

Two different situations and corresponding sets of measurements
are relevant:
\begin{itemize}
\item
measurements that would be performed
by running at $\rts= 500\gev$ 
at the NLC (or in NLC-like running at the FMC) ---
the production modes of interest are 
$\epem\to Z\hsm$, $\epem\to\epem\hsm$ ($ZZ$-fusion)
and $\epem\to\nu\anti\nu\hsm$ ($WW$-fusion);~\footnote{In the following,
we will consistently use the notation $\epem\hsm$ and $\nu\anti\nu\hsm$
for the $ZZ$ fusion and $WW$ fusion contributions to these final state channels
only. The contributions to these same final states from $Z\hsm$ with
$Z\to\epem$ and $Z\to\nu\anti\nu$, respectively, and interference
at the amplitude level with the $ZZ$ and $WW$ fusion graphs is
presumed excluded by appropriate cuts requiring that the $\epem$
or $\nu\anti\nu$ reconstructed mass not be near $\mz$.}
\item
measurements performed in $s$-channel production at the FMC ---
the production mode being $\mupmum\to\hsm$.
\end{itemize}
In the first case, we presume that $L=200\fbi$ is available
for the measurements at $\rts=500\gev$. (Such operation at a FMC,
would only be appropriate if the NLC has not been constructed
or is not operating at expected instantaneous luminosity.)
Many new strategies developed~\cite{rickjack} for $\rts=500\gev$ running
are detailed in the Higgs96 report~\cite{snowmasssummary}
and are very briefly reviewed here.
In the second case, we implicitly presume that the NLC is
already in operation, so that a repetition 
of $\rts=500\gev$ data collection would not be useful and
devoting all the FMC luminosity to $s$-channel Higgs
production would be entirely appropriate.  The errors we quote
in this second case will be those obtained if $L=200\fbi$
is devoted to a scan of the Higgs peak (in the $s$-channel)
that is optimized for the crucial measurement
of $\gamhsm$; this scan requires
devoting significant luminosity to the wings of the peak (see later
discussion). Results presented in this case are largely from
the FMC report.~\cite{bbgh}

\subsubsection{Measuring 
{\boldmath $\sigma\br(\hsm\to c\anti c, b\anti b,W\wstar)$} at
the NLC}\label{sssnlcccbbww}

The accuracy with which cross section times branching ratio
can be measured in various channels will prove to be vitally important
in determining the branching ratios themselves and, ultimately,
the total width and partial widths of the Higgs boson, which are 
its most fundamental properties.  In addition, the ratios
\begin{equation}
{\sigma\br(\hsm\to c\anti c)\over \sigma\br(\hsm\to b\anti b)}\,,~~~
{\sigma\br(\hsm\to W\wstar)\over \sigma\br(\hsm\to b\anti b)}
\label{ratios}
\end{equation}
will themselves be a sensitive probe of deviations from SM predictions
to the extent that SM values for these branching ratios can
be reliably computed (see later discussion).
It should be noted that the $c\anti c$ and $W\wstar$ modes are
complementary in that for $\mhsm\lsim 130\gev$ only the $c\anti c$
mode will have good measurement accuracy, while for $\mhsm\gsim 130\gev$
accuracy in the $W\wstar$ mode will be best.

\begin{figure}[p]
\leavevmode
\begin{center}
\centerline{\psfig{file=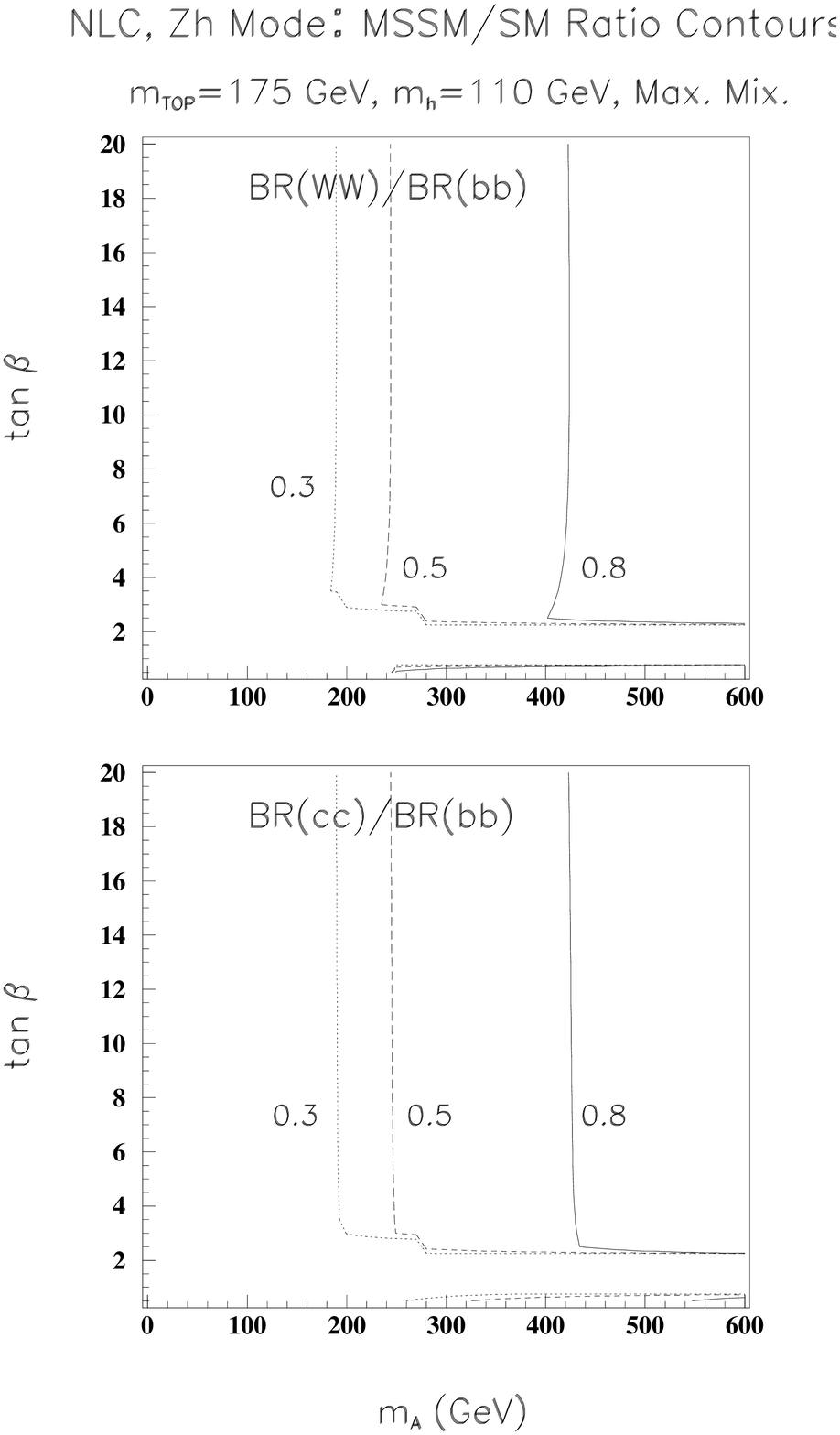,width=3.1in}}
\caption{Constant value contours in $(\mha,\tanb)$ parameter space
for the ratios $[W\wstar/b\anti b]_{\hl}/[W\wstar/b\anti b]_{\hsm}$  and
$[c\anti c/b\anti b]_{\hl}/[c\anti c/b\anti b]_{\hsm}$.
We assume ``maximal mixing'' in the squark sector and present
results for the case of fixed $\mhl=\mhsm=110\gev$. The band extending
out to large $\mha$ at $\tanb\sim 2$ is where $\mhl=110\gev$
is theoretically disallowed in the case of maximal mixing.
For no mixing, see DPF95,~\protect\cite{dpfreport} the vertical contours
are essentially identical --- only the size of the disallowed band
changes.}
\label{figdevsm}
\end{center}
\end{figure}

The $\hl$ of the MSSM
provides a particularly useful testing ground for the accuracy
with which the above ratios must be determined
in order that such deviations be detectable.  As $\mha$ increases,
the $\hl$ becomes increasingly SM-like.  
The DPF95 Higgs survey~\cite{dpfreport} and further work performed
for Higgs96,~\cite{snowmasssummary,gdev}
shows that the $c\anti c$, $b\anti b$
and $WW^\star$ partial widths and ratios of branching ratios
provide sensitivity to $\hl$ vs. $\hsm$ deviations out to higher
values of $\mha$ than any others.  In particular, the $c\anti c/b\anti b$
and $W\wstar/b\anti b$ ratio deviations essentially depend only upon $\mha$
for $\mha\gsim \mz$,
and are quite insensitive to details of squark mixing and so forth.
To illustrate, we present in Fig.~\ref{figdevsm} 
the ratio of the MSSM prediction to the SM prediction
for these two ratios taking $\mhl=110\gev$ (held fixed, implying
variation of stop masses as $\mha$ and $\tanb$ are changed)
and assuming ``maximal mixing'' in the stop sector
(as defined in the European $\epem$ study~\cite{eurostudy}
and the DPF95 report~\cite{dpfreport}). Results 
are presented using contours in
the $(\mha,\tanb)$ parameter space. Aside from an enlargement
of the allowed parameter space region, the ``no mixing'' scenario
contours are essentially the same. Results for larger $\mhl$ are very similar
in the allowed portion of parameter space. We observe that 
it is necessary to detect deviations in the ratios at the level
of 20\% in order to have sensitivity at the $>1\sigma$ level
up to $\mha\sim 400\gev$.
For a Higgs mass as small as $\mhl=110\gev$, only the $c\anti c$
branching ratio has a chance of being measured with reasonable accuracy
at the NLC. The $W\wstar$ branching ratio will inevitably 
be poorly measured for the $\hl$ of the MSSM if
stop squark masses are $\lsim 1\tev$, implying $\mhl\lsim 130\gev$.
In non-minimal supersymmetric models the lightest Higgs can, however,
be heavier and the $W\wstar$ branching ratio would then prove useful.

There are both experimental and theoretical
sources of uncertainty for the branching-ratio ratios
of Eq.~(\ref{ratios}). The primary theoretical uncertainty
is that associated with knowing the running $b$ and $c$ quark masses
at the Higgs mass scale. As reviewed in Higgs96,~\cite{snowmasssummary}
errors for masses
obtained via QCD sum rules and lattice calculations are
getting small and will improve 
significantly by the time the NLC is operating.~\cite{narison,shigemitsu}
For given input masses, the running-mass and other QCD corrections
to decay widths are under good control.~\cite{djouadi}
It now seems reasonable to suppose
that an accuracy of better  than $\pm 10\%$ can be achieved for the theoretical
computations of the ratios of Eq.~(\ref{ratios}).

New estimates~\cite{snowmasssummary}
for the experimental accuracy with which the separate event
rates for $Z\hsm$ production with $\hsm$ decaying to $b\anti b$,
$c\anti c$ and $W\wstar$ have been made based on
employing topological jet tagging in which
$b$-jets are identified by a secondary and a tertiary vertex (in addition
to the primary event vertex), while a $c$-jet should have only a secondary
vertex and a primary vertex, and a light quark or gluon jet only the primary
vertex. Extraordinary purities and efficiencies for each class of events
are possible for a typical NLC detector.~\cite{nlcreport,nlc} Errors
for individual channel event rates and ratios are remarkably small.
Similar results for the errors for individual $\hsm$ decay channel
rates are obtained~\cite{snowmasssummary} in the $WW$-fusion
$\nu\anti\nu\hsm$ and $ZZ$-fusion $\epem\hsm$ production modes. 
Errors obtained by combining results from all three production modes
are tabulated later in our final summary table, Table~\ref{nlcerrors}.

For $\mhsm\lsim 130\gev$, only
the $c\anti c$ and $b\anti b$ channel rates are measured with high accuracy.
If the net statistical error (from $Z\hsm$, $\epem\hsm$ and $\nu\anti\nu\hsm$
production) for $c\anti c/b\anti b$ is combined
in quadrature with a
$\lsim\pm 10\%$ systematic error in the theoretical
calculation, we arrive at a net 
error of $\lsim 12\%$. Fig.~\ref{figdevsm} shows that this would allow 
differentiation of the $\hl$ from the $\hsm$
at the $2\sigma$ level out to $\mha\sim 450\gev$. This is a very encouraging
result.  The dominance of the theoretical error 
indicates the high priority of obtaining theoretical predictions for 
$c\anti c/b\anti b$ that are as precise as possible. 
Overall, precision $\hl$ measurements at $\rts=500\gev$ with $L=200\fbi$
appear to have a good chance of probing the heavier Higgs
mass scale (which is related to important SUSY-breaking parameters)
even when the heavier Higgs bosons cannot be (pair) produced
without going to higher energy.

Moving to higher masses, 
we~\cite{rickjack} combine the $Z\hsm$ and $\nu\anti\nu\hsm$ channel 
results, and obtain accuracies for $\br(W\wstar)/\br(b\anti b)$
as given in Table~\ref{nlcerrors}. (We have not
pursued the degree to which these errors would be further reduced by including
the $\epem\hsm$ channel determination of this ratio.)
Fig.~\ref{figdevsm} (which is fairly
independent of the actual $\mhl$ value aside from the extent
of the allowed parameter region) implies that a $\lsim 10\%$ error,
as achieved for $\mhsm$ in the $140-150\gev$ mass range, would be a very
useful level of accuracy in the MSSM should stop masses (contrary
to expectations based on naturalness) be sufficiently above 1 TeV
to make $\mhl=140-150\gev$ possible.
In the NMSSM, where the lightest Higgs (denoted $\h_1$)
can have mass $\mhi\sim 140-150\gev$, even if stop masses
are substantially below 1 TeV, deviations from SM expectations
are typically even larger.  In general, the $W\wstar/b\anti b$ ratio will
provide an extremely important probe of a non-minimal Higgs sector
whenever the $b\anti b$ and $W\wstar$ decays of the Higgs both have substantial
branching ratio.

\subsubsection{Measuring {\boldmath $\sigma(\mupmum\to\hsm)\br(\hsm\to b\anti
b,W\wstar,Z\zstar)$} in {\boldmath $s$}-channel FMC production}\label{sssfmcccbbww}

The accuracies expected for these measurements were determined~\cite{bbgh} 
under the assumption
that the relevant detector challenges associated with detecting
and tagging final states in the potentially harsh FMC environment can be met.
If $L=200\fbi$ is used so as to optimize the Higgs peak scan
determination of $\gamhsm$, then the equivalent $\rts=\mhsm$ Higgs 
peak luminosity accumulated 
for measuring $\sigma(\mupmum\to\hsm)\br(\hsm\to X)$
in various channels is less, roughly of order $L=50\fbi$. The associated errors 
expected for $\sigma(\mupmum\to\hsm)\br(\hsm\to
b\anti b, W\wstar,Z\zstar)$ are summarized as a function of $\mhsm$
in Table~\ref{fmcsigbrerrors}. As is apparent from the
table, the errors are remarkably small for $\mhsm\lsim 150\gev$.
As already stated, detector performance in the FMC environment
will be critical to whether or not such small errors can be achieved
in practice. As an example, to achieve the good $b$-tagging
efficiencies and purities employed in obtaining the NLC detector errors
given in this report, a relatively clean environment is required
and it must be possible to get as close as 1.5 cm to the beam.
FMC detectors discussed to date do not allow for instrumentation this
close to the beam. More generally, in all the channels
it is quite possible that the FMC
errors will in practice be at least in the few per cent range.

The errors summarized in Table~\ref{fmcsigbrerrors} lead to the
errors for coupling-squared ratios later summarized in Table~\ref{fmcerrors}.
The level of precision achieved would be very valuable
for distinguishing between the $\hsm$ and a supersymmetric $\hl$.  
Note, in particular, that the $W\wstar/b\anti b$ branching-ratio ratio
is well-measured for Higgs masses even as low as $100\gev$.
For $\mh=110\gev$, Fig.~\ref{figdevsm} shows that even if
we triple the $W\wstar/b\anti b$ 
error of Table~\ref{fmcerrors} to $\sim\pm 5\%$,
the $\hl$ of the MSSM can be distinguished from the SM $\hsm$
at the $ \geq 4\sigma$ level for $\mha\leq 400\gev$.

\begin{table}[h]
\caption[fake]{Summary of approximate errors for
$\Gamma(\hsm\to\mupmum)\br(\hsm\to b\anti b, W\wstar, Z\zstar)
\propto (\mupmum\hsm)^2\br(\hsm\to b\anti b, W\wstar, Z\zstar)$,
assuming $L=50\fbi$ devoted to $\rts=\mhsm$ and beam energy resolution
of $R=0.01\%$.}
\footnotesize
\begin{center}
\small
\begin{tabular}{|c|c|c|c|c|c|}
\hline
 Channel & \multicolumn{5}{c|}{Errors} \\
\hline
\hline
{$\bf\mhsm$}{\bf (GeV)} & {\bf 80} & {\bf 90} & {\bf 100} & {\bf 110} & {\bf 120} \\
\hline
$b\anti b $ & 
$\pm 0.2\%$ & $\pm 1.6\%$ & $\pm 0.4\%$ & $\pm 0.3\%$ & $\pm 0.3\%$ \\
\hline
$W\wstar $ &
$-$ & $-$ & $\pm 3.5\%$ & $\pm 1.5\%$ & $\pm 0.9\%$ \\
\hline
$Z\zstar $ &
$-$ & $-$ & $-$ & $\pm 34\%$ & $\pm 6.2\%$ \\
\hline
\hline
{$\bf\mhsm$}{\bf (GeV)} & {\bf 130} & {\bf 140} & {\bf 150} & {\bf 160} & {\bf 170} \\
\hline
$b\anti b $ & 
$\pm 0.3\%$ & $\pm 0.5\%$ & $\pm 1.1\%$ & $\pm 59\%$ & $-$ \\
\hline
$W\wstar $ &
$\pm 0.7\%$ & $\pm 0.5\%$ & $\pm 0.5\%$ & $\pm 1.1\%$ & $\pm 9.4\%$ \\
\hline
$Z\zstar $ &
$\pm 2.8\%$ & $\pm 2.0\%$ & $\pm 2.1\%$ & $\pm 22\%$ & $\pm 34\%$ \\
\hline
\hline
{$\bf\mhsm$}{\bf (GeV)} & {\bf 180} & {\bf 190} & {\bf 200} & {\bf 210} & {\bf 220} \\
\hline
$W\wstar $ &
 $\pm 18\%$ & $\pm 38\%$ & $\pm 58\%$ & $\pm 79\%$ & $-$ \\
\hline
$Z\zstar $ &
$\pm 25\%$ & $\pm 27\%$ & $\pm 35\%$ & $\pm 45\%$ & $\pm 56\%$ \\
\hline
\end{tabular}
\end{center}
\label{fmcsigbrerrors}
\end{table}

\subsubsection{Measuring {\boldmath $\sigma\br(\hsm\to\gam\gam)$}
at {\boldmath $\protect\rts=500\gev$}}\label{sssbrgamgam}

It turns out that a determination of $\br(\hsm\to\gam\gam)$
is required for extracting $\gamhsm$ in the $\mhsm\lsim 130\gev$
mass range in the absence
of a direct scan determination at the FMC.~\cite{dpfreport}
Of course, $\br(\hsm\to\gam\gam)$ and especially
$\Gamma(\hsm\to\gam\gam)$ are of special interest themselves in that
the $\gam\gam\hsm$ coupling is sensitive to one-loop graphs 
involving arbitrarily heavy states (that get their mass from the $\hsm$
sector vev --- to be contrasted with, for example, heavy SUSY partner states 
which decouple since they get mass from explicit SUSY breaking).

At the NLC, the only means of getting at $\br(\hsm\to\gam\gam)$
is to first measure $\sigma\br(\hsm\to\gam\gam)$ in all accessible
production modes. This has been studied for the $Z\hsm$ and
$\nu\anti\nu\hsm$ ($WW$-fusion) production modes.~\cite{gm}
The best errors for $\rts=500\gev$ running are obtained in the $WW$-fusion
mode, but $Z\hsm$ mode errors are not so much larger.
Since errors for $\sigma(Z\hsm)$ and $\sigma(\nu\anti\nu\hsm)$ are much smaller
than the $\sigma\br(\hsm\to\gam\gam)$ errors, it is appropriate to
combine the $\sigma\br(\hsm\to\gam\gam)$ statistical errors in the two
channels to obtain the net, or effective, 
error expected for $\br(\hsm\to\gam\gam)$.
Assuming a calorimeter at the
optimistic end of current plans for the NLC detector, 
the net $\br(\hsm\to\gam\gam)$ error ranges from
$\sim\pm 22\%$ at $\mhsm=120\gev$
to $\sim\pm 35\%$ ($\sim\pm 53\%$) at $\mhsm=150\gev$ ($70\gev$).
In the $100\lsim \mhsm\lsim 140\gev$ mass 
region, the error is smallest and lies in the $\pm22\%-\pm 27\%$ range.

Due to these large errors, we will combine the NLC determination
of $\br(\hsm\to\gam\gam)$ with that available via
an indirect procedure in which LHC $\sigma\br(\hsm\to\gam\gam)$
measurements are combined with NLC measurements of
the couplings entering into the corresponding LHC $\sigma$'s.
The indirect determination of $\br(\hsm\to\gam\gam)$ turns out
to be substantially more accurate than the direct measurement
at the NLC. Quoted
errors in the summary Table~\ref{nlcerrors} will reflect the
combined error. This is important since the errors for $\br(\hsm\to\gam\gam)$
will dominate in computing some important quantities that
potentially allow discrimination between the SM Higgs boson and a SM-like
Higgs boson of an extended model.

\subsubsection{Determining the {\boldmath $ZZ\hsm$}
 coupling at the NLC}\label{ssszzh}

Determination of the $(ZZ\hsm)^2$ coupling-squared is possible in two modes.
These are (using $\epem$ collision notation):
\begin{itemize}
\item
$\epem\to Z\hsm$, where $Z\to \ell^+\ell^-$ ($\ell=e,\mu$);
\item
$\epem\to\epem \hsm$ (via $ZZ$-fusion).~\cite{ghs}
\end{itemize}
Results presented here for the $ZZ$-fusion channel are preliminary.
It is convenient to separate $Z\hsm$ and $ZZ$-fusion
for the purposes of discussion
even though in the $\epem\hsm$ final state there is some interference
between the $ZZ$-fusion and $Z\hsm$ diagrams. Experimentally this
separation is easily accomplished by an appropriate cut on the $\epem$
pair mass.~\footnote{When $ZZ$-fusion dominates the 
$\zstar\to Z\hsm$ diagrams, such a cut, requiring $M_{\epem}\slash\sim\mz$,
usually improves $S/\sqrt B$ and reduces the $\sqrt{S+B}/S$ error.}
In both channels, the $\hsm$ is inclusively isolated by examining the recoil
mass spectrum computed using the incoming $\epem$ momentum
and the momenta of the outgoing leptons. 
The error estimates of Higgs96~\cite{snowmasssummary} summarized
here will assume momentum resolution such
that the recoil mass peak is sufficiently narrow that backgrounds
are small and can be neglected in the limit of large luminosity.

\begin{figure}[htb]
\leavevmode
\begin{center}
\centerline{\psfig{file=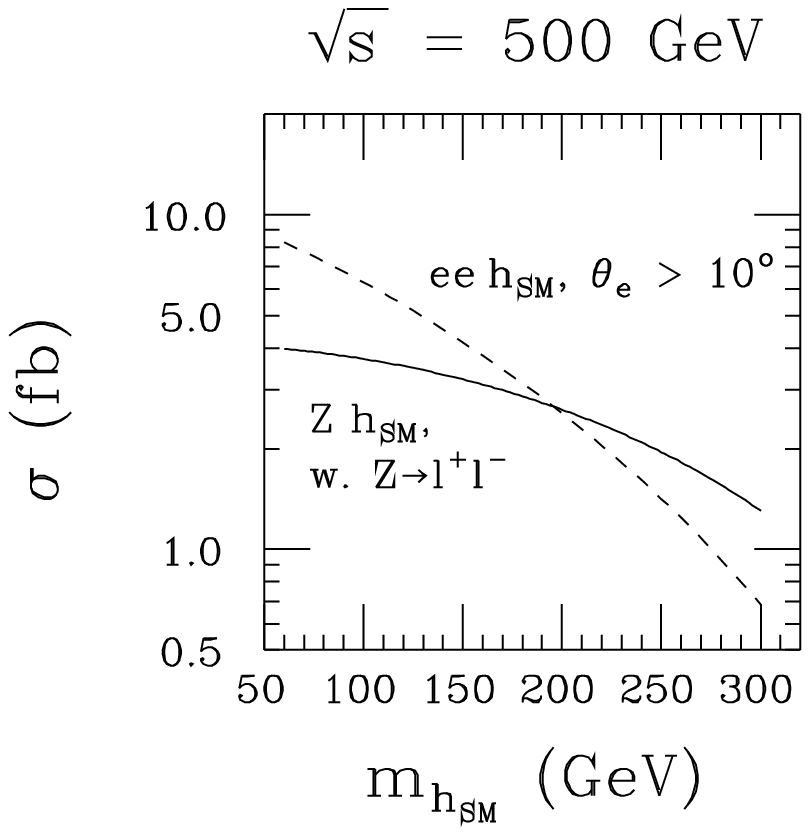,width=2.9in}}
\end{center}
\caption{$\sigma(Z\hsm)\br(Z\to \ell^-\ell^+)$ ($\ell=e,\mu$, no cuts)
and $\sigma(\epem\hsm)$ (with a cut of $\theta>10^\circ$ on the $e^+$
and $e^-$ in the final state) as a function of $\mhsm$ for 
$\protect\rts=500\gev$.~\protect\cite{ghs}}
\label{figzheeh}
\end{figure}

The relative value of the two production modes depends
upon many factors, including $\rts$. In Fig.~\ref{figzheeh},
we plot $\sigma(Z\hsm)\br(Z\to \ell^-\ell^+)$ ($\ell=e,\mu$, no cuts)
and $\sigma(\epem\hsm)$ (with a $\theta>10^\circ$
cut~\footnote{Assuming coverage down to such angles
is optimistic, but not unrealistic.}
on the angles of the final state $e^+$ and $e^-$)
as a function of $\mhsm$ for $\rts=500\gev$. 
We observe a cross-over such that, for $\mhsm\lsim 200\gev$, a
higher raw event rate for the recoil spectrum is obtained using $ZZ$ fusion.
Combining~\cite{rickjack,ghs} the $\rts=500\gev$
errors for the two processes gives an error on the $(ZZ\hsm)^2$
coupling-squared that ranges from $\sim 3\%$ to 
$\sim 6\%$ to $\sim 9\%$ for $\mhsm=60$, 200, and $300\gev$, respectively.
A more detailed listing appears in the final summary Table~\ref{nlcerrors}.

Since excellent accuracy can be achieved for measuring the $ZZ$ coupling
of a SM-like Higgs boson with mass below $150\gev$, it might
be supposed that discrimination between the $\hl$ of the MSSM and the $\hsm$
would be possible.  Unfortunately, one finds that the $(ZZ\hl)^2/(ZZ\hsm)^2$
ratio exhibits very small deviations from unity once $\mha\gsim 150\gev$.
However, measurable deviations emerge for large regions of NMSSM
parameter space.  These same statements apply to the $WW$ coupling,
the determination of which is discussed in the next subsection.

\subsubsection{Determining {\boldmath $\hsm$} branching ratios 
and the {\boldmath $WW\hsm$} coupling at the NLC}\label{ssshbrwwh}

A determination
of $\br(\hsm\to X)$ requires measuring $\sigma(\hsm)\br(\hsm\to X)$
and $\sigma(\hsm)$ for some particular production mode,
and then computing
\begin{equation}
\br(\hsm\to X)={\sigma(\hsm)\br(\hsm\to X)\over \sigma(\hsm)}\,.
\label{brform}
\end{equation}
In $\epem$ collisions, the $\epem\to Z\hsm$
and $\epem\to\epem\hsm$ ($ZZ$-fusion) modes just discussed
are the only ones for which the absolute magnitude of $\sigma(\hsm)$ 
can be measured, inclusively summing over all final states $X$.
The $WW$-fusion $\epem\to \nu\anti\nu\hsm$ cross section must be determined
by the procedure of first measuring $\sigma\br(\hsm\to X)$
in some mode $X$ and then dividing by $\br(\hsm\to X)$ as determined
from the $ZZ$-fusion or $Z\hsm$ channels.

\smallskip
\centerline{\underline{$\br(\hsm\to b\anti b)$ and $\br(\hsm\to c\anti c)$}}
\smallskip

We combine the earlier-discussed determination of
$\sigma(Z\hsm)\br(\hsm\to b\anti b)$
with the just-discussed measurement of $\sigma(Z\hsm)$ to obtain via
Eq.~(\ref{brform}) one determination of 
$\br(\hsm\to b\anti b)$. A second determination results from
combining the $\sigma(\epem\hsm)\br(\hsm\to b\anti
b)$ measurement~\cite{ghs} as summarized in
Higgs96,~\cite{snowmasssummary} 
with the $\sigma(\epem\hsm)$ measurement.
By combining~\cite{rickjack,ghs}
the $Z\hsm$ and $\epem\hsm$ determinations,
we find that $\br(\hsm\to b\anti b)$ can be measured with good accuracy
for $\mhsm\lsim 150\gev$; see Table~\ref{nlcerrors}.

An entirely similar procedure is employed for obtaining $\br(\hsm\to c\anti
c)$. For instance, in the $Z\hsm$ mode we
start with the topological tagging measurement of 
$\sigma(Z\hsm)\br(\hsm\to c\anti c)$ and divide by $\sigma(Z\hsm)$.
The analogous procedure is employed for
the $\epem\hsm$ production mode. The final error
for $\br(\hsm\to c\anti c)$ is estimated to be of order $\sim\pm 9\%$
for relatively light masses. A more detailed summary appears
in Table~\ref{nlcerrors}.

\smallskip
\centerline{\underline{$\br(\hsm\to W\wstar)$}}
\smallskip

The possible procedures are:~\cite{rickjack}
\begin{itemize}
\item
\begin{sloppypar}
Measure $\sigma(Z\hsm)\br(\hsm\to W\wstar)$ and $\sigma(Z\hsm)$
and compute $\br(\hsm\to W\wstar)$ by dividing. 
\end{sloppypar}
\item
Measure $\sigma(\epem\hsm)\br(\hsm\to W\wstar)$ and $\sigma(\epem\hsm)$
(the $ZZ$-fusion processes) and again compute $\br(\hsm\to W\wstar)$
by dividing.
\end{itemize}
Errors on $\br(\hsm\to W\wstar)$ in the $\epem\hsm$ production channel
will be close to those in the $Z\hsm$
channel for $\mhsm$ in the $130-200\gev$ mass range.
If we combine~\cite{rickjack} the above two determinations, 
one obtains $\br(\hsm\to W\wstar)$ errors below $10\%$ for $\mhsm\lsim
200\gev$; a full summary appears in Table~\ref{nlcerrors}.

\smallskip
\centerline{\underline{$WW\hsm$ coupling and testing custodial SU(2)}}
\smallskip

The goal will be to determine 
$\sigma(\nu\anti\nu\hsm)$ which is proportional to
the the $(WW\hsm)^2$ coupling-squared. 
The best procedure~\cite{rickjack} depends upon $\mhsm$:
\begin{itemize}
\item 
\begin{sloppypar}
If $\mhsm\lsim 140\gev$, then good accuracy is attained by
measuring $\sigma(\nu\anti\nu\hsm)\br(\hsm\to b\anti b)$ and then dividing
by $\br(\hsm\to b\anti b)$. 
\end{sloppypar}
\item 
\begin{sloppypar}
If $\mhsm\gsim 150\gev$, then good accuracy is achieved by measuring
$\sigma(\nu\anti\nu\hsm)\br(\hsm\to W\wstar)$ (in $WW$-fusion)
and dividing by $\br(\hsm\to W\wstar)$
to get $\sigma(\nu\anti\nu\hsm)$. 
\end{sloppypar}
\end{itemize}
At $\mhsm=140\gev$, the $W\wstar$ mode accuracy
is poorer than that obtained in the $b\anti b$ mode, but by $\mhsm=150\gev$
the $W\wstar$ mode determination has become comparable,
and for higher masses is distinctly superior.
If we combine the $b\anti b$ and $W\wstar$ mode determinations,
we get errors for $(WW\hsm)^2$ of order $\pm 5\%$ for $\mhsm\lsim 140\gev$,
worsening to about $\pm 8\%$ for $\mhsm\gsim 150\gev$.
For a full summary, see Table~\ref{nlcerrors}.

It is, of course, of great interest to test the custodial SU(2) symmetry
prediction for the coupling-squared ratio $(WW\hsm)^2/(ZZ\hsm)^2$.
Using the errors estimated above for these two squared couplings,
we obtain the results tabulated in Table~\ref{nlcerrors}.
For extended Higgs sectors containing only doublets and singlets
(such as those of the MSSM and NMSSM),
this ratio is predicted to have the SM value.  However, if
there are higher Higgs representations (\eg\ triplets), deviations
from the SM value would be expected.

\smallskip
\centerline{\underline{$\br(\hsm\to \gam\gam)$}}
\smallskip

We focus on $\mhsm\lsim 130\gev$.
The methods to determine 
$\br(\hsm\to\gam\gam)$ are detailed in Higgs96.~\cite{snowmasssummary}
\begin{itemize}
\item
The first involves measuring
$\sigma(pp\to W\hsm)\br(\hsm\to \gam\gam)$ 
and $\sigma(pp\to t\anti t\hsm)\br(\hsm\to\gam\gam)$
at the LHC. 
These measurements can be employed in two ways.  
\begin{itemize}
\item 
\begin{sloppypar}
In the first approach
one also measures $\sigma(pp\to t\anti t\hsm)\br(\hsm\to b \anti b)$ 
at the LHC and then computes 
$\br(\hsm\to\gam\gam)$ as $\br(\hsm\to b\anti b)\times 
[\sigma(pp\to t\anti t\hsm)\br(\hsm\to\gam\gam)]$
divided by
$[\sigma(pp\to t\anti t\hsm)\br(\hsm\to b\anti b)]$,
using $\br(\hsm\to b\anti b)$ determined at the NLC as described earlier.
\end{sloppypar}
\item
\begin{sloppypar}
In the second approach, one uses only $\sigma(pp\to W\hsm)\br(\hsm\to\gam\gam)$
from the LHC, and then divides
by the $\sigma(pp\to W\hsm)$ cross section as
computed (including systematic errors) using the $(WW\hsm)^2$ coupling-squared
determination from the NLC. 
\end{sloppypar}
\end{itemize}
To the extent that determinations from these 
two ways of getting at $\br(\hsm\to \gam\gam)$ are statistically
independent, they can be combined to yield
statistical accuracy of $\lsim \pm 16\%$ in the $\mhsm\lsim 130\gev$ range.
\item
There are two independent techniques~\cite{gm} for using
the $\sigma\br(\hsm\to\gam\gam)$ measurements at the NLC,
discussed earlier, to determine $\br(\hsm\to\gam\gam)$.
\begin{itemize}
\item Measure $\sigma(\epem\to Z\hsm) \br(\hsm\to \gam\gam)$ and compute
$\br(\hsm\to\gam\gam)$ as
${[\sigma( Z\hsm)\br(\hsm\to
\gam\gam)]/\sigma( Z\hsm)}\,;$
\item 
\begin{sloppypar}
Measure $\sigma(\epem\to \nu\anti\nu \hsm) \br(\hsm\to \gam\gam)$ 
and $\sigma(\epem\to \nu\anti\nu \hsm) \br(\hsm\to b\anti b)$ 
(both being $WW$-fusion processes) and compute $\br(\hsm\to\gam\gam)$
as
$[\sigma(\nu\anti\nu\hsm)\br(\hsm\to \gam\gam)]\br(\hsm\to
b\anti b)$ divided by $[\sigma(\nu\anti\nu\hsm)\br(\hsm\to b\anti b)]\,.$
\end{sloppypar}
\end{itemize}
(The $\epem\hsm$ final state from $ZZ$-fusion is a third
alternative, but does not yield errors competitive with
the above two techniques.)
The error on $\br(\hsm\to\gam\gam)$ 
would be of order $\pm 22\%$ at the best case $\mhsm=120\gev$.
\end{itemize}
Of course, the NLC-based and LHC-based methods can be combined.
The net error is tabulated in the summary Table~\ref{nlcerrors}.

\subsubsection{Determining {\boldmath $\gamhsm$}}\label{sssgam}

The most fundamental properties of the Higgs boson
are its mass, its total width and its partial widths.
Discussion of the mass determination will be left till the next subsection.
The total Higgs width, while certainly important in its own right,
becomes doubly so since it is required in order to compute many
important partial widths.
The partial widths, being directly proportional to the underlying
couplings, provide the most direct means of verifying that the observed
Higgs boson is or is not the $\hsm$. Branching ratios, being
the ratio of a partial width to the total width can not be unambiguously
interpreted. In contrast,
a partial width is directly related to the corresponding
coupling-squared which, in turn, is directly determined in the SM or any
extension thereof without reference to mass scales for possibly
unexpected (\eg\ SUSY) decays.
Any deviations of partial widths from SM predictions 
can be directly compared to predictions of alternative models
such as the MSSM, the NMSSM, or the general \thdm.  The more accurately
the total width and the various branching ratios can be measured, 
the greater the sensitivity
to such deviations and the greater our ability to recognize and constrain
the alternative model.

The rapid variation of $\gamhsm$ is well-known: \eg\
$\gamhsm\sim 17\mev$, $32\mev$, $400\mev$, $1\gev$, $4\gev$, $10\gev$ for 
$\mhsm\sim 150$, 155, 170, 190, 245, $300\gev$, respectively.
For $\mhsm\gsim 180-245\gev$, determination of $\gamhsm$
via final state resonance peak reconstruction is possible, the 
exact $\mhsm$ above which reasonable errors are achieved depending
upon the resolution as determined by the machine/technique
and detector characteristics.
For lower $\mhsm$, and certainly for $\mhsm< 2\mw$ (as relevant
for the SM-like MSSM $\hl$),
there are only two basic possibilities for determining $\gamhsm$.
\begin{itemize}
\item
The first is to employ FMC $\mupmum$ collisions at $\rts\sim \mhsm$ and
directly measure $\gamhsm$ by scanning.  In this case, 
the FMC determination of $\gamhsm$
can be used to compute the partial width for any channel with a 
branching ratio measured at the NLC:
\begin{equation}
\Gamma(\hsm\to X)=\gamhsm\br(\hsm\to X)\,.
\label{partialw}
\end{equation}
\item
If there is no muon collider, then $\gamhsm$ must be determined indirectly
using a multiple step process; the best process depends upon
the Higgs mass. $\gamhsm$ is ultimately computed as:
\begin{equation}
\gamhsm={\Gamma(\hsm\to X)\over\br(\hsm\to X)}\,,
\label{partialwi}
\end{equation}
where $X=\gam\gam$ ($W\wstar$) gives the best error for $\mhsm\lsim 130\gev$
($\gsim140\gev$).
In this case, $\gamhsm$ can be used to compute partial widths
via Eq.~(\ref{partialw}) only for channels other than those
used in the determination of $\gamhsm$ via Eq.~(\ref{partialwi}).
\end{itemize}
In what follows we outline the errors anticipated in the ultimate
determination of $\gamhsm$ in the $\mhsm\leq 2\mw$
mass region, and then discuss implications for
the errors in partial widths, both with and without combining NLC
and FMC data. We also discuss the determination of $\gamhsm$ by
final state mass peak reconstruction.

Before proceeding, we make a few
remarks regarding the use of the total width, per se, as a means
for discriminating between models. Certainly, the Higgs 
total width will exhibit
deviations from $\gamhsm$ if there is an extended Higgs sector.
However, these deviations turn out to be model-dependent.  For instance,
even restricting to the case of the MSSM
and assuming that there are no supersymmetric decays of the $\hl$, 
the ratio $\gamhl/\gamhsm$ depends strongly
on the squark-mixing scenario; for a fixed $\mhl$,
``no mixing'' constant value contours for this ratio 
in the $(\mha,\tanb)$ parameter space differ very substantially
in shape and location 
from those obtained for ``maximal mixing'', regardless of how large
$\mha$ is. Thus, the exact value of $\gamhl/\gamhsm$ does not pin 
down any one parameter of the model; instead, it constrains a very
complicated combination of parameters.  As already noted,
partial widths will prove to be much more valuable.

\bigskip
\begin{center}
\underline{FMC-scan determination of $\gamhsm$}
\end{center}
\smallskip

Only the $\mupmum$ collider
can have the extremely precise energy resolution ($R\sim 0.01\%$) 
and energy setting (1 part in $10^6$)
capable of measuring $\gamhsm$ by scanning in the $\mhsm\leq 2\mw$
mass region where $\gamhsm$ is of order tens of MeV.~\cite{bbgh}
The most difficult case is if $\mhsm\sim \mz$,
implying a large $Z$ background to $\hsm$ production in the $s$-channel.
We assume that since the mass of the Higgs boson will be 
relatively precisely known from the LHC (see next subsection)
the FMC would be designed to have optimal luminosity at
$\rts\sim\mhsm$, so that accumulation of $L=200\fbi$ for
scanning the Higgs peak would be possible. 
A complete listing of $L=200\fbi$ $\gamhsm$ errors appears 
in Table~\ref{fmcerrors}.
For $\mhsm\not\sim\mz$, the $s$-channel 
FMC accuracy would be much superior to that achievable
on the basis of NLC data alone, and would provide
an extremely valuable input to precision tests of the Higgs sector.

\bigskip
\begin{center}
\underline{Indirect determination of $\gamhsm$}
\end{center}
\smallskip

If there is no $\mupmum$ collider, then $\gamhsm$ must be determined
indirectly.  The best procedure for doing
so depends upon the Higgs mass. If $\mhsm\lsim 130\gev$,
then one must make use of $\gam\gam$ Higgs decays. If $\mhsm\gsim 140\gev$,
$W\wstar$ Higgs decays will be most useful. In both cases, we ultimately
employ Eq.~(\ref{partialwi}) to obtain $\gamhsm$.

Since the $\Gamma(\hsm\to\gam\gam)$ partial width plays a crucial
role in the $\mhsm\leq 130\gev$ procedure, it is convenient
to discuss it first. This partial width
is obtained by first measuring the rate for $\gam\gam\to\hsm \to b\anti b$ 
at the NLC photon-photon collider facility 
by tuning the beam energy so that the $\gam\gam$ luminosity peak at 
$\sim 0.8\rts$ coincides with $\mhsm$.~\cite{ghgamgam,borden}
The statistical and systematic errors 
for $\Gamma(\hsm\to\gamgam)\br(\hsm\to b\anti b)$ for $L=50\fbi$
(we presume that this is the maximal luminosity that might be
devoted to NLC running in the photon-photon collider mode)
were discussed in Higgs96.~\cite{snowmasssummary}
The net error in the $\mhsm\lsim 120\gev$ mass region
will be in the 8\%-10\% range, rising to 15\% by $\mhsm=140\gev$
and peaking at 30\% at $\mhsm=150\gev$.
To get the $\Gamma(\hsm\to \gam\gam)$ partial width itself,
we divide by $\br(\hsm\to b\anti b)$, adding in the errors of
the latter by quadrature. The resulting errors for $\Gamma(\hsm\to\gam\gam)$
are summarized in Table~\ref{nlcerrors}.

We now give the procedures for determining $\gamhsm$.
\begin{itemize}
\item
For $\mhsm\leq 130\gev$ (\ie\ in the MSSM $\mhl$ range), 
the only known procedure for determining $\gamhsm$ is that outlined
in DPF95.~\cite{dpfreport} NLC data is required.
\begin{itemize}
\item
As described above, measure
$\Gamma(\hsm\to \gam\gam)\br(\hsm\to b\anti b)$ and then compute
$\Gamma(\hsm\to\gam\gam)$ by dividing by the value of 
$\br(\hsm\to b\anti b)$.
\item
\begin{sloppypar}
Compute $\gamhsm=\Gamma(\hsm\to\gam\gam)/\br(\hsm\to\gam\gam)$.
We employ the earlier-described determination of $\br(\hsm\to\gam\gam)$ based
on combining NLC and LHC data.
\end{sloppypar}
\end{itemize}

\item
For $\mhsm\gsim 130\gev$, a second possible procedure based on
$\hsm\to W\wstar$ decays emerges.
Use $(WW\hsm)^2$ to compute $\Gamma(\hsm\to W\wstar)$ and then compute
$\gamhsm=\Gamma(\hsm\to W\wstar)/\br(\hsm\to W\wstar)$.

\end{itemize}
In Table~\ref{nlcerrors}, we tabulate the errors for $\gamhsm$ 
obtained by using both the $\gam\gam$ and the $W\wstar$
techniques, and including the LHC determination of $\br(\hsm\to\gam\gam)$
in the former. For more details, see Higgs96.~\cite{snowmasssummary}

As apparent from Tables~\ref{fmcerrors} and \ref{nlcerrors},
for $\mhsm\leq 130\gev$ (and $\mhsm\not\sim\mz$) the FMC-scan
determination of $\gamhsm$ is very much superior to the NLC determination.
The superiority is still significant at $\mhsm=140\gev$ while  errors
are similar at $\mhsm=150\gev$. 
For $\mhsm\geq 160\gev$, FMC $s$-channel detection
of the $\hsm$ becomes difficult, and only the NLC allows a reasonable
determination of $\gamhsm$.

\begin{center}
\underline{Final-state mass peak determination of $\gamhsm$: NLC and LHC}
\end{center}
\smallskip

\def\gamr{\Gamma_{\rm R}}
\def\gameff{\Gamma_{\rm eff}}
Direct measurement of $\gamhsm$ from the shape of the Higgs mass
peak becomes possible when $\gamhsm$ is not
too much smaller than $\gamr$, the relevant final state 
mass resolution.
Detailed results are given in Higgs96.~\cite{snowmasssummary}
The $Z\hsm$ production mode was studied for NLC operation at $\rts=500\gev$.
For `super' detector performance, it was found that 
the direct measurement errors for $\gamhsm$ are only competitive
with those from the indirect determination for $\mhsm\gsim 180\gev$.
For `standard' tracking/calorimetry, the direct measurement errors only 
become competitive with indirect errors for $\mhsm\gsim 250\gev$.  
Measurement of $\gamhsm$ in the $gg\to \hsm\to ZZ^{(*)}\to 4\ell$ mode
at the LHC was also studied, but only becomes competitive with
the best NLC errors for $\mhsm\gsim 280\gev$.
Thus, direct final state measurement of $\gamhsm$ will not be 
possible at either the NLC or the LHC if the SM-like Higgs boson
has mass in the $< 150\gev$ region expected in supersymmetric models.

\subsubsection{Partial widths using {\boldmath $\gamhsm$}}\label{ssspw}

In this section, we focus on results obtained in 
Higgs96~\cite{snowmasssummary}
using NLC data, FMC data, or a combination thereof. (It is important to recall
our convention that the notation NLC means $\rts=500\gev$ running
in $\epem$ or $\mupmum$ collisions, while FMC refers explicitly
to $s$-channel Higgs production in $\mupmum$ collisions.)
Due to lack of time, LHC data was not generally incorporated.
The only exception is that the error on $\br(\hsm\to\gam\gam)$
is estimated after including the determinations that
employ LHC data via the procedures outlined earlier. This is
particularly crucial in obtaining a reasonable error for
the indirect determination of $\gamhsm$ when $\mhsm\leq 130\gev$.

\smallskip
\centerline{\underline{$(b\anti b\hsm)^2$ and $(c\anti c\hsm)^2$: 
NLC only or NLC+FMC data}}
\smallskip

Given a determination of $\gamhsm$, we employ 
Eq.~(\ref{partialw}) and the determination
of $\br(\hsm \to b\anti b)$ to determine
$\Gamma(\hsm\to b\anti b)$, \ie\
the $(b\anti b\hsm)^2$ squared-coupling. 
The $(c\anti c\hsm)^2$ squared-coupling is best computed 
from $(b\anti b\hsm)^2$ and the $(c\anti c\hsm)^2/(b\anti b\hsm)^2$
measurement. For both squared-couplings, the errors
are ultimately dominated by those for $\gamhsm$, and are thus
greatly improved for $\mhsm\lsim 140\gev$ by including the
FMC-scan determination of $\gamhsm$.
The resulting errors are those tabulated as $(b\anti b\hsm)^2|_{\rm NLC+FMC}$ 
and $(c\anti c\hsm)^2|_{\rm NLC+FMC}$
in Table~\ref{nlcfmcerrors}. 

\begin{figure}[h]
\leavevmode
\begin{center}
\centerline{\psfig{file=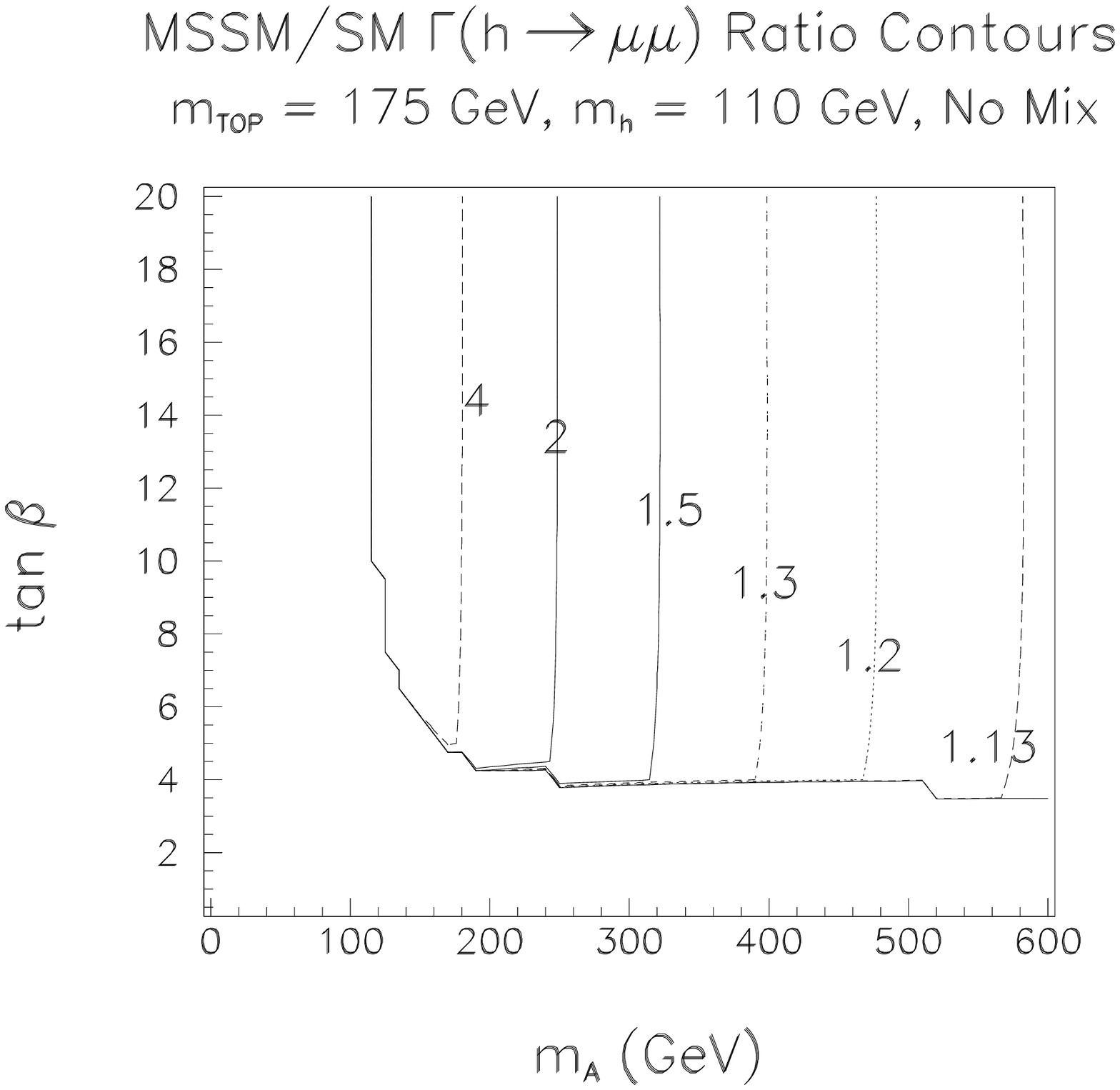,width=3.1in}}
\caption{Constant value contours in $(\mha,\tanb)$ parameter space
for the ratio $\Gamma(\hl\to\mupmum)/\Gamma(\hsm\to\mupmum)$.
We assume ``no mixing'' in the squark sector and present
results for the case of fixed $\mhl=\mhsm=110\gev$. 
For ``maximal mixing'', the vertical contours
are essentially identical --- only the size of the allowed parameter
range is altered. Contours for $\Gamma(\hl\to b\anti b)/\Gamma(\hsm\to b\anti
b)$ are identical.}
\label{mumucontours}
\end{center}
\end{figure}

A deviation of the squared $b\anti b$ coupling from the SM prediction
is a sure indication of an extended Higgs sector and is potentially
very useful in the case of the MSSM for determining $\mha$. 
The fixed $\mhl=\mhsm$ contours in $(\mha,\tanb)$
parameter space for $(b\anti b\hl)^2/(b\anti b\hsm)^2$ are 
exactly the same as those for $(\mupmum\hl)^2/(\mupmum\hsm)^2$,
which will be illustrated in Fig.~\ref{mumucontours}, and are independent
of squark mixing scenario. However, similarly to the case of the 
$\br(c\anti c)/\br(b\anti b)$ ratio discussed earlier, systematic uncertainty
in $\mb(\mb)$ as determined from QCD sum rules and/or lattice calculations
leads to a certain level of uncertainty in the $(b\anti b\hsm)^2$
prediction and, therefore, in our ability to employ an experimental
determination of the $b\anti b$ partial width to 
determine $\mha$. The $\mupmum$ partial width avoids this problem,
and, in addition, has smaller experimental error, as we now discuss.

\smallskip
\centerline{\underline{$(\mupmum\hsm)^2$: NLC+FMC data}}
\smallskip

The very small errors for the FMC $s$-channel measurements of 
$\Gamma(\hsm\to\mupmum)\br(\hsm\to b\anti b,W\wstar,W\zstar)$~\cite{bbgh}
are summarized in Table~\ref{fmcsigbrerrors}.~\footnote{Recall
that the FMC $s$-channel errors quoted are for $L=50\fbi$, the
amount of luminosity exactly on the $\rts=\mhsm$ Higgs peak
that is roughly equivalent to the on-peak and off-peak luminosity
accumulated in performing the $L=200\fbi$ scan determination of $\gamhsm$.}
Given these measurements, there are four independent ways of combining
NLC data with the $s$-channel FMC
data to determine $\Gamma(\hsm\to\mupmum)$.~\cite{bbghnew}
\sloppy
\begin{description}
\item{ 1)} compute $\Gamma(\hsm\to\mupmum)=[\Gamma(\hsm\to\mupmum) \br(\hsm\to
b\anti b)]_{\rm FMC} / \br(\hsm\to b\anti b)_{\rm NLC}$;
\item{ 2)} compute $\Gamma(\hsm\to\mupmum)=[\Gamma(\hsm\to\mupmum) \br(\hsm\to
W\wstar)]_{\rm FMC} / \br(\hsm\to W\wstar)_{\rm NLC}$;
\item{ 3)} compute $\Gamma(\hsm\to\mupmum)=[\Gamma(\hsm\to\mupmum) \br(\hsm\to
Z\zstar)]_{\rm FMC}\gamhsm / \Gamma(\hsm\to Z\zstar)_{\rm NLC}$, where
the combined direct FMC plus indirect NLC determination of $\gamhsm$ can
be used since the NLC $(Z\zstar\hsm)^2$ determination was not used in
the indirect NLC determination of $\gamhsm$;
\item{ 4)} compute $\Gamma(\hsm\to\mupmum)=[\Gamma(\hsm\to\mupmum)\br(\hsm\to
W\wstar)\gamhsm]_{\rm FMC}/\Gamma(\hsm\to W\wstar)_{\rm NLC}$, where
we can only employ $\gamhsm$ as determined at the FMC since $(W\wstar\hsm)^2$
is used in the NLC indirect determination of $\gamhsm$.
\end{description}
\fussy
The resulting (very small) errors for $(\mupmum\hsm)^2$ 
obtained by combining determinations from all four techniques are 
labelled $(\mupmum\hsm)^2|_{\rm NLC+FMC}$ and tabulated
in Table~\ref{nlcfmcerrors}.

The $\mupmum$ coupling-squared provides the best opportunity
for distinguishing between the $\hl$ of the MSSM and the $\hsm$ of the SM
and for determining $\mha$.
Contours of constant $(\mupmum\hl)^2/(\mupmum\hsm)^2$ in $(\mha,\tanb)$
parameter space are illustrated for $\mhl=\mhsm=110\gev$
in Fig.~\ref{mumucontours}, assuming
``no mixing'' in the squark sector; the ``maximal mixing'' contours are
the same in the portion of parameter space where $\mhl=110\gev$
is theoretically allowed.
Note that $\geq 13\%$ deviations are predicted for $\mha\leq 600\gev$.
The $4\%$ statistical error for $\Gamma(\mupmum)$
(Table~\ref{nlcfmcerrors}) implies the ability to distinguish
between the $\hl$ and the $\hsm$ at the $\geq 3\sigma$ level for $\mha\leq
600\gev$. If a deviation is observed, $\mha$ will be determined
with a $1\sigma$ error of roughly $\pm 50\gev$.
Depending upon the then-prevailing systematic
theoretical uncertainty in $\mb(\mb)$, the $b\anti b$ and $\mupmum$
partial width measurements can be combined to improve the accuracy
of the $\mha$ determination.

\smallskip
\centerline{\underline{$(WW\hsm)^2$ and $(\gam\gam\hsm)^2$: NLC+FMC data}}
\smallskip

In Table~\ref{nlcerrors} we summarized the errors for the $(WW\hsm)^2$
coupling squared coming from determining the $\nu\anti\nu\hsm$
cross section from $\rts=500\gev$ running at the NLC.
We can obtain a second independent determination of $(WW\hsm)^2$
by taking $\br(\hsm\to W\wstar)$ (as determined in $Z\hsm$ and $\epem\hsm$
production at the NLC) and multiplying by $\gamhsm$ as determined 
by $s$-channel scanning at the FMC.
These errors are summarized in Table~\ref{nlcfmcerrors}
using the notation $(WW\hsm)^2|_{\rm FMC}$. If we combine the
two different determinations, then we get the errors denoted
$(WW\hsm)^2|_{\rm NLC+FMC}$. 
(For $\mhsm\leq
130\gev$, $\br(\hsm\to W\wstar)$ is too poorly
measured for this procedure to yield any improvement
over the errors of Table~\ref{nlcerrors}.)

In close analogy to the $W\wstar$ procedure given above, we can
determine $(\gam\gam\hsm)^2$
by taking $\br(\hsm\to \gam\gam)$ (as determined using LHC and
$\nu\anti\nu\hsm$ NLC data) and multiplying by $\gamhsm$ as determined 
by $s$-channel scanning at the FMC.
The resulting errors are summarized in Table~\ref{nlcfmcerrors}
using the notation $(\gam\gam\hsm)^2|_{\rm FMC}$. If we combine this
determination with the independent determination
from $\rts=500\gev$ NLC running (see Table~\ref{nlcerrors}), 
then we get the errors denoted
$(\gam\gam\hsm)^2|_{\rm NLC+FMC}$.

\begin{table}[p]
\caption[fake]{Summary of approximate errors for branching ratios,
coupling-squared ratios, couplings-squared and $\gamhsm$ as
determined using only data accumulated in $\protect\rts=500\gev$
running at the NLC, assuming $L=200\fbi$ is accumulated.
For $\br(\hsm\to\gam\gam)$ we have combined the NLC $\rts=500\gev$ results
with results obtained using LHC data; the net accuracy so obtained
for $\br(\hsm\to\gam\gam)$ is also reflected in the
determination of $\gamhsm$ following the indirect procedure.
The errors for $\Gamma(\hsm\to\gam\gam)$ quoted are for $L=50\fbi$
accumulated in $\gam\gam$ collider running at $\rts\sim \mhsm/0.8$,
and are those employed in the indirect $\gamhsm$ determination.
A $-$ indicates large error and a $?$ indicates either that a reliable
simulation or estimate is not yet available or that the indicated
number is a very rough estimate.}
\footnotesize
\begin{center}
\begin{tabular}{|c|c|c|c|c|}
\hline
 Quantity & \multicolumn{4}{c|}{Errors} \\
\hline
\hline
{$\bf\mhsm$}{\bf (GeV)} & { \bf80} & { \bf100} & { \bf 110} & {\bf 120} \\
\hline
 $(c\anti c\hsm)^2/(b\anti b\hsm)^2$ & \multicolumn{4} {c|}{$\sim\pm7\%$} \\
\hline
 $(WW\hsm)^2/(b\anti b\hsm)^2$ & $-$ & $-$ & $-$  & $\pm 23\%$ \\
\hline
 $(\gam\gam \hsm)^2/(b\anti b\hsm)^2$ & $\pm 52\%$ & $\pm 33\%$ & $\pm 29\%$ &
 $\pm 26\%$ \\
\hline
 $(ZZ\hsm)^2$ & \multicolumn{4} {c|}{$\pm 3\%-\pm 4\%$} \\
\hline
 $\br(\hsm\to b\anti b)$ & \multicolumn{4}{c|}{$\pm5\%$} \\
\hline
 $\br(\hsm\to c\anti c)$ & \multicolumn{4}{c|}{$\sim\pm9\%$} \\
\hline
 $\br(\hsm\to W\wstar)$ & \multicolumn{4}{c|}{$-$} \\
\hline
 $(WW\hsm)^2$ & \multicolumn{4}{c|}{$\pm5\%$} \\
\hline
 $(ZZ\hsm)^2/(WW\hsm)^2$ & \multicolumn{4}{c|}{$\pm 6\%-\pm 7\%$}\\
\hline
 $\br(\hsm\to\gam\gam)$ & $\pm 15\%$ & $\pm 14\%$ & $\pm 14\%$ & $\pm 14\%$ \\
\hline
 $(\gam\gam\hsm)^2$ & \multicolumn{4}{c|}{$\sim \pm 12\%$}\\
\hline
 $\gamhsm$ (indirect) & $\pm 19\%$ & $\pm 19\%$ & $\pm 19\%$ & 
 $\pm 18\%$ \\
\hline
 $(b\anti b\hsm)^2$ & $\pm 20\%$ & $\pm 19\%$ & $\pm 19\%$ & 
 $\pm 19\%$ \\
\hline
\hline
{$\bf\mhsm$}{\bf (GeV)} & {\bf 130} & {\bf 140} & {\bf 150} & {\bf 170} \\
\hline
 $(c\anti c\hsm)^2/(b\anti b\hsm)^2$ & $\pm7\%$ & 
\multicolumn{3} {c|}{$?$} \\
\hline
 $(WW\hsm)^2/(b\anti b\hsm)^2$ & $\pm 16\%$ & $\pm 8\%$ & $\pm 7\%$ &
 $\pm 16\%$ \\
\hline
 $(\gam\gam \hsm)^2/(b\anti b\hsm)^2$ & $\pm 27\%$ & $\pm 30\%$ & $\pm 41\%$ &
 $-$ \\
\hline
 $(ZZ\hsm)^2$ & \multicolumn{4} {c|}{$\pm 4\%$} \\
\hline
 $\br(\hsm\to b\anti b)$ & \multicolumn{2}{c|}{$\pm6\%$} & $\pm 9\%$ & $\sim
 20\%?$ \\
\hline
 $\br(\hsm\to c\anti c)$ & $\sim\pm 9\%$ & \multicolumn{3}{c|}{$?$} \\
\hline
 $\br(\hsm\to W\wstar)$ & $\pm 16\%$ & $\pm 8\%$ & $\pm 6\%$ & $\pm 5\%$ \\
\hline
 $(WW\hsm)^2$ & $\pm 5\%$ & $\pm 5\%$ & $\pm 8\%$ & $\pm 10\%$ \\
\hline
 $(ZZ\hsm)^2/(WW\hsm)^2$ & $\pm 7\%$ & $\pm 7\%$ & $\pm 9\%$ & $\pm 11\%$ \\
\hline
 $\br(\hsm\to\gam\gam)$ & $\pm 14\%$ & $\pm 20\%?$ & $\pm 41\%$ & $-$ \\
\hline
 $(\gam\gam\hsm)^2$ & $\pm 15\%$ & $\pm 17\%$ & $\pm 31\%$ & $-$ \\
\hline
 $\gamhsm$ (indirect) & $\pm 13\%$ & $\pm 9\%$ & $\pm 10\%$ & 
 $\pm 11\%$ \\
\hline
 $(b\anti b\hsm)^2$ & $\pm 14\%$ & $\pm 11\%$ & $\pm 13\%$ & 
 $\pm 23\%$ \\
\hline
\hline
{$\bf\mhsm$}{\bf (GeV)} & {\bf 180} & {\bf 190} & {\bf 200} & {\bf 300} \\
\hline
 $(ZZ\hsm)^2$ & \multicolumn{2} {c|}{$\pm 4\%-\pm5\%$}& $\pm 6\%$ & $\pm 9\%$ \\
\hline
 $(WW\hsm)^2$ & $\pm 11\%$ & $\pm 12\%$ & $\pm 13\%$ & $\pm 24\%$ \\
\hline
 $(ZZ\hsm)^2/(WW\hsm)^2$ & $\pm 12\%$ & $\pm 13\%$ & $\pm 14\%$ & $\pm 25\%$ \\
\hline
 $\br(\hsm\to WW)$ & $\pm 6\%$ & $\pm 7\%$ & $\pm 8\%$ & $\pm 14\%?$ \\
\hline
 $(\gam\gam\hsm)^2$ & $\pm 13\%$ & $\pm 12\%$ & $\pm 12\%$ &  $\pm 22\%$ \\
\hline
 $\gamhsm$ (indirect) & $\pm 13\%$ & $\pm 14\%$ & $\pm 15\%$ & 
 $\pm 28\%$ \\
\hline
\end{tabular}
\end{center}
\label{nlcerrors}
\end{table}

\begin{table}[h]
\caption[fake]{Summary of approximate errors for 
coupling-squared ratios and $\gamhsm$ in the case of 
$s$-channel Higgs production at the FMC, assuming $L=200\fbi$
total scan luminosity (which for rate
measurements in specific channels is roughly equivalent to $L=50\fbi$ 
at the $\rts=\mhsm$ peak). Beam resolution of $R=0.01\%$ is assumed.
A $-$ indicates large error and a $?$ indicates either that a reliable
simulation or estimate is not yet available or that the indicated
number is a very rough estimate.}
\footnotesize
\begin{center}
\begin{tabular}{|c|c|c|c|c|}
\hline
 Quantity & \multicolumn{4}{c|}{Errors} \\
\hline
\hline
{$\bf\mhsm$}{\bf (GeV)} & {\bf 80} & {\bf $\mz$} & {\bf 100} & {\bf 110} \\
\hline
$(W\wstar\hsm)^2/(b\anti b\hsm)^2$ & $-$ & $-$ & $\pm 3.5\%$ & $\pm 1.6\%$ \\
\hline
$(Z\zstar\hsm)^2/(b\anti b\hsm)^2$ & $-$ & $-$ & $-$ & $\pm 34\%$ \\
\hline
$(Z\zstar\hsm)^2/(W\wstar\hsm)^2$ & $-$ & $-$ & $-$ & $\pm 34\%$ \\
\hline
 $\gamhsm$ & $\pm 2.6\%$ & $\pm 32\%$ & $\pm 8.3\%$ & 
  $\pm 4.2\%$ \\
\hline
\hline
{$\bf\mhsm$}{\bf (GeV)} & {\bf 120} & {\bf 130} & {\bf 140} & {\bf 150} \\
\hline
$(W\wstar\hsm)^2/(b\anti b\hsm)^2$ & 
 $\pm 1\%$ & $\pm 0.7\%$ & $\pm 0.7\%$ & $\pm 1\%$ \\
\hline
$(Z\zstar\hsm)^2/(b\anti b\hsm)^2$ & 
 $\pm 6\%$ & $\pm 3\%$ & $\pm 2\%$ & $\pm 2\%$ \\
\hline
$(Z\zstar\hsm)^2/(W\wstar\hsm)^2$ & 
 $\pm 6\%$ & $\pm 3\%$ & $\pm 2\%$ & $\pm 2\%$ \\
\hline
 $\gamhsm$ & $\pm 3.6\%$ & $\pm 3.6\%$ & $\pm 4.1\%$ &
  $\pm 6.5\%$ \\
\hline
\end{tabular}
\end{center}
\label{fmcerrors}
\end{table}

\begin{table}[h]
\caption[fake]{Summary of approximate errors for branching ratios,
coupling-squared ratios, couplings-squared 
and $\gamhsm$ obtained by combining the results of Tables~\ref{nlcerrors}
and \ref{fmcerrors}. See text for further discussion.
A $-$ indicates large error and a $?$ indicates either that a reliable
simulation or estimate is not yet available or that the indicated
number is a very rough estimate.}
\footnotesize
\begin{center}
\begin{tabular}{|c|c|c|c|c|}
\hline
 Quantity & \multicolumn{4}{c|}{Errors} \\
\hline
\hline
{$\bf\mhsm$}{\bf (GeV)} & {\bf 80} & {\bf 100} & {\bf 110} & {\bf 120} \\
\hline
 $(b\anti b\hsm)^2|_{\rm NLC+FMC} $ & $\pm6\%$ & $\pm 9\%$ & $\pm 7\%$ &
  $\pm6\%$ \\
\hline
 $(c\anti c\hsm)^2|_{\rm NLC+FMC} $ & $\pm9\%$ & $\pm 10\%$ & $\pm 10\%$ &
  $\pm9\%$ \\
\hline
 $(\mupmum\hsm)^2|_{\rm NLC+FMC}$ & 
$\pm 5\%$ & $\pm 5\%$ & $\pm 4\%$ & $\pm 4\%$ \\
\hline
 $(\gam\gam\hsm)^2|_{\rm FMC}$ & $\pm 16\%$ & $\pm 17\%$ & $\pm 15\%$ &
 $\pm 14\%$ \\
\hline
 $(\gam\gam\hsm)^2|_{\rm NLC+FMC}$ & $\pm 9\%$ & $\pm 10\%$ & $\pm 9\%$ &
 $\pm 9\%$ \\
\hline
\hline
{$\bf\mhsm$}{\bf (GeV)} & {\bf 130} & {\bf 140} & {\bf 150} & {\bf 170} \\
\hline
 $(b\anti b\hsm)^2|_{\rm NLC+FMC}$ & $\pm7\%$ & $\pm7\%$ & $\pm10\%$ &
 $\pm23\%$ \\
\hline
 $(c\anti c\hsm)^2|_{\rm NLC+FMC} $ & $\pm10\%$ & \multicolumn{3}{c|}{$?$} \\
\hline
 $(\mupmum\hsm)^2|_{\rm NLC+FMC}$ & 
$\pm 4\%$ & $\pm 3\%$ & $\pm 4\%$ &  $\pm 10\%$ \\
\hline
 $(W\wstar\hsm)^2|_{\rm FMC}$ & $\pm 16\%$ & $\pm 9\%$ & $\pm 9\%$ &
 $-$ \\
\hline
 $(W\wstar\hsm)^2|_{\rm NLC+FMC}$ & $\pm 5\%$ & $\pm 4\%$ & $\pm 6\%$ &
 $\pm 10\%$ \\
\hline
 $(\gam\gam\hsm)^2|_{\rm FMC}$ & $\pm 14\%$ & $\pm 20\%$ & $\pm 41\%$ & $-$ \\
\hline
 $(\gam\gam\hsm)^2|_{\rm NLC+FMC}$ & 
    $\pm 10\%$ & $\pm 13\%$ & $\pm 24\%$ & $-$ \\
\hline
\end{tabular}
\end{center}
\label{nlcfmcerrors}
\end{table}

\subsubsection{Summary Tables}\label{ssssumtabs}

Employing SM notation,
we present in Tables~\ref{nlcerrors}, \ref{fmcerrors}, and \ref{nlcfmcerrors} 
a final summary of the errors that can be achieved for the fundamental
properties (other than the mass) of a SM-like Higgs boson,
in three different situations:
\begin{itemize}
\item
$L=200\fbi$ devoted to $\rts=500\gev$ running at the NLC
supplemented with $L=50\fbi$ of $\gam\gam$
collider data obtained by running at $\rts_{\epem}\sim \mhsm/0.8$;
\item
A total $L=200\fbi$ of luminosity devoted to scanning the Higgs peak
to determine $\gamhsm$ --- as explained earlier, specific channel rate
errors are equivalent to those that would be obtained by devoting
$L=50\fbi$ to the Higgs peak at $\rts=\mhsm$;
\item
combining the above two sets of data.
\end{itemize}
The results we have obtained depend strongly on detector
parameters and analysis techniques and in some cases (those marked
by a ?) were obtained by extrapolation rather than full simulation.
Nonetheless, these results should serve as a reasonable estimate
of what might ultimately be achievable on the basis of NLC 
$\rts=500\gev$ running and/or FMC $s$-channel data.
Results for FMC $s$-channel errors assume very excellent $0.01\%$ beam energy
resolution and the ability to measure the beam energy with
precision on the order of 1 part in $10^6$.
Except for the determination
of $\br(\hsm\to\gam\gam)$ and implications for $\gamhsm$,
the undoubted benefits
that would result from combining NLC/FMC data with LHC data
have not yet been explored.

Of course, it should not be forgotten that the
$\rts=500\gev$ data could also be obtained by running an FMC with a final
ring optimized for this energy. (Confirmation 
that the FMC can achieve the same precisions as the NLC when
run at $\rts=500\gev$ must await a full machine and detector design;
it could be that the FMC backgrounds and detector design will
differ significantly from those employed in the $\rts=500\gev$ studies reported
here.) However,
it should be apparent from comparing Tables~\ref{nlcerrors}, \ref{fmcerrors}
and \ref{nlcfmcerrors} that if there is a SM-like
Higgs boson in the $\mhsm\lsim 2\mw$ mass region (as
expected in supersymmetric models) then 
it is very advantageous to have $L=200\fbi$
of data from both $\rts=500\gev$ running and from an FMC $s$-channel scan
of the Higgs resonance. Thus, the importance of
obtaining a full complement of Higgs boson data on a reasonable time scale
argues for having either an NLC plus a FMC or two FMC's. A single FMC
with two final rings --- one optimized for $\rts=\mhsm$
and one for $\rts=500\gev$ --- would suffice, but take twice
as long (8 years at $L_{\rm year}=50\fbi$) to accumulate the necessary data.

\subsubsection{Measuring {\boldmath $\mhsm$}
 at TeV33, LHC and NLC}\label{sssmh}

In our discussion, we will focus on the $\mhsm\leq 2\mw$ mass region,
but give some results for higher masses.
In the $\mhsm\leq 2\mw$ region,
measurement of the Higgs boson mass at the LHC and/or NLC
will be of great practical importance for the FMC since it
will enable a scan of the Higgs resonance with minimal luminosity wasted on
locating the center of the peak. Ultimately the accuracy
of the Higgs mass measurement will impact precision tests of
loop corrections, both in the SM and in extended models such as the MSSM.
For example, in the minimal supersymmetric standard model,
the prediction for
the mass of the light SM-like $\hl$ to one loop is:~\cite{haber}
\begin{equation}
\mhl^2={1\over
2}\Biggl[\mha^2+\mz^2
 -\biggl\{(\mha^2+\mz^2)^2 
-4\mha^2\mz^2\cos^22\beta\biggr\}^{1/2}~\Biggr]
  + \Delta\mhl^2\,,
\label{mhlform}
\end{equation}
where
$\Delta\mhl^2=3g^2\mt^4\ln\left(\mstop^2/\mt^2\right)/[8\pi^2\mw^2]$. Here,
$\mstop$ is the top-squark mass and
we have simplified by neglecting top-squark mixing and non-degeneracy.
From Eq.~(\ref{mhlform}), one can compute $d\mhl/d\mha$, $d\mhl/d\tanb$,
$d\mhl/d\mt$, and $d\mhl/d\mstop$ for a given choice of input parameters.
These derivatives determine the sensitivity of these parameters to the error
in $\mhl$.  For example, for $\mha=200\gev$, $\mstop=260\gev$, $\tanb=14$
and $\mt=175\gev$, for which $\mhl=100\gev$, we find that
a $\pm 100\mev$ measurement of
$\mhl$ (a precision that should be easily achieved, as discussed below)
would translate into constraints (for variations of one variable
at a time) on $\mha$, $\tanb$, 
$\mt$ and $\mstop$ of about $\pm 37\gev$, $\pm 0.7$, $\pm 670\mev$
and $\pm 1\gev$, respectively.  Since $\mt$ will be known to much
better accuracy than this and (for such low $\mha$)
the $\ha$ would be observed and its
mass measured with reasonable accuracy, 
the determination of $\mhl$ would be used as a joint constraint on
$\mstop$ and $\tanb$. More generally, 
squark mixing parameters should be included in the analysis.
The challenge will be to compute higher loop corrections to $\mhl$
to the $\pm 100\mev$ level.

Determination of $\mhsm$ will proceed
by examining a peaked mass distribution 
constructed using the measured momenta of particles appearing
in the final state. At TeV33 and the LHC, these will be the particles
into which the Higgs boson decays. For $Z\hsm$ production at the NLC,
there are two possibilities; 
we may employ the $Z\to \ell^+\ell^-$ decay products and
reconstruct the recoil mass peak or we may directly reconstruct
the Higgs mass from its decay products.
The accuracy of the Higgs boson mass
determination will depend upon the technique/channel, 
the detector performance and
the signal and background statistics.  Details are presented
in Higgs96.~\cite{snowmasssummary} Here, we give only
a bare outline of the results.

At LEP2, the accuracy of the $\mhsm$ measurement
will be limited by statistics. For
a conservative resolution of $\gamr\sim 3\gev$,~\cite{janotcom} one finds
the errors quoted in Table~\ref{dmhsm}.

At the Tevatron, the primary discovery mode is $W\hsm$ with $\hsm\to b\anti b$.
Assuming an ultimate integrated
luminosity of $L=60\fbi$ (3 years for two detectors) we find
the statistical errors quoted in Table~\ref{dmhsm}.
Allowing for systematic effects
at the level of $\Delta\mh^{\rm syst}=0.01\mh$, added in quadrature,
already increases these errors substantially. 
It is crucial that systematic effects be well controlled.

At the LHC, the  excellent $\gam\gam$ mass resolution planned
by both the ATLAS and CMS detectors implies that the best mass measurement 
in the $\mhsm\lsim 150\gev$ range will come from detection modes in
which $\hsm\to \gam\gam$; the production modes for which detection
in the $\gam\gam$ final state is possible are $gg\to\hsm$ inclusive
and $W\hsm,t\anti t\hsm$ associated production. After combining
the results for the two modes and the two detectors,
and including a systematic error (in quadrature)
given by $\delmhsm^{\rm syst}=0.001\mhsm$ (the ATLAS estimate).
The resulting net error $\delmhsm$ is given as 
a function of $\mhsm$ in Table~\ref{dmhsm}
For $\mhsm\gsim 130\gev$, $\mhsm$ can also be determined at the LHC
using the inclusive
$\hsm\to ZZ^{(*)}\to 4\ell$ final state.  
After including a 1 per mil systematic uncertainty in the overall mass scale,
we obtain the $\delmhsm$ values given in Table~\ref{dmhsm}.

\begin{table}[h]
\caption[fake]{Summary of approximate errors, $\Delta\mhsm$, for $\mhsm\leq
300\gev$. LEP2 errors are for $L=600\pbi$. Tev33 errors are for $L=60\fbi$. 
LHC errors are for $L=600\fbi$ for ATLAS+CMS.
NLC errors are given for a luminosity times efficiency 
of $L\eps=200\fbi\times 0.6$ at $\rts=500\gev$ and 
`standard' NLC~\cite{nlc} hadronic calorimetry.
NLC threshold results are for $L=50\fbi$ at $\rts=\mz+\mhsm+0.5\gev$, \ie\
just above threshold, and are quoted before including beamstrahlung,
bremsstrahlung and beam energy smearing.
FMC scan errors are for $L=200\fbi$ devoted to the scan
with beam energy resolution of 0.01\%. TeV33 and NLC errors are
statistical only. Systematic FMC error is neglected assuming
extremely accurate beam energy determination.}
\small
\begin{center}
\begin{tabular}{|c|c|c|c|c|}
\hline
 Machine/Technique & \multicolumn{4}{c|}{$\Delta\mhsm$ (MeV)} \\
\hline
\hline
{$\bf\mhsm$}{\bf (GeV)} & {\bf 80} & {\bf $\mz$} & {\bf 100} & {\bf 110} \\
\hline
LEP2 & 250 & 400 & $-$ & $-$ \\
\hline
TeV33 & 960 & ? & 1500 & 2000 \\
\hline
LHC/$\gam\gam$ (stat+syst) & 90 & 90 & 95 & 100 \\
\hline
NLC/hadronic $\rts=500$ & 51 & ? & 51 & 51 \\
\hline
NLC/threshold & 40 & 70 & 55 & 58 \\
\hline
FMC/scan & 0.025 & 0.35 & 0.1 & 0.08 \\
\hline
\hline
{$\bf\mhsm$}{\bf (GeV)} & {\bf 120} & {\bf 130} & {\bf 140} & {\bf 150} \\
\hline
TeV33 & 2700 & $-$ & $-$ & $-$ \\
\hline
LHC/$\gam\gam$ (stat+syst) & 105 & 110 & 130 & 150 \\
\hline
LHC/$4\ell$ (stat+syst) & $-$ & 164 & 111 & 90 \\
\hline
NLC/hadronic $\rts=500$ & 52 & 52 & 53 & 55 \\
\hline
NLC/threshold & 65 & 75 & 85 & 100 \\
\hline
FMC/scan & 0.06 & 0.12 & 0.20 & 0.49 \\
\hline
\hline
{$\bf\mhsm$}{\bf (GeV)} & {\bf 170} & {\bf 190} & {\bf 200} & {\bf 300} \\
\hline
LHC/$4\ell$ (stat+syst) & 274 & 67 & 56 & 90 \\
\hline
NLC/hadronic $\rts=500$ & 58 & 62 & 65 & 113 \\
\hline
NLC/threshold & 120 & 150 & 170 & ? \\
\hline
\end{tabular}
\end{center}
\label{dmhsm}
\end{table}

At the NLC, $\delmhsm$ depends upon
the tracking/calorimeter performance and the technique employed.
Assuming that $L=200\fbi$ is accumulated
at $\rts=500\gev$, for $\mhsm\lsim 2\mw$ the best technique is to reconstruct
the Higgs peak using the $b\anti b$
and $W\wstar$ hadronic final state decay channels.
Results appear in Table~\ref{dmhsm}.
One finds that distinctly greater accuracy at the NLC is possible 
in the final state hadronic decay channel than
by using the $\gam\gam$ mode at the LHC.

Another technique that is available at the NLC is to employ
a threshold measurement of the $Z\hsm$ cross section.~\cite{bbghzh}
The ratio of the cross section at $\rts=\mz+\mhsm+0.5\gev$
to that at $\rts=500\gev$ is insensitive to systematic effects and
yields a rather precise $\mhsm$ determination.
The expected precisions for the Higgs mass, assuming
that $L=50\fbi$ is accumulated at 
$\rts=\mz+\mhsm+0.5\gev$,~\footnote{We deem it unlikely
that more than $L=50\fbi$ would be devoted to this special purpose
energy.}~are tabulated in Table~\ref{dmhsm}.
Bremsstrahlung, beamstrahlung and
beam energy smearing yield an increase in the tabulated errors
of 15\% at a muon collider and 35\% at an $e^+e^-$ collider. 
From Table~\ref{dmhsm}, we see that the threshold
measurement errors would be quite competitive with the NLC
errors if $\mhsm\lsim 120\gev$ and $\mhsm\not\sim\mz$.

The ultimate in $\mhsm$ accuracy is that which can be achieved
at a muon collider by scanning the Higgs mass peak in the $s$-channel.
The scan was described earlier. For $L=200\fbi$ devoted to the scan
and a beam energy resolution of $0.01\%$,
one finds~\cite{bbgh} the extraordinarily small $\Delta\mhsm$
values given in Table~\ref{dmhsm}.

\subsection{Verifying the spin, parity and CP of a Higgs boson}\label{sscp}

Much of the following material is summarized in more detail
and with more referencing in DPF95.~\cite{dpfreport}  We present
here only a very rough summary. We often focus on 
strategies and results for a relatively light SM-like Higgs boson.

If the $\hsm$ is seen in the $\gam\gam$ decay mode (as possible
at the LHC and at the NLC or FMC with sufficient luminosity
in mass regions M1, M2 and M3) or produced at the LHC via
gluon fusion (as presumably could be verified for
all mass regions) or produced
in $\gam\gam$ collisions at the NLC, then Yang's theorem
implies that it must be a scalar and not a vector, and, of course,
it must have a CP$=+$ component (C and P can no longer be regarded
as separately conserved once the Higgs is allowed to have fermionic
couplings). If the Higgs is observed with substantial rates in production
and/or decay channels that require it
to have $ZZ$ and/or $WW$ couplings, then it is very likely
to have a significant CP-even component given that the $ZZ/WW$ coupling
of a purely CP-odd Higgs boson arises only at one-loop.
Thus, if there is a Higgs boson with anything like SM-like couplings
it will be evident early-on that it
has spin-zero and a large CP$=+$ component.
Verifying that it is purely CP-even as predicted
for the $\hsm$ will be much more challenging. 

As we have discussed in earlier sections, 
observation of a Higgs boson in the $Z\h$ and/or $\epem\h$ mode
at LEP2 or the NLC via the missing-mass technique yields
a direct determination of the squared coupling $(ZZ\h)^2$.
Other techniques allow determination of $(WW\h)^2$.
At LEP2, only $Z\h$ production is useful; for
a SM-like Higgs boson its reach will be confined to $\mhsm\lsim 95\gev$ and
the accuracy of the $(ZZ\hsm)^2$ determination is quite limited ($\sim\pm 26\%$
at $\mhsm\sim\mz$).  Errors in the case of $L=200\fbi$ at the NLC
for a SM-like Higgs boson were quoted in Table~\ref{nlcerrors}
--- for $\mhsm\lsim 2\mw$, $(ZZ\hsm)^2$
can be measured to $\pm3\%-\pm4\%$ and $(WW\hsm)^2$ to $\pm 5\%-\pm8\%$.  
If the measurement yields the SM value to this accuracy, 
then the observed Higgs must be essentially
purely CP-even unless there are Higgs representations
higher than doublets.  This follows from the sum rule 
\begin{equation}
\sum_i (ZZ\h_i)^2=\sum_i (WW\h_i)^2=1
\label{srsat}
\end{equation}
(where the $(VV\h_i)^2$ -- $V=W,Z$ -- are defined relative to the SM-values)
that holds when all Higgs bosons are in singlet or doublet representations.
However, even if a single $\h$ appears to
saturate the coupling strength sum-rule, the possibility remains
that the Higgs sector is exotic and that saturation
of the sum rule by a single $\h$ is purely accidental.
Further, even if the $ZZ\h$ coupling is not full strength the $\h$
could still be purely CP-even.  To saturate the sum rule of Eq.~(\ref{srsat}), 
one need only have other Higgs bosons with appropriate
CP-even components; such Higgs bosons are present in the many attractive
models (including the minimal supersymmetric model) 
that contain additional doublet and/or 
some number of singlet Higgs representations beyond the single doublet
Higgs field of the SM.

When the $Z\h$ rate is significant, as particularly true
at the NLC, it will be possible to
cross check that there is a large CP-even component by examining
the angular distribution in $\theta$, the polar angle
of the $Z$ relative to the $\epem$ beam-axis
in the $Z\h$ (\ie\ $\epem$) center of mass.
(For a brief summary, see DPF95.~\cite{dpfreport}) 
However, the $Z\h$ rate is adequate
to measure the $\theta$ distribution only if the $\h$ has significant
$ZZ\h$ coupling, which in most models is only possible if the $\h$
has a significant CP-even component (since
only the CP-even component has a tree-level $ZZ\h$ 
coupling).  Further, if the CP-even component dominates the $ZZ\h$ coupling,
it will also dominate the angular distribution which will then 
not be sensitive to any CP-odd component of the $\h$ that might be present.
Thus, we arrive at the unfortunate conclusion
that whenever the rate is adequate for the angular distribution measurement,
the angular distribution will appear to be that for a purely CP-even
Higgs, namely $d\sigma/d\cos\theta\propto 8\mz^2/s+\beta^2\sin^2\theta$,
even if it contains a very substantial CP-odd component. Thus,
observation of the above $\theta$ distribution only implies
that the $\h$ has spin-0 and that it is not {\it primarily} CP-odd.

At machines other than the NLC, measurement of the $\theta$
distribution for $Z\h$ events will be substantially more difficult.
Rates for $Z\h$ production will be at most just adequate for detecting
the $\h$ at LEP2,
TeV33 and the LHC.  Further, at TeV33 (in the $\h\to b\anti b$ channel)
and at the LHC (in the $\h\to \gam\gam$ channel) background rates
are substantial (generally larger than the signal). Further, 
$W\h$ production at TeV33 and the LHC cannot be employed because of inability
to reconstruct the $W\h$ center of mass (as required to determine $\theta$)
in the $W\to \ell\nu$ decay mode.

The $\tauptaum$ decays of the $\h$ provide a more
democratic probe of its CP-even vs. CP-odd components~\cite{kksz,ggcp}
than does the $\theta$ angular distribution.
Further, the $\taup$ and $\taum$ decays are
self analyzing. The distribution in the azimuthal angle
between certain effective `spin' directions that can be defined
for these decays depends upon the CP mixture for the $\h$ eigenstate.
However, LEP2 is unlikely to produce
the large number of events required for decent statistical precision
for this measurement. 
Expectations at the NLC~\cite{kksz,ggcp} or FMC~\cite{ggcp}
are much better. Particularly valuable
would be a combination of $Z\h$ with $\h\to\tauptaum$
measurements at $\rts=500\gev$ at the NLC and $\mupmum\to\h\to\tauptaum$
measurements in the $s$-channel mode at the FMC. Relatively
good verification of the CP-even nature of a light SM-like $\h$ is possible.
At higher Higgs masses (and higher machine energies) the self-analyzing
nature of the $t\anti t$ final states of Higgs decay can be
exploited in analogous fashion at the two machines.

One should not give up on a direct CP determination at the LHC.
There is one technique that shows real promise.
The key is the ability to observe the Higgs in the $t\anti t\h$
production channel with $\h\to \gam\gam$ or $\h\to b\anti b$.  
We saw earlier that separation of the $t\anti t\h$ from the $W\h$
channel at the LHC can be performed with good efficiency and purity.
A procedure for then determining the CP nature of the $\h$ was
developed.~\cite{ghcp}  The $\gam\gam$ decay mode shows the greatest
promise because of a much smaller background. It is possible
to define certain projection operators that do not require knowledge
of the $t\anti t\h$ center of mass and yet are are sensitive to the angular
distributions of the $t$ and $\anti t$ relative to the $\h$.
Assuming $\mh=100\gev$ and $L=600\fbi$ for ATLAS+CMS combined, 
these projection operators distinguish between a SM-like (purely CP-even) 
Higgs boson and a purely CP-odd Higgs boson at roughly
the $6\sigma$ to  $7\sigma$ statistical level. For $\mh=100\gev$,
discrimination between a SM-like Higgs boson and a Higgs which is an equal 
mixture of CP-even and CP-odd is possible 
at the $2\sigma$ to $3\sigma$ level. (These statements assume
that the CP-even coupling squared plus CP-odd coupling squared
for $t\anti t\h$ is equal to the SM coupling-squared.)
Of course, rates are only adequate for relatively light Higgs bosons.
Verification of the efficiencies assumed in this analysis by full simulation
will be important. The projection operator technique (but not
the statistical significance associated with its application) is independent
of the overall event rate.

There is also a possibility that polarized beams at the LHC could be
used to look for spin asymmetries in the $gg\to\h$ production rate
that would be present if the $\h$ is a CP-mixed state.~\cite{gycp}

Angular distributions in the $t\anti t\h$ final state in $\epem$
collisions at the NLC or $\mupmum$ collisions at the FMC
are even more revealing than those in the $t\anti t\h$ final state
at the LHC.~\cite{gghcp,ghe} By combining $Z\h$ measurements
with $t\anti t\h$ measurements, verification of the $t\anti t$ and $ZZ$
couplings of a SM-like $\h$ will be possible at a remarkable level
of accuracy.~\cite{ghe}  For instance, for 
$\rts=1\tev$ (we must be substantially above $t\anti t\h$
threshold), 2 1/2 years of running is expected to yield $L=500\fbi$
and in the case of $\mhsm=100\gev$ we can achieve
a determination of the CP-even $t\anti t\hsm$ coupling magnitude at the
$\sim\pm 3\%$ level, the (CP-even) 
$ZZ\hsm$ coupling magnitude at the $\sim\pm 2\%$ level, and a meaningful
limitation on the CP-odd $t\anti t\hsm$ coupling magnitude.

The most elegant determination of the CP nature of Higgs boson
is probably that possible in $\gam\gam\to\h$ production
at the $\gam\gam$ collider facility of the NLC.~\cite{ggcpgamgam}
Since the CP-even and CP-odd components
of a Higgs boson couple with similar strength to $\gam\gam$ 
(via one-loop graphs),
there is no masking of the CP-odd component such as
occurs using probes involving $ZZ\h$ or $WW\h$ couplings.
The most useful measurement depends upon whether the Higgs is a pure or a mixed
CP eigenstate.
\begin{itemize}
\item
The most direct probe of a CP-mixed state is provided by
comparing the Higgs boson production rate
in collisions of two back-scattered-laser-beam 
photons of different helicities.~\cite{ggcpgamgam}
The difference in rates for photons colliding with $++$ vs. $--$ 
helicities is non-zero only if CP violation is present.
A term in the cross section changes sign when
both photon helicities are simultaneously flipped.
Experimentally, this is achieved by 
simultaneously flipping the helicities of both of the initiating
back-scattered laser beams. One finds that the asymmetry 
is typically larger than 10\% and is 
observable if the CP-even and CP-odd components of the $\h$
are both substantial.
\item
In the case of a CP-conserving Higgs sector, 
one must have colliding photons with substantial transverse polarization.
The difference in rates for parallel vs. perpendicular
polarizations divided by the sum is $+1$ ($-1$) for a CP-even (CP-odd)
Higgs boson. The statistical accuracy with which this ratio can
be measured is strongly dependent upon the degree of transverse polarization
that can be achieved for the energetic colliding photons.
The most obvious means
of achieving transverse polarization for the colliding
photons is by transversely polarizing the incoming
back-scattered laser beams (while maintaining the ability
to rotate these polarizations relative to one another) and optimizing
the laser beam energy.  This optimization has been 
discussed.~\cite{gkgamgamcp,kksz} The transverse
polarization achieved is not large, but still
statistics are predicted to be such 
that, with not unreasonable integrated luminosity,
one could ascertain that a SM-like $\h$ is CP-even vs. CP-odd.
A new proposal~\cite{kotkinserbo} has recently appeared
that could potentially result in
nearly 100\% transverse polarization for the colliding photons.
This would allow excellent statistical accuracy for the
transverse-polarization cross sections and a high degree of statistical
discrimination between CP-even vs. CP-odd.

\end{itemize}

A $\mupmum$ collider might provide an analogous opportunity
for directly probing the CP properties of any Higgs boson that
can be produced and detected in the $s$-channel mode.~\cite{atsoncp,dpfreport}
However, it must be possible to transversely
polarize the muon beams.  Assume that
we can have 100\% transverse polarization and that 
the $\mu^+$ transverse polarization is rotated with respect
to the $\mu^-$ transverse polarization by an angle $\phi$.  The production
cross section for a $\h$ with coupling of a mixed
CP nature exhibits a substantial asymmetry of the form~\cite{atsoncp}
\begin{equation}
A_1\equiv {\sigma(\pi/2)-\sigma(-\pi/2)\over \sigma(\pi/2)+\sigma(-\pi/2)}\,.
\end{equation}
For a pure CP eigenstate, the asymmetry~\cite{dpfreport}
\begin{equation}
A_2\equiv {\sigma(\pi)-\sigma(0) \over \sigma(\pi)+\sigma(0)}
\end{equation}
is $+1$ or $-1$ for a CP-even or CP-odd $\h$, respectively.
Of course, background processes in the final states where
a Higgs boson can be most easily observed ({\it e.g.} $b\anti b$
for the MSSM Higgs bosons) will typically dilute these asymmetries
substantially. Whether or not they will prove useful depends even more 
upon our very uncertain ability to transversely polarize the muon
beams while maintaining high luminosity.

\section{Non-minimal Higgs bosons}\label{snonsm}

The most attractive non-minimal Higgs
sectors are those containing extra doublet and/or singlet fields.
In this section, we will focus on the
particularly attractive MSSM and NMSSM supersymmetric models
in which the Higgs sector consists of exactly two doublets or two doublets
plus one singlet, respectively. Some of the material presented
is extracted from the DPF95~\cite{dpfreport} and
Higgs96~\cite{snowmasssummary} reports. 

\subsection{Branching ratios}

Higgs branching ratios are crucial in determining the modes and channels
for Higgs boson discovery and study. A detailed review is not possible here.
Branching ratio graphs for the MSSM appear in many 
places.~\cite{perspectivesi,dpfreport,gunionerice,djouadi,brdjouadi,brkunszt,brbartl,brdjouadinew}
Only a few broad discussions 
of NMSSM branching ratios are available.~\cite{hhg,eghrz}
In preparation for the following discussions
we mention some of the general features of the MSSM
branching ratios. For simplicity, the outline presented focuses
on the portion of parameter space where $\mha> 2\mz$, for which
the $\hl$ is SM-like and the $\hh$ has largely decoupled from $WW,ZZ$.
\begin{itemize}
\item
The largest decay modes for the $\hl$ are the $b\anti b$
and $\tauptaum$ channels. 
The $\gam\gam$ decay branching ratio is small
but crucial (just as for the $\hsm$). For squark mixing
scenarios such that $\mhl\geq 130\gev$, then $\hl\to W\wstar,Z\zstar$
can also become significant. If the lightest neutralino $\cnone$
has low enough mass, $\hl\to\cnone\cnone$ can be an important decay mode.
\item
The possibilities for $\hh$ decays are very numerous.  At large $\tanb$,
$\hh\to b\anti b,\tauptaum$ are the dominant decays (due to
their enhanced couplings) regardless of
what other channels are kinematically allowed. At low to moderate $\tanb$,
many modes compete. The most important and/or useful are 
$b\anti b$, $\tauptaum$, $\hl\hl$, $ZZ^{(*)}$, $WW^{(*)}$, and
$t\anti t$.
Among these, the $\hl\hl$ channel is dominant below $t\anti t$ threshold.
At low to moderate $\tanb$, superparticle pair channels can also be
important when kinematically allowed. These include $\chitil\chitil$
chargino and neutralino pairs, $\slep{\slep},\snu{\snu}$ 
slepton and sneutrino
pairs, and $\sq{\sq}$ squark pairs; in this last category,
$\stopone{\stopone},\stoptwo{\stoptwo}$
pairs could be particularly important.~\cite{brbartl} 
\item
The decays of the $\ha$ are equally varied.  At large $\tanb$, $b\anti b$
and $\tauptaum$ are the dominant channels (just as for the $\hh$). At low
to moderate $\tanb$, the most important non-superparticle competing modes are
$b\anti b$, $\tauptaum$, $Z\hl$ and $t\anti t$. Potentially important
superparticle pair channels are $\chitil\chitil$ and
$\stopone{\stoptwo}$
(large mixing being required for scalar sparticle pair channels
in the case of the $\ha$).
\item
Potentially important decays of the $\hp$ include the $\tau^+\nu_\tau$, $t\anti
b$, $\wp\hl$, $c\anti s$ non-supersymmetric channels, and the
$\chitil^+\chitil^0$, $\stop\,{\sbot}$ supersymmetric channels.
\end{itemize}
In what follows,
further details regarding branching ratios will be noted where necessary.

\subsection{The MSSM at the LHC}\label{ssmssmlhc}

\begin{figure}[htbp]
\leavevmode
\begin{center}
\centerline{\psfig{file=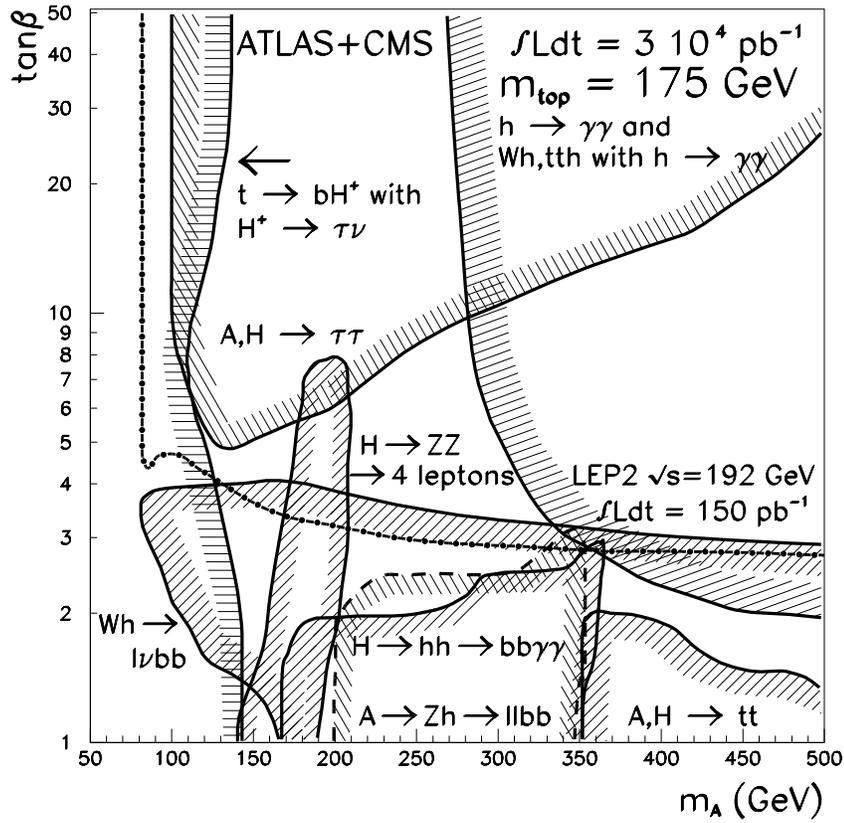,width=12.2cm}}
\end{center}
\caption{Discovery contours ($5\sigma$) in the parameter space of the 
minimal supersymmetric model for ATLAS+CMS at the 
LHC:~\protect\cite{latestplots} $L=30\fbi$. \Twoloop\ radiative corrections 
to the MSSM Higgs sector are included
assuming $\mstop=1\tev$ and no squark mixing.}
\label{mssmlolum}
\end{figure}

\begin{figure}[htbp]
\leavevmode
\begin{center}
\centerline{\psfig{file=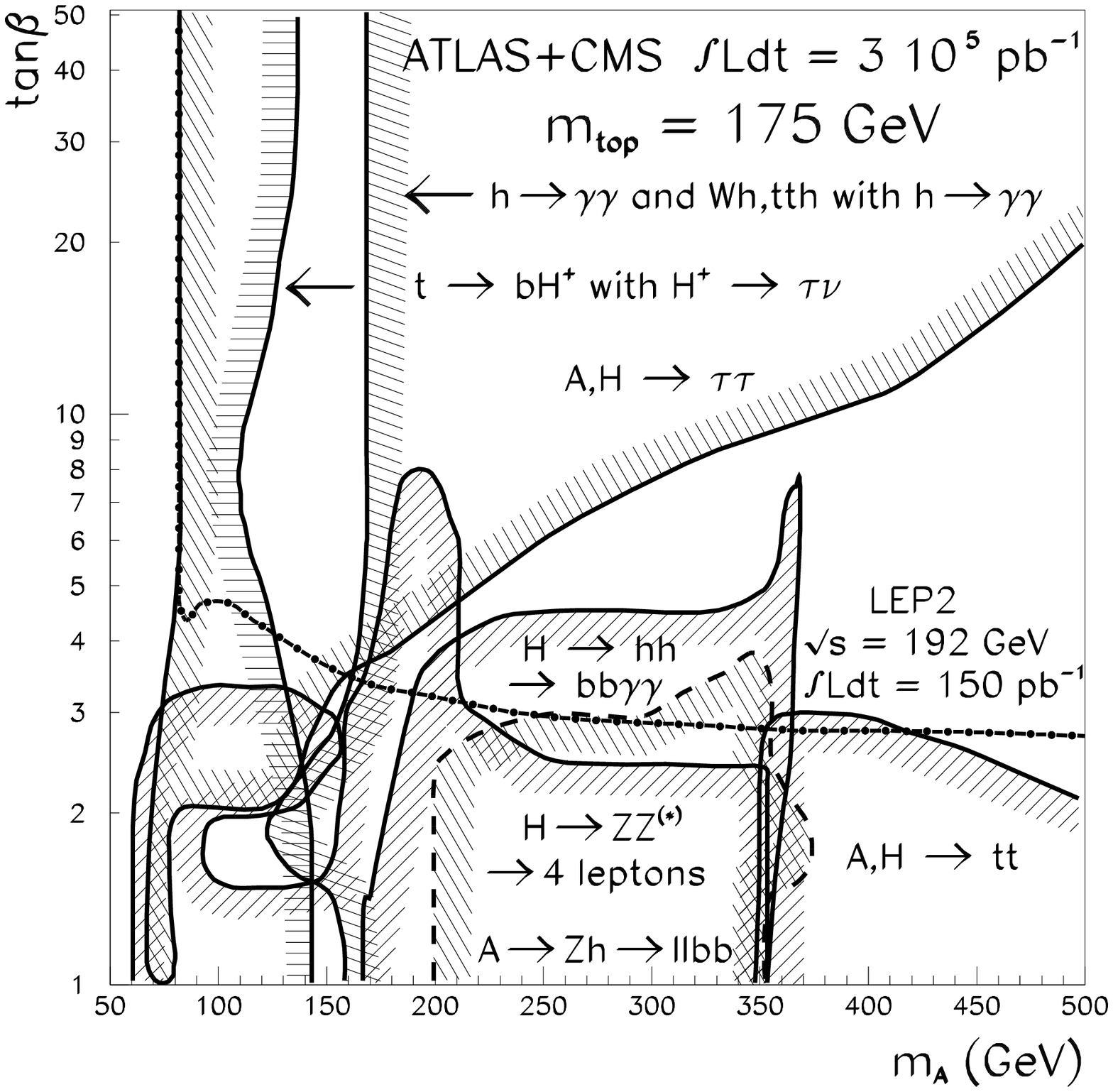,width=12.2cm}}
\end{center}
\caption{Discovery contours ($5\sigma$) in the parameter space of the 
minimal supersymmetric model for ATLAS+CMS at the
LHC:~\protect\cite{latestplots} $L=300\fbi$. \Twoloop\ radiative corrections 
to the MSSM Higgs sector are included
assuming $\mstop=1\tev$ and no squark mixing.}
\label{mssmhilum}
\end{figure}

In the extreme decoupling limit of very large $\mha$, the $\hl$ will
be very SM-like and is guaranteed to be visible in either the $\gam\gam$
or $Z\zstar$ decay modes. The $\hh$, $\ha$ and $\hpm$ are unlikely
to be detected. However, for moderate to low $\mha$ the phenomenology is
far more complex. We review the situation assuming 
a top squark mass of order $1\tev$.  First,
there is very little of the standard $(\mha,\tanb)$
parameter space in which all four SUSY 
Higgs bosons are observable; rather, one asks if at least one SUSY Higgs 
boson can be detected over the entire parameter space.  This appears to be
the case, using a combination of detection modes. The
early theoretical studies of this issue~\cite{KZ,BBKT,GO,BCPS} 
and newer ideas (to be referenced below) have
been confirmed and extended in detailed studies by
the ATLAS and CMS detector groups.
Surveys of the experimental studies are 
available.~\cite{fgianotti,latestplots}
Figures for ATLAS+CMS at low ($L=30\fbi$) luminosity and
high ($L=300\fbi$) luminosity~\cite{fgianotti,latestplots}
are included below as Figs.~\ref{mssmlolum} and \ref{mssmhilum},
respectively. Note that the ATLAS+CMS notation
means that the $L=30\fbi$ or $L=300\fbi$ 
signals from the two detectors are combined
in determining the statistical significance of a given signal.
All results discussed in the following are
those obtained without including higher order QCD ``$K$'' factors
in the signal and background cross sections.  The $K$ factors
for both signal and background are presumably significant; if
they are similar in size, then statistical significances would
be enhanced by a factor of $\sqrt K$. Full \twoloop\ radiative
corrections to the Higgs masses and couplings~\cite{haber} have
been included. Supersymmetric decays of the Higgs bosons were assumed
to be kinematically forbidden in obtaining the results discussed.

In the limit $\mha \to \infty$, the $\hh$, $\ha$, and $\hpm$ are all heavy, 
and decouple from the weak bosons. The lightest neutral scalar Higgs boson, 
$\hl$, approaches its upper bound, and behaves like a 
standard Higgs boson. Since this bound (for pole mass $\mt=175\gev$)
is about 113 GeV
(assuming small stop-squark mixing and $\mstop\leq 1\tev$), the primary
channels for $\hl$ detection will be those based on
the $\gam\gam$ decay mode.  The $5\sigma$ contours are
shown in Figs.~\ref{mssmlolum} and \ref{mssmhilum}. At high luminosity
$\hl$ discovery in its $\gam\gam$ decay mode is possible
for $\mha\gsim 170$. For low luminosity the coverage of the $\gam\gam$
mode decreases substantially, reaching only down as far
as $\mha\sim 270\gev$ at high $\tanb$ with no coverage for any $\mha$
if $\tanb\lsim 2$. For large stop mixing, 
the maximum $\mhl$ mass increases to about $130\gev$,
and the $\hl$ will also be observable via $\hl \to Z\zstar \to 4\ell$ over 
an overlapping part of the parameter space.

For $\tanb \sim 1$, the lightest scalar Higgs is observable at LEP2 via
$e^+e^- \to \ha\hl,Z\hl$.  
Including \twoloop\ corrections ($\mstop=1\tev$, no squark mixing)
for $\mt=175\gev$ the LEP-192 discovery region asymptotes 
at $\tanb\lsim 3$, assuming $L=150\pbi$ per detector,
as shown in Figs.~\ref{mssmlolum} and \ref{mssmhilum}.~\cite{janot}

For $60\lsim \mha \lsim 2\mt$ the heavy scalar Higgs has high enough mass and 
for $\tanb\lsim 3$ maintains enough of a 
coupling to weak vector bosons to allow its discovery via $\hh \to ZZ^{(*)} 
\to 4\ell$ at high luminosity, as shown in Fig.~\ref{mssmhilum}. 
The height in $\tanb$ as a function of $\mha$ 
of the $\hh\rta 4\ell$ discovery region varies significantly for $\mha\lsim
2\mt$ due to large swings in the branching ratio for $\hh\rta\hl\hl$ decays,
rising as high as $\tanb\sim 8$ for $\mha\sim 190\gev$ (where
$BR(\hh\rta\hl\hl)$ actually has a zero).
The importance of the $\hh\rta\hl\hl$ decays makes the $4\ell$
mode of marginal utility at low luminosity except for $\mha\sim 190\gev$,
see Fig.~\ref{mssmlolum}.
At high luminosity, the $\hh\rta 4\ell$ 
contour is cut off for $\mha \approx \mhh > 2 m_t$ due to the dominance
of the decay $\hh\to t\anti t$.
The $\hh\rta\hl\hl$ and $\hh,\ha\rta t\anti t$ 
channels can also provide Higgs signals. The key ingredient in employing
these channels is efficient and pure $b$-tagging. We will discuss
these modes shortly.

For $\mha \approx 70$ GeV and $\tanb >$ 3 (CMS) or 5 (ATLAS), the heavy 
scalar Higgs is observable in its two-photon decay mode. 
This is not indicated in the plots given here.

These ``standard'' modes are not enough to cover the entire SUSY 
parameter space, so others must be considered. The uncovered region 
is for large $\tanb$ and moderate $\mha$. In this region,
the $\hl$ has suppressed $ZZ^*$ and $\gam\gam$ branching ratios
compared to the $\hsm$ and must be sought in its decay
to $b\anti b$ or $\taup\taum$ in this region. 
Since $\mhl < 113$ GeV for $\mt=175\gev$
(taking $\mstop=1\tev$ and assuming no squark mixing) the $\hl$
is too close to the $Z$ peak to be observed.  
Thus, observation of the $\hh,\ha,\hpm$ will be crucial.  The ATLAS
and CMS detector groups have found that observation of $\hh,\ha\to\tauptaum$
will be viable.  The masses $\mha\sim\mhh$
are generally sufficiently above $\mz$ to avoid
being swamped by the $Z\to\tauptaum$ background and, for large $\tanb$,
the cross section for the production of these particles in association with
$b\anti b$ is greatly enhanced; it is the dominant production 
mechanism.~\cite{DW} Further, at large $\tanb$ one
finds $\br(\ha,\hh\to \tauptaum)\sim 0.1$, even when $t\anti t$ or SUSY
decay modes are allowed.
The region in the ($\mha,\tanb$) plane which is covered by the 
$\hh,\ha \to \taup
\taum$ channel is shown in Figs.~\ref{mssmlolum} and \ref{mssmhilum}. 
For $L=300\fbi$ and $\mt=175\gev$ the region over
which the $\ha,\hh\rta \tau\tau$ discovery channel  is viable
extends all the way down to $\tanb=1$ for $\mha\sim 70\gev$, but
is limited to
$\tanb> 20$ by $\mha\sim 500\gev$. (For $\tanb\lsim 2$, the 
$gg\rta \ha\rta \tau\tau$ reaction provides the crucial contribution to
this signal.) This, in particular, means that discovery
of the $\hh,\ha$ will be possible for $80\gev\lsim\mha\lsim 160\gev$ 
and $\tanb\gsim 4$
where the $\hl\to\gam\gam$ modes are not viable and $Z\hl$
production cannot be observed at LEP2.

CMS has explored the decay modes $\hl,\hh,\ha \to \mu^+\mu^-$ for large $\tanb$.
Although the branching ratio is very small, about $3 \times 10^{-4}$, the 
large enhancement of the cross section for $b\anti b\ha$ 
and either $b\anti b \hh$ (high $\mha$) or $b\anti b \hl$ (low $\mha$)
compensates.
The main background is Drell-Yan production of $\mu^+\mu^-$.  
Very roughly, $\tanb> 10$
is required for this mode to be viable
if $\mha\sim 100\gev$, rising to $\tanb> 30$ by
$\mha\sim500\gev$. The $\mupmum$ contours
are close to the $\tauptaum$ contour obtained with $L=10 \fbi$, but the
$\mu^+\mu^-$ channel yields a cleaner signal identification and better mass 
resolution. Nonetheless, the $\taup\taum$ mode will
probe to lower $\tanb$ values at any given $\mha$. At high $\tanb$,
both the $\tauptaum$ and $\mupmum$ signals can be enhanced by tagging
the $b$ jets produced in association with the Higgs bosons.
It will be interesting to see how the $\mupmum$ and $\tauptaum$
modes compare once $b$-tagging is required.

The charged Higgs boson of the minimal supersymmetric model is best sought in
top-quark decays, $t\to \hp b$. For $\tanb >1$, the branching ratio 
of $\hp \to \taup\nu_{\tau}$ exceeds $30\%$, and is nearly unity for 
$\tanb > 2$.  CMS and ATLAS have studied the
signal from $t\anti t$ events with one semileptonic top decay and one top
decay to a charged Higgs, followed by $\hp \to \taup\nu_{\tau}$. 
After accounting for backgrounds,
CMS and ATLAS find that, with $30 \fbi$ of 
integrated luminosity each, a charged Higgs of mass less than about 140 GeV can be
detected for all values of $\tanb$, extending to $\lsim 160\gev$ 
at low or high $\tanb$ values,
in the case of a top-quark pole mass of $175\gev$.  This is
indicated by the approximately vertical contour that begins
at $\mha=140\gev$ at $\tanb=1$ in Fig.~\ref{mssmlolum};
the coverage expands slightly for $L=300\fbi$, 
as shown in Fig.~\ref{mssmhilum}.

These processes combined are enough to guarantee
detection of at least one MSSM Higgs boson throughout
the entire SUSY parameter space at high luminosity but not at low
luminosity. 
This is because of the much more extensive coverage of the $\hl\to\gam\gam$
and $\hh,\ha\to\taup\taum$ modes at high luminosity.
The observability of the $\hh,\ha\to \tauptaum$
modes is such that $L=600\fbi$ (combining ATLAS+CMS) provides more
than adequate coverage of the entire $(\mha,\tanb)$
parameter plane.  At $L=100\fbi$ coverage is already complete.

We shall now turn to a discussion of additional detection modes
that rely on $b$-tagging.  Not only might these modes provide
backup in this `hole' region, they also expand the portions
of parameter space over which more than one of the MSSM Higgs 
bosons can be discovered. Equally important, they allow a direct
probe of the often dominant $b\anti b$ decay channel.
Expectations for $b$-tagging efficiency and purity have improved
dramatically since these modes were first examined.~\cite{DGV1}
For current estimates, ATLAS and CMS employ
efficiency (purity) of 60\% (99\%) for $p_T\gsim 15\gev$
for low luminosity running 
and 50\% (98\%) for $p_T\gsim 30\gev$ 
for high luminosity running, obtained solely from
vertex tagging; high-$p_T$ lepton tags
could further improve these efficiencies.

The most direct way to take advantage of $b$-tagging is to employ
$W$+Higgs~\cite{SMW2} and $t\anti t$+Higgs~\cite{tthg,DGV2}
production, where the Higgs boson decays to $b\anti b$.
As already discussed, these modes are marginal
for the SM Higgs, but in the MSSM
both have the potential to contribute in the hole region
since the $WW$ and $t\anti t$ couplings can be of roughly standard
model strength~\footnote{The $\hl$ has approximately SM strength couplings
once $\mha\gsim \mz$, while the $\hh$ has roughly SM-like strength couplings
when $\mha\lsim\mz$ and $\mhh$ approaches its lower bound
(a more precise discussion appears in DPF95~\cite{dpfreport}).}
while the $b\anti b$ branching ratio can be somewhat enhanced.
The impact of the $W\hl$ (with $\hl\rta b\anti b$) mode has been 
examined.~\cite{fgianotti,latestplots}
The coverage provided by this mode for $L=30\fbi$
after combining the ATLAS signal with a presumably equal
signal from CMS is illustrated in Fig.~\ref{mssmlolum}.
There, the $W\hl$ mode is shown to cover most of the $\mha\gsim 100\gev$,
$\tanb\lsim 4$ region, where $\mhl\lsim 105\gev$.
Unfortunately, it seems that the experimental analysis does
not find enough enhancement for the $\hl$ rate relative
to the $\hsm$ rate in this channel to
provide backup in the `hole' region of the low-luminosity figure.
It should be noted, however, that the boundary of $\tanb\lsim 4$
is almost certainly a very soft one, depending delicately
on the exact luminosity assumed, precise \twoloop\
radiative corrections employed, and so forth. For instance, as $\mstop$
is lowered below $1\tev$, the upper bound on $\mhl$ decreases rapidly,
and the region of viability for this mode would expand dramatically.
At high luminosity, event pile-up, makes isolation of the $W$+Higgs mode
difficult. However, the $t\anti t$+Higgs mode is likely to be viable.
The experimental studies of the $t\anti t\hl$ (with $\hl\to b\anti b$)
mode have not been refined
to the point that a corresponding contour has been included
in Figs.~\ref{mssmlolum} and \ref{mssmhilum}. 
The theoretical results~\cite{DGV2} claim substantial 
coverage in the hole region.

\begin{figure}[htbp]
\leavevmode
\begin{center}
\centerline{\psfig{file=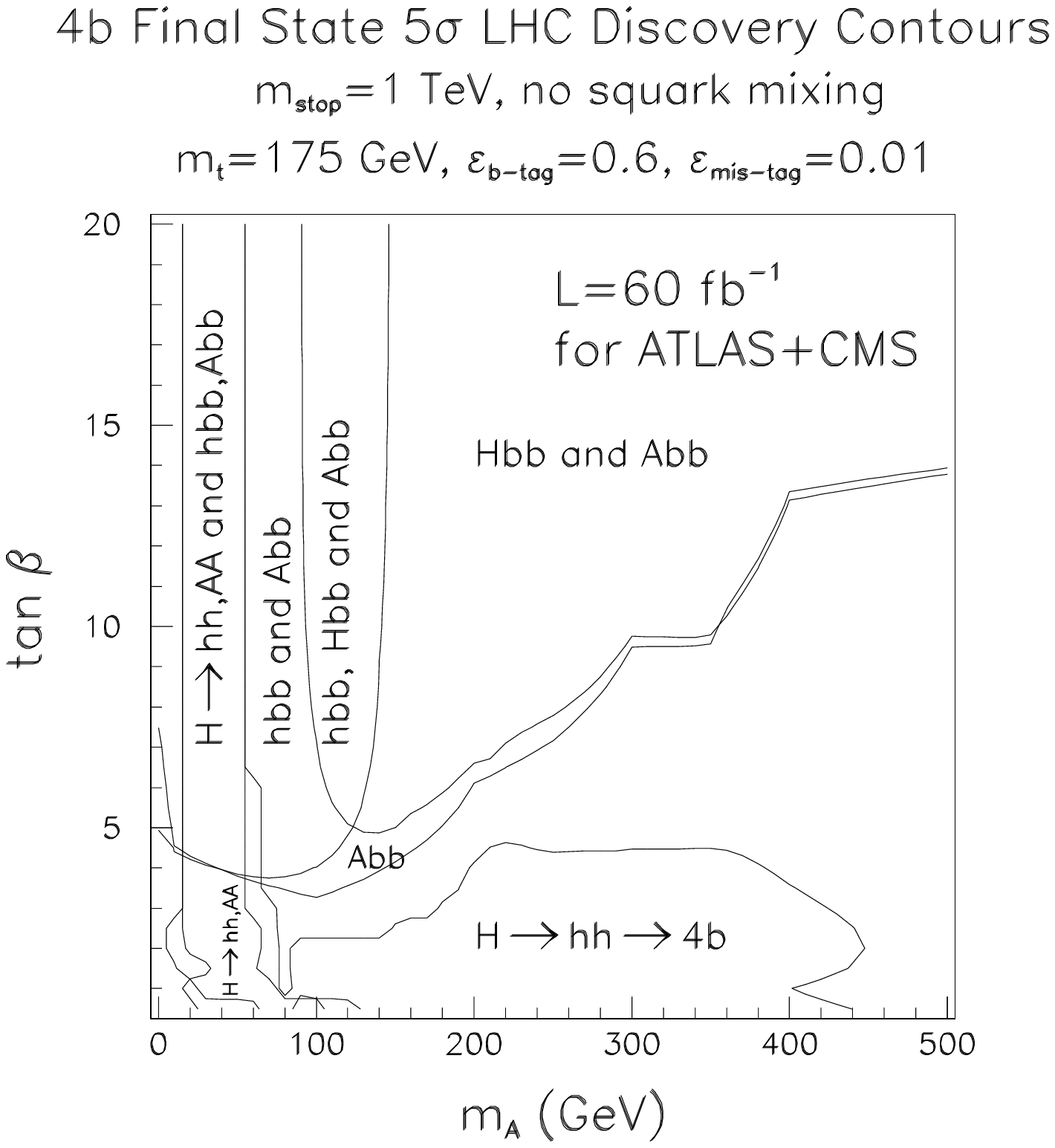,width=12.2cm}}
\end{center}
\caption[fake]{$4b$ final state $5\sigma$ discovery regions
for $\hh b\anti b$, $\ha b\anti b$, $\hh\rta\hl\hl$ and $\hh\rta\ha\ha$
in $(\mha,\tanb)$ parameter space at the LHC for 
combined ATLAS+CMS luminosity of $L=60\fbi$, assuming
that an efficient $4b$ trigger can be developed.
\Twoloop\ radiative corrections to the MSSM Higgs sector
are included assuming $\mt=175\gev$,
$\mstop=1\tev$ and no squark mixing.}
\label{4bfigure}
\end{figure}

For large $\tanb$, the enhanced cross section 
for associated production of SUSY Higgs bosons with $b\anti b$, 
followed by Higgs decay to $b\anti b$, yields a
four $b$-jet signal.~\cite{DGV3} Tagging at least three $b$ jets 
with $p_T>15\gev$ is required
to reduce backgrounds.  It is necessary to establish an efficient
trigger for these events in order to observe this signal;
this is currently being studied by the ATLAS and CMS collaborations.  
The dominant 
backgrounds are $gg \to b\anti bb\anti b$, and $gg\to b\anti bg$ with a 
mis-tag of the gluon jet. Assuming the latest $60\%$ efficiency
and 99\% purity for $b$-tagging at $L=30\fbi$, the parameter
space regions for which 
the $b\anti b\hl$,  $b\anti b\hh$ and $b\anti b\ha$
reactions yield a viable signal in the $4b$ final state
are displayed in Fig.~\ref{4bfigure}.~\cite{dgv3update}
They imply that the $4b$ final state
could be competitive with the $\taup\taum$ final state modes
for detecting the $\hh$ and $\ha$ if an efficient trigger can be
developed for the former. Even if a full $5\sigma$ signal cannot be
seen in the $4b$ final states, once the $\hh,\ha$ are observed in
the $b\anti b\tauptaum$ decay mode
and $\mhh,\mha$ determined, the $b\anti b b\anti b$ final states
will allow a determination of $\br(\hh,\ha\to b\anti b)/\br(\hh,\ha\to
\tauptaum)$ of reasonable accuracy.  This will be a very important
test of our understanding of the couplings of the heavier MSSM
Higgs bosons.

For $\mhpm > m_t + m_b$, one can consider searching for the decay of the 
charged Higgs to $t\anti b$.  This signal is most promising when used in 
conjunction with the production processes $gg \to t\anti b\hm,b\anti t\hp$, 
and tagging 
several of the four $b$ jets in the final state.~\cite{GBPR} For moderate
$\tanb$, the production cross section is suppressed such that the signal is
not observable above the irreducible $t\anti tb\anti b$ background.  The 
potential of this process is therefore limited to small and large values of
$\tanb$.  With $200 \fbi$, a signal may be observable for $\tanb< 2$ and
$\mhpm < 400$ GeV, and for $\tanb>20$ and $\mhpm<300$ GeV. 

CMS and ATLAS have 
considered the process $gg \to \ha \to Z\hl \to \ell^+\ell^-b\anti 
b$.  For $\tanb < 2$, the branching ratio of $\ha \to Z\hl$ is about 
$50\%$.  They have demonstrated an observable signal with single and double
$b$ tagging. In Fig.~\ref{mssmhilum} one finds a discovery region
for $200\lsim\mha\lsim 2\mt$ and $\tanb\lsim 3$ for $L=600\fbi$ (\ie\
$L=300\fbi$ for ATLAS and CMS separately),
reduced to $\tanb\lsim 2$ for $L=60\fbi$, Fig.~\ref{mssmlolum}.

Recent results from CMS and ATLAS for the mode
$\hh,\ha\rta t\anti t$ also appear in Figs.~\ref{mssmlolum} and \ref{mssmhilum}.
Even with good $b$-tagging,
the decays $\hh,\ha \to t\anti t$ are challenging to detect at 
the LHC due to the large background from $gg \to t\anti t$.
Nonetheless, the preliminary studies indicate that,
for the anticipated $b$-tagging capability, ATLAS and CMS
can detect $\ha,\hh\rta t\anti t$ for 
$\tanb\lsim 2-1.5$ with $L=60\fbi$ and for $\tanb\lsim 3-2.5$ with $L=600\fbi$.
Caution in accepting these preliminary results is perhaps warranted
since they have been obtained by
simply comparing signal and background cross section levels;
excellent knowledge of the magnitude of the 
$t\anti t$ background will then be required since $S/B\sim 0.02-0.1$.

The $\hh\rta\hl\hl$ mode can potentially be employed in the channels
$\hl\hl\rta b\anti b b\anti b$ and $\hl\hl\rta b\anti b \gam\gam$.
The former mode has been explored;~\cite{DGV4} using 4 $b$-tagging
(with 50\% efficiency and 98\% purity for $p_T>30\gev$ at $L=600\fbi$)
and requiring that there be two $b\anti b$ pairs of mass $\sim \mhl$
yields a viable signal for $170\lsim\mha\lsim 500\gev$ and $\tanb\lsim 5$.
For $L=60\fbi$, $b$-tagging cuts can be softened and one
can be sensitive to lower masses.  Using 60\% efficiency
and 99\% purity for $p_T\gsim 15\gev$, one finds that $\hh\rta\hl\hl$
and/or $\hh\rta\ha\ha$ can also be detected in the region 
$\mha\lsim 60\gev,\tanb\gsim 1$. This is illustrated in
Fig.~\ref{4bfigure}. Note from this figure that
the ATLAS+CMS $b\anti b\hl$, $b\anti b\hh$, $b\anti b\ha$, $\hh\rta\hl\hl$
and $\hh\rta\ha\ha$ $4b$ final state signals at combined $L=60\fbi$ 
yield a signal for one or more of the MSSM Higgs bosons over a very
substantial portion of parameter space.

Because of uncertainty concerning the ability to trigger
on the $4b$ final state, ATLAS and CMS have examined the 
$\hh\rta\hl\hl\rta b\anti b \gam\gam$
final state.  This is a very clean channel (with $b$ tagging),
but is rate limited. For $L=600\fbi$ (Fig.~\ref{mssmhilum}) a discovery
region for ATLAS+CMS of $175\gev\lsim\mha\lsim 2\mt$, $\tanb\lsim 4-5$ is 
found (using the 50\% efficiency and 98\% purity for $b$-tagging
claimed by ATLAS
at high luminosity); the region is substantially reduced for $L=30\fbi$
(Fig.~\ref{mssmlolum}). It is important to note
that when both $\hh\rta\hl\hl\rta 2b2\gam$ and $4b$ can be observed,
then it will be possible to determine the very important
ratio $BR(\hl\rta b\anti b)/BR(\hh\rta \gam\gam)$.

Putting together all these modes, we can summarize by saying that
for moderate $\mha\leq 200\gev$ there is an excellent chance
of detecting more than one of the MSSM Higgs bosons.  However, for large
$\mha\geq 200\gev$ (as preferred in the GUT scenarios)
only the $\hl$ is certain to be found. For $\mha\geq 200\gev$,
the $\hl$ modes that are guaranteed to be observable are the 
$\hl\rta\gam\gam$ production/decay modes ($gg\rta\hl$,
$t\anti t\hl$, and $W\hl$, all with $\hl\rta\gam\gam$).
Even for $\mha$ values as large as $400-500\gev$, it is also likely
that the production/decay mode $t\anti t\hl$ with $\hl \rta
b\anti b$ can be observed, especially if $\mstop$
is sufficiently below $1\tev$ that $\mhl$ is $\lsim 100\gev$.  
For high enough $\mstop$ or large stop mixing, 
$\hl\rta Z\zstar$ might also be detected.
Whether or not it will be possible to
see any other Higgs boson depends on $\tanb$. There are basically
three possibilities when $\mha\geq 200\gev$.
i) $\tanb\lsim 3-5$, in which case $\hh\rta \hl\hl\rta b\anti b \gam\gam,4b$
or (for $\mha\geq 2\mt$) $\ha,\hh\rta t\anti t$  will be observable;
ii) $\tanb\gsim 5$ (increasing as $\mha$ increases above $200\gev$),
for which $\ha,\hh\rta \taup\taum$ (and at larger $\tanb$, $\mu^+\mu^-$)
will be observable, supplemented by $b\anti b\ha,b\anti b\hh\rta 4b$
final states;
and iii) $3-5\lsim \tanb\lsim 6$ at $\mha\sim 250\gev$, increasing
to $3-5\lsim \tanb\lsim 13$ by $\mha\sim500\gev$,
which will be devoid of $\ha,\hh$ signals.
Further improvements in $b$-tagging efficiency and purity
would lead to a narrowing of this latter wedge of parameter space.

We must emphasize that the above results have assumed
an absence of SUSY decays of the Higgs bosons. For a light ino sector
it is possible that $\hl\rta \cnone\cnone$ will be the dominant decay.
Detection of the $\hl$ in the standard modes becomes difficult or impossible.
However, it has been demonstrated that detection 
in  $t\anti t\hl$~\cite{guninvis} and $W\hl$~\cite{fjk,cr} production
will be possible after employing cuts requiring large missing energy.
Assuming universal gaugino masses at the GUT scale, our first
warning that we must look in invisible modes would be the observation
of $\cpone\cmone$ production at LEP2.
The $\ha,\hh,\hp$ could all also
have substantial SUSY decays, especially if $\mha$ is large.  
Such decays will not be significant
if $\tanb$ is large since the $b\anti b,\taup\taum,
\mu^+\mu^-$ modes are enhanced, but would generally severely reduce signals
in the standard channels when $\tanb$ is in the small to moderate range.

Finally, we note that our discussion has focused on `direct' production
of Higgs bosons.  Substantial indirect production of 
the Higgs bosons is also possible via decay chains of abundantly
produced superparticles.  For larger $\mha$ values,
it will be mainly the $\hl$ that can have a large indirect
production rate. In particular, a strong signal for the $\hl$
in its primary $b\anti b$ decay mode can emerge from gluino pair
production in some scenarios.~\cite{hinch}

\subsection{The NMSSM at the LHC}

As summarized above, at least one of
the Higgs bosons of the MSSM can be discovered either at LEP2 or at the LHC
throughout all of the standard $(\mha,\tanb)$ parameter space.
This issue has been re-considered in the context of the NMSSM.~\cite{ghm}
In the NMSSM there is greater freedom. Assuming CP conservation
(which is not required in the NMSSM Higgs sector) there are three
instead of two CP-even Higgs bosons (denoted $\h$) 
and two CP-odd Higgs bosons
(denoted $\a$), and correspondingly greater
freedom in all their couplings. It is found that there are regions
of parameter space for which none of the NMSSM Higgs bosons can be detected
at either LEP2 or the LHC.  This result is to be contrasted with
the NLC or FMC no-lose theorem~\cite{kimoh,kot,KW,ETS} to be
discussed later, according to which at least one
of the CP-even Higgs bosons of the NMSSM
will be observable in $\zstar\to Z\h$ production.

The detection modes considered for the NMSSM are the same as those
employed in establishing the LEP2 plus LHC no-lose theorem
for the MSSM:
1) $\zstar\to Z\h$ at LEP2; 2) $\zstar\to \h\a$ at LEP2;
3) $gg\to \h\to\gam\gam$ at LHC; 4) $gg\to\h\to Z\zstar~{\rm or}~ZZ\to 4\ell$
at LHC; 5) $t\to\hp b$ at LHC;
6) $gg\to b\anti b \h,b\anti b\a \to b\anti b \tauptaum$ at LHC;
7)  $gg\to\h,\a\to\tauptaum$ at LHC.
Additional Higgs decay modes that could be considered at the LHC include:
a) $\a\to Z\h$; b) $\h\to\a\a$;
c) $\h_j\to\h_i\h_i$; d) $\a,\h\to t\anti t$.  Because
of the more complicated Higgs self interactions,
b) and c) cannot be reliably computed in the NMSSM without
additional assumptions. The Higgs mass values for
which mode a) is kinematically allowed can be quite different
than those relevant to the MSSM and thus there are
uncertainties in translating ATLAS and CMS results for the MSSM
into the present more general context. Finally, mode d) is currently
of very uncertain status and might turn out to be either more
effective or less effective than current estimates.
Thus, to be conservative, any choice
of NMSSM parameters for which the modes a)-d) might be relevant is excluded.
Even over this restricted region of parameter space, 
NMSSM parameter choices can be found such that there are
no observable Higgs signatures at either LEP2 or the LHC.

The free parameters of the model can be chosen
to be $\tanb$, $\mhi$, $\lam$, $\alpha_{1,2,3}$, and $\ma$.
Here, $\hi$ is the mass of the lightest CP-even Higgs mass eigenstate,
$\a$ is the lightest CP-odd scalar (for the present demonstration,
the 2nd CP-odd scalar can be taken to be much heavier), and
$\lam$ appears in the superpotential in the term $W\ni \lam\hat H_1\hat H_2\hat
N$. A crucial ingredient in constraining the model is that $\lam\lsim 0.7$
is required if $\lam$ is to remain perturbative during evolution
from scale $\mz$ to the Planck scale. This limitation on $\lam$ implies
a $\tanb$-dependent upper limit on $\mhi$ in the range $\lsim 140\gev$.
The angles $\alpha_{1,2,3}$
are those parameterizing the orthogonal matrix which diagonalizes the CP-even
Higgs mass-squared matrix. 
All couplings and cross sections are determined once the above parameters
are specified. Details regarding the procedure for scanning the NMSSM parameter
space and assessing observability of the various Higgs
bosons are given elsewhere.~\cite{ghm}  A choice
of parameters such that none of the Higgs bosons
$\h_{1,2,3}$, $\a$ or $\hpm$ are observable at LEP2 or the LHC is declared to
be a ``point of unobservability'' or a ``bad point''.

The results obtained are the following.
If $\tanb\lsim 1.5$
then all parameter points that are included in the search are observable
for $\mhi$ values up to the maximum allowed ($\mhi^{\rm max}\sim 137\gev$
for $\lam_{\rm max}=0.7$, after including radiative corrections).
For such low $\tanb$, the LHC $\gam\gam$ and $4\ell$ modes allow
detection if LEP2 does not. For high $\tanb\gsim 10$, the
parameter regions where points of unobservability are found
are also of very limited extent, disappearing as the $b\anti b\h_{1,2,3}$
and/or $b\anti b\a$ LHC modes allow detection where LEP2 does not.
However, significant portions of
searched parameter space contain points of unobservability
for moderate $\tanb$ values. For moderate $\tanb$, $b\anti b\h_i$
processes \etc\ are not observable at the LHC,
$\mhi$ and $\ma$ can be chosen 
so that $\mhi+\mz$ and $\mhi+\ma$ are close to or above the $\rts$ of LEP2,
and the $\h_{1,2,3}\to \gam\gam$ modes can be suppressed
in the NMSSM by  parameter choices such that the
$WW\h_{1,2,3}$ couplings (and thus the $W$-boson loop
contribution to the $\gam\gam\h_{1,2,3}$ couplings)
are all reduced relative to SM strength.

\begin{figure}[htb]
\leavevmode
\begin{center}
\centerline{\psfig{file=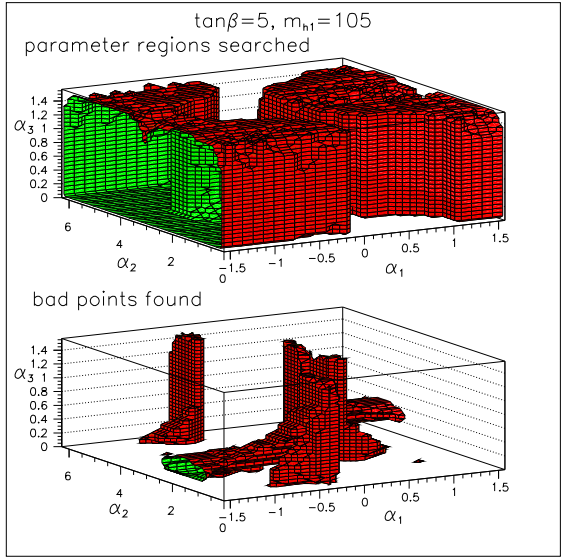,width=2.9in}}
\end{center}
\caption{For $\tanb=5$ and $\mhi=105\gev$, we display in three dimensional
%$\alpha_{1,2,3}$
$(\alpha_1,\alpha_2,\alpha_3)$ parameter space the parameter regions
searched (which lie within the surfaces shown), and the
%`regions of (Higgs boson) unobservability' (see text) therein
regions therein for which the remaining model parameters can
be chosen so that no Higgs boson is observable
(interior to the surfaces shown).}
\label{tanb5}
\end{figure}

To illustrate, we shall discuss results for $\tanb=3$, $\tanb=5$ and
$\tanb=10$ (for which $\mhi^{\rm max}\sim 124\gev$, $118\gev$ and $114\gev$,
respectively) and $\mhi=105\gev$.
\begin{itemize}
\item
In Fig.~\ref{tanb5}, we display for $\tanb=5$ both the
portions of $(\alpha_1,\alpha_2,\alpha_3)$
parameter space that satisfy our search restrictions,
and the regions (termed ``regions of unobservability'')
within the searched parameter space such that,
for {\it some} choice of the remaining parameters ($\lam$ and $\ma$),
no Higgs boson will be detected
using any of the techniques discussed earlier.~\footnote{For a 
given $\alpha_{1,2,3}$ value
such that there is a choice of $\lam$ and $\ma$ for which no Higgs
boson is observable, there are generally
other choices of $\lam$ and $\ma$ for which at least one
Higgs boson {\it is} observable.} Relatively large regions of unobservability
within the searched parameter space are present.
\item
At $\tanb=3$, The search region that satisfies our
criteria is nearly the same; the regions of unobservability lie mostly within
those found for $\tanb=5$, and are about 50\% smaller.
\item
For $\tanb=10$, the regions of unobservability comprise only a
very small portion of those found for $\tanb=5$.
This reduction is due to the increased $b\anti b$ couplings
of the $\h_i$ and $\a$, which imply increased $b\anti b\h_i,b\anti b\a$
production cross sections. As these cross sections become large, detection
of at least one of the $\h_i$ and/or the $\a$ in the
$b\anti b\taup\taum$ final state becomes
increasingly difficult to avoid. For values of 
$\tanb\gsim 10$,~\footnote{The precise value 
of the critical lower bound on $\tanb$
depends sensitively on $\mhi$.}
%In the present example, no region of observability is left for
%values of $\tanb$ above 11 when $\mhi=105\gev$.}
we find that one or
more of the $\h_i,\a$ should be observable regardless of location in
$(\alpha_1,\alpha_2,\alpha_3,\lam,\ma)$ parameter space (within
the somewhat restricted search region that we explore).
\end{itemize}
Details of what goes wrong at a typical point are summarized
elsewhere.~\cite{ghm}

Supersymmetric decays of the Higgs bosons are neglected in the above.
If these decays are important,
the regions of unobservability found without using the SUSY final states will
increase in size.  However, Higgs masses in the
regions of unobservability
are typically modest in size ($100-200\gev$), and
as SUSY mass limits increase with LEP2 running this additional
concern will become less relevant.  Of course, if SUSY
decays are significant, detection of the Higgs
bosons in the SUSY modes might be possible, in which case
the regions of unobservability might decrease in size. Assessment of this
issue is dependent upon a specific model for soft SUSY breaking.

Although it is not possible to establish a no-lose
theorem for the NMSSM Higgs bosons by combining data from LEP2 and the LHC
(in contrast to the no-lose theorems applicable to
the NLC Higgs search with $\rts\gsim 300\gev$), the regions of complete
Higgs boson unobservability appear to constitute a small fraction of the
total model parameter space.
It would be interesting to see whether or not these
regions of unobservability correspond to unnatural choices
for the Planck scale supersymmetry-breaking parameters.

\subsection{The MSSM at the NLC}

In what follows, we shall use NLC language when referring
to operation of a lepton-lepton collider at $\rts=500\gev$. 
The reader should keep
in mind that the FMC can also be run in an NLC-like mode.
Results for $s$-channel Higgs discovery in the MSSM at the FMC
will be summarized separately.

In the MSSM, the important production mechanisms are mainly determined
by the parameters $\mha$ and $\tanb$. In general, we shall
try to phrase our survey in terms of the parameter space
defined by $(\mha,\tanb)$, keeping fixed
the other MSSM parameters such as the masses of the top squarks.
The most important general point
is the complementarity of the $\epem\rta Z\hl$ and $\epem\rta\ha\hh$ 
cross sections, which are proportional to $\sin^2(\beta-\alpha)$,
and the $\epem\rta Z\hh$ and $\epem\rta \ha\hl$ cross sections,
proportional to $\cos^2(\beta-\alpha)$.
(Here, $\alpha$ is the mixing angle that emerges in diagonalizing the
CP-even Higgs sector.)
Since $\cos^2(\beta-\alpha)$ and $\sin^2(\beta-\alpha)$ cannot simultaneously
be small, if there is sufficient energy
then there is a large production cross section for all three of the
neutral Higgs bosons.  In the more likely case that $\mha$ is large,
$\sin^2(\beta-\alpha)$ will be large  and the $\hl$ will be
most easily seen in the $Z\hl$ channel, while the $\ha$ and $\hh$
will have a large production rate in the $\ha\hh$ channel if
$\sqrt s$ is adequate. If $\cos^2(\beta-\alpha)$
is large, then $\mha$ must be small and $\ha\hl$ and $Z\hh$ will
have large rates.

\subsubsection{Detection of the {\boldmath$\hl$}}

In the MSSM, the lightest Higgs boson is accessible in the
$Z\hl$ and $WW$-fusion modes for all {\it but} the $\mha\leq \mz$, 
$\tanb\geq 7-10$ corner of parameter space.  Outside this
region the $\hl$ rapidly becomes SM-like.
A figure illustrating this and detailed discussions 
can be found in DPF95~\cite{dpfreport} and in many other reviews and 
reports.~\cite{gunionerice,desyworkshopsusy,janotlep,janothawaii,gunionhawaii,eurostudy}
As noted earlier, 
event rates in the $Z\hl$ mode will be sufficient even if the $\hl$
decays mostly invisibly to a $\cnone\cnone$ pair.~\cite{guninvis,fjk,cr}

If such SUSY decays are not important for the $\hl$,
then, to a very good approximation, the entire SM discussion 
can be taken over for $Z\hl$ for large $\mha$.  As we emphasized
earlier, the interesting question becomes for what portion
of moderate-$\mha$ parameter space can
the cross sections, branching ratios and/or couplings be
measured with sufficient precision to distinguish an approximately SM-like
$\hl$ from the $\hsm$ at the NLC and/or FMC. 
Expectations based on the accuracies tabulated in 
Tables~\ref{nlcerrors}, \ref{fmcerrors}, and \ref{nlcfmcerrors}
were given in Sec.~\ref{ssnlcfmc}.   The conclusions depend upon
whether only NLC data, or both NLC {\it and} $s$-channel FMC data, 
are available. For an NLC running at $\rts=500\gev$ with $L=200\fbi$
of integrated luminosity and a vertex tagger good enough to topologically
separate $b$ from $c$ jets, we will be able to tell the difference between
the $\hsm$ and the $\hl$ at the $\sim 3\sigma$ level if $\mha\lsim 400\gev$.  
If both NLC and $s$-channel FMC data ($L=200\fbi$ each) are available,
then discrimination between the $\hsm$ and the $\hl$ at the $3\sigma$ level
should be possible up to $\mha\sim 600\gev$ (based upon the determination
of the $\mupmum$ partial width).

Although strongly disfavored by model building and increasingly
restricted by LEP2 data, there is
still a possibility that $\mha\lsim \mz$. The important production
process will then be $\epem\rta \zstar\rta \hl\ha$, the $\hl ZZ$ and $\hl WW$
couplings being suppressed for such $\mha$ values.
Once LEP2 runs at $\sqrt s=192\gev$ 
$\hl\ha$ associated production will be kinematically allowed and detectable for
all but very large $\mstop$ values.
The NLC could be run at a $\sqrt s$ value optimized for detailed studies
of the $\hl\ha$ final state. At $\sqrt s=500\gev$, at least 1500 events are 
predicted for precisely the $\mha\lsim 100\gev$, $\tanb\gsim 2-5$ section
of parameter space for which fewer events than this would be found
in the $Z\hl$ mode.  Thus, $\epem\rta\hl\ha$ will give
us a large number of $\hl$'s if $Z\hl$ does not.
In what follows, our focus will be on large $\mha$; we
will confine ourselves to a only few general remarks regarding
the small-$\mha$ scenario.

\subsubsection{Detection of the 
{\boldmath $\hh$, $\ha$ and $\hpm$}}

For $\mha\lsim \mz$, the $\ha$, along with the $\hl$,
can be easily detected in the $\hl\ha$ mode, as discussed above.
If 50 events are adequate, detection of both the $\hl$ and $\ha$
in this mode will even be possible for $\mha$ up to $\sim 120\gev$.
In this same region
the $\hh$ will be found via $Z\hh$ and $WW$-fusion 
production.~\cite{desyworkshopsusy,janothawaii,gunionhawaii,eurostudy}
In addition, $\hp\hm$ pair production will be kinematically allowed and easily 
observable.~\cite{desyworkshopsusy,janothawaii,gunionhawaii,eurostudy}
In this low to moderate $\mha$ region, the only SUSY decay mode that
has a real possibility of being present is the invisible $\cnone\cnone$
mode for the neutral Higgs bosons.  If this mode were to dominate
the decays of all three neutral Higgs bosons, then only the
$\hh$ could be detected, 
using the recoil mass technique in the $Z\hh$ channel.
However, in the $\hp\hm$ channel the final states would probably
not include SUSY modes and $\hp$ discovery would be straightforward.
If a light $\hpm$ is detected, then one would know that the Higgs
detected in association with the $Z$ was most likely the $\hh$
and not the $\hl$.  A dedicated search for the light $\ha$ and $\hl$
through (rare) non-invisible decays would then be appropriate.

For $\mha\geq 120\gev$, $\epem\rta\hh\ha$ and $\epem\rta\hp\hm$ must be
employed for detection of the three heavy Higgs bosons.
Assuming that SUSY decays are not dominant, and
using the 50 event criterion, the mode $\hh\ha$ is observable up to
$\mhh \sim \mha \sim 240$ GeV, and $\hp\hm$ can be detected up to
$\mhpm =230$ GeV,~\cite{desyworkshopsusy,janothawaii,gunionhawaii,eurostudy}
assuming $\rts=500\gev$. For $\rts_{\epem}=500\gev$,
the $\gamma\gamma$ collider mode could potentially
extend the reach for the $\hh,\ha$ bosons up to 400 GeV, especially if $\tanb$
is not large. This is discussed in several 
places~\cite{dpfreport,ghgamgam,gunionerice}
and will not be reviewed further here.

The upper limits in the $\hh\ha$ and $\hp\hm$
modes are almost entirely a function of the
machine energy (assuming an appropriately higher integrated luminosity
is available at a higher $\sqrt s$).
Two recent studies~\cite{gk,fengmoroi} show that
at $\sqrt s = 1$ TeV, with an integrated luminosity of $200 \fbi$, 
$\hh\ha$ and $\hp\hm$ detection would extend
to $\mhh\sim\mha\sim\mhpm\sim 450$ GeV even if substantial
SUSY decays of these heavier Higgs are present.
As frequently noted, models in 
which the MSSM is implemented in the coupling-constant-unification,
radiative-electroweak-symmetry-breaking context often
predict masses above 200 GeV, suggesting that
this extension in mass reach over that for $\sqrt s =500$ GeV
might be crucial. If a high luminosity high energy
muon collider with $\rts\sim 4\tev$ proves feasible, pair production
could be studied for $\mha\sim\mhh\sim\mhpm$ as large as $\sim 1.8\tev$,
which certainly would include any reasonable model.

It is crucial that the $\hh,\ha,\hpm$ be found if supersymmetric particles
are observed.  Only in this way can we verify directly that
the MSSM Higgs sector includes at least the minimal content required.
Once discovered, 
an important question is how much can be learned from $\hh\ha$ and
$\hp\hm$ pair production regarding the detailed structure of the MSSM.
The recent investigations~\cite{gk,fengmoroi} indicate that 
a full study of these pair production final states will place extremely
powerful constraints on the GUT boundary conditions underlying the MSSM
model. Alternatively, the Higgs studies could reveal inconsistencies
between the minimal two-doublet Higgs sector predictions and constraints
from direct supersymmetry production studies (for any choice
of GUT boundary conditions).  Then, searches for additional heavy Higgs
would become a priority.

It is useful to illustrate just how powerful the details of $\hh,\ha,\hpm$
decays are as a
test of the model and for determining the underlying GUT boundary conditions
at the GUT/Planck mass scale.  In one study,~\cite{gk}, 
this has been illustrated by examining six not
terribly different GUT-scale boundary condition scenarios in
which there is universality for the soft-SUSY-breaking parameters
$\mhalf$, $m_0$ and $A_0$ associated with soft gaugino masses, soft scalar
masses and soft Yukawa coefficients, respectively.~\cite{susyref}
After requiring that the electroweak symmetry breaking generated
as a result of parameter evolution yield the correct $Z$ boson mass,
the only other parameters required to fully specify a model 
in this universal-boundary-condition class are $\tanb$ and the
sign of the $\mu$ parameter (appearing in the superpotential $W\ni \mu \hat
H_1\hat H_2$). The six models considered~\cite{gk} are
denoted \DM, \DP, \NSM, \NSP, \HSM, \HSP, where the superscript indicates
$\sign(\mu)$. Each is specified by a particular 
choice for $m_0:\mhalf:A_0$, thereby leaving
only $\mhalf$, in addition to $\tanb$,
as a free parameter in any given model. Pair production is
then considered in the context of each model as a function
of location in the kinematically and constraint allowed
portion of $(\mhalf,\tanb)$ parameter space.

It is found~\cite{gk} that event rates for anticipated machine luminosities
are such that $\hh\ha$ and $\hp\hm$ pair
production can be detected in final state modes where $\hh,\ha\to b\anti b$
or $t\anti t$ and $\hp\to t\anti b,\hm\to b\anti t$ even when the branching
ratios for SUSY decays are substantial. Further, the mass of the $\hh$
or $\ha$ can be determined with substantial accuracy using the fully
reconstructable all jet final states associated with these modes.
Most importantly, in much of the kinematically and phenomenologically allowed
parameter space Higgs branching ratios for 
a variety of different decay channels can be measured by ``tagging''
one member of the Higgs pair in a fully reconstructable all jet decay mode
and then searching for different types of final states in the decay of the 
second (recoiling) Higgs boson.

The power of Higgs pair observations for determining the GUT
boundary conditions is most simply illustrated by an example.
Let us suppose that the \DM\ model with $\mhalf=201.7\gev$ and $\tanb=7.5$
is nature's choice. This implies that $\mha=349.7\gev$ and
$\mcpmone=149.5\gev$. Experimentally, one would measure $\mha$ as above
and $\mcpmone$ (the lightest chargino) mass in the usual way and then
infer the required parameters for a given model. For the six models
the parameters are given in Table~\ref{mhalftanbtable}.
Note that if the correct GUT scenario can be ascertained experimentally, then
$\tanb$ and $\mhalf$ will be fixed.

\begin{table}[hbt]
\caption[fake]{We tabulate the values of $\mhalf$ (in GeV)
and $\tanb$ required in each of our six scenarios in order
that $\mha=349.7\gev$ and $\mcpmone=149.5\gev$.
Also given are the corresponding values of $\mhh$. Masses are in GeV.}
\begin{center}
\begin{tabular}{|c|c|c|c|c|c|c|}
\hline
 & \DM\ & \DP\ & \NSM\ & \NSP\ & \HSM\ & \HSP\ \\
\hline
\hline
$\mhalf$ & 201.7 & 174.4 & 210.6 & 168.2 & 203.9 & 180.0 \\
$\tanb$ & 7.50 & 2.94 & 3.24 & 2.04 & 12.06 & 3.83 \\
$\mhh$ & 350.3 & 355.8 & 353.9 & 359.0 & 350.1 & 353.2 \\
\hline
\end{tabular}
\end{center}
\label{mhalftanbtable}
\end{table}

Determination of the GUT scenario proceeds as follows.
Given the parameters required for the observed $\mha$
and $\mcpmone$ for each model, as tabulated in Table~\ref{mhalftanbtable},
the rates for different final states of the recoil (non-tagged) Higgs boson
in pair production can be computed. Those for the input
\DM\ model are used to determine the statistical accuracy
with which ratios of event numbers in different types of final states
can be measured.~\footnote{We focus on ratios in order to
be less sensitive to systematic uncertainties in efficiencies \etc;
however, absolute rates will also be useful in some instances.~\cite{gk}}
The ratios predicted in
the \DP, \NSM, \NSP, \HSM, and \HSP models will be different
from those predicted for the input \DM\ model.  Thus,
the statistical uncertainty predicted for the various ratios in the input \DM\
model can be used to compute the $\chisq$ by which the predictions of
the other models differ from the central values of the input \DM\ model.
The results for a selection of final state ratios are given
in Table~\ref{chisqtable}. The final states considered are:
$b\anti b$ and $t\anti t$ 
for the $\hh,\ha$; $\hl\hl$ (light Higgs pair, with $\hl\to b\anti b$) for
the $\hh$; $\hl\wp$ and $\tau^+\nu_\tau$ for the $\hp$ (or
the charge conjugates for the $\hm$); and SUSY modes
(experimentally easily identified by the presence of missing energy)
classified according to the number of charged leptons summed over
any number of jets (including 0). All branching ratios and reasonable
efficiencies are incorporated in the statistical errors employed
in constructing this table. The effective luminosity $\leff=80\fbi$ 
is equivalent to
an overall tagging and reconstruction efficiency for events
of $\eps=0.4$ at a total integrated luminosity of $L=200\fbi$.
Results presented are for $\rts=1\tev$.

\begin{table}[hbt]
\caption[fake]{We tabulate $\Delta\chi^2_i$ 
(relative to the \DM\ scenario) for the indicated branching
fraction ratios as a function of scenario,
assuming the measured $\mha$ and $\mcpmone$ values are $349.7\gev$
and $149.5\gev$, respectively. The SUSY channels have been resolved into 
final states involving a fixed number of leptons.  
The error used in calculating each $\Delta\chi^2_i$ is the approximate
$1\sigma$ error with which the given ratio could be measured
for $\leff=80\fbi$ at $\rts=1\tev$ {\it assuming that the \DM\
scenario is the correct one}.
}
\begin{center}
{%\footnotesize
\begin{tabular}{|c|c|c|c|c|c|}
\hline
Ratio & \DP\ & \NSM\ & \NSP\ & \HSM\ & \HSP\ \\
\hline
\multicolumn{6}{|c|} {$\langle\hh,\ha\rangle$} \\
\hline
$ [0\ell][\geq0 j]/b\anti b,t\anti t$ 
 & 12878 & 1277 & 25243 & 0.77 & 10331 \\
$ [1\ell][\geq0 j]/b\anti b,t\anti t$ 
  & 13081 & 2.41 & 5130 & 3.6 & 4783 \\
$ [2\ell][\geq0j]/b\anti b,t\anti t$ 
  & 4543 & 5.12 & 92395 & 26.6 & 116 \\
$ \hl\hl/ b\anti b$  & 109 & 1130 & 1516 & 10.2 & 6.2 \\
\hline
\multicolumn{6}{|c|} {$\hp$} \\
\hline
$ [0\ell][\geq0j]/t\anti b$ 
 & 12.2 & 36.5 & 43.2 & 0.04 & 0.2 \\
$ [1\ell][\geq0j]/t\anti b$ 
 & 1.5 & 0.3 & 0.1 & 5.6 & 0.06  \\
$\hl W/ t\anti b$ 
 & 0.8 & 0.5 & 3.6 & 7.3 & 0.3 \\
$\tau\nu/ t\anti b$ 
 & 43.7 & 41.5 & 47.7 & 13.7 & 35.5 \\
\hline
$\sum_i\Delta\chi^2_i$ & 30669 & 2493 & 124379 & 68 & 15272 \\
\hline
\end{tabular}
}
\end{center}
\label{chisqtable}
\end{table}

From Table~\ref{chisqtable}
it is clear that the five alternative models can be discriminated
against at a high (often very high) level of confidence.  Further
subdivision of the SUSY final states into states containing
a certain number of jets yields even more discrimination power.~\cite{gk}
Thus, not only will detection
of Higgs pair production in $\epem$ or $\mupmum$
collisions (at planned luminosities) be 
possible for most of the kinematically accessible
portion of parameter space in a typical GUT model, but also the detailed rates
for and ratios of different neutral and charged Higgs decay final states
will very strongly constrain the possible GUT-scale boundary condition
scenario and choice of parameters, \eg\ $\tanb$ and $\mhalf$, therein.

\subsection{The MSSM in $s$-channel collisions at the FMC}

We have seen that other colliders offer various mechanisms
to directly search for the $\ha,\hh$, but have significant limitations:
\begin{itemize}
\item The LHC has ``$\hl$-only'' regions at moderate
$\tan \beta$, $\mha\geq 200\gev$.
\item At the NLC one can use the mode $\ee\to \zstar\to \hh\ha$ 
(the mode $\hl\ha$ is suppressed for
large $\mha$), but it is limited to $\mhh\sim \mha\lsim \sqrt{s}/2$.
\item A $\gamma \gamma$ collider could probe heavy Higgs up to masses of
$\mhh\sim \mha\sim 0.8\sqrt{s}_{\epem}$, but this would quite likely require
$L\sim 100-200{\fb}^{-1}$, especially if the Higgs bosons have
masses near $400\gev$ and $\tanb$ 
is large.~\cite{ghgamgam,dpfreport,gunionerice}
\end{itemize}
In contrast, there is an excellent chance of being able to detect
the $\hh,\ha$ at a $\mupmum$ collider provided only that $\mha$ is smaller
than the maximal $\rts$ available. This could prove to be very important
given that GUT MSSM models usually predict $\mha> 200\gev$.

A detailed study of $s$-channel production
of the $\hh,\ha$ has been made.~\cite{bbgh} The optimal strategy
for their detection and study depends upon the circumstances.
First, it could be that the $\hh$ and/or $\ha$ will already have been
discovered at the LHC.  With $L=300\fbi$ of integrated
luminosity for ATLAS and CMS (each), 
this would be the case if $\tanb\leq 3$ (or $\tanb$ is large);
see Fig.~\ref{mssmhilum}. 
Even if the $\hh,\ha$ have not been detected,
then, as described earlier,
strong constraints on $\mha$ are possible through precision measurements
of the properties of the $\hl$ at the NLC and/or FMC. 
For example, if no deviation is observed
at the NLC in $L=200\fbi$ running at $\rts=500\gev$, 
then we would know that $\mha>400\gev$. 
If a statistically significant deviation from SM predictions is observed,
then an approximate determination of $\mha$ would be possible.
Either way, we could limit the $\rts$ scan for the $\ha$
in the $s$-channel to the appropriate mass region and thereby 
greatly facilitate direct observation of the $\ha$
and $\hh$ since it would allow us to devote more luminosity per
scan point than if $\mha$ is not constrained.

With such pre-knowledge
of $\mha$, it will be possible to detect and perform detailed
studies of the $\hh,\ha$ for essentially all $\tanb\geq 1$
provided only that $\mha\leq \rts_{\rm max}$.~\footnote{We
assume that a final ring optimized for maximal luminosity at $\rts\sim \mha$
would be constructed.} If $\tanb\leq 3$, then
excellent resolution, $R\sim 0.01\%$, will be necessary for detection
since the $\ha$ and $\hh$
become relatively narrow for low $\tanb$ values. 
For higher $\tanb$ values $R\sim 0.1\%$ is adequate
for $\hh,\ha$ detection, but $R\sim 0.01\%$ would be required in order
to separate the rather degenerate $\hh$ and $\ha$ peaks (as a function
of $\rts$) from one another.

Even without pre-knowledge of $\mha$, 
there would be an excellent chance for discovery of the $\ha,\hh$
Higgs bosons in the $s$-channel at a $\mupmum$ collider if they
have not already been observed at the LHC, given that the latter implies 
that $\tanb> 3$. Indeed, detection of the $\ha,\hh$ is possible~\cite{bbgh}
in the mass range 
from 200 to 500 GeV via a $s$-channel scan in $\mupmum$ collisions 
provided $\tanb\geq 3$ and $L=200\fbi$ of
luminosity is devoted to the scan. (A detailed strategy as to how
much luminosity to devote to different $\rts$ values 
in the $200-500\gev$ range during
this scan must be employed.~\cite{bbgh})
That the signals become viable when $\tanb>1$
(as favored by GUT models) is due to the fact that the couplings
of $\ha$ and (once $\mha\geq 150\gev$) 
$\hh$ to $b\anti b$ and, especially to $\mu^+\mu^-$,  are
proportional to $\tanb$, and thus increasingly enhanced as
$\tanb$ rises. In the $\tanb\geq 3$ region, a beam energy resolution of
$R\lsim 0.1\%$ is adequate for the scan to be successful in discovery,
but as already noted $R\sim 0.01\%$ is needed to separate the $\hh$ and $\ha$
peaks from one another.
That the LHC and the FMC are complementary in this
respect is a very crucial point. Together, the LHC and FMC
essentially guarantee discovery of the $\ha,\hh$
so long as they have masses less than the maximum $\rts$ of the FMC.

In the event that the NLC has not been constructed, it could be
that the first mode of operation of the FMC would be to optimize
for and accumulate luminosity at $\rts=500\gev$ (or whatever the
maximal value is).  In this case, there is still a significant chance
for detecting the $\hh,\ha$ even if $\mha\geq \rts/2$. 
Although reduced in magnitude compared to an electron
collider, there is a long low-energy bremsstrahlung tail 
at a muon collider that provides a
self-scan over the range of energies below 
the design energy, and thus can 
be used to detect $s$-channel resonances. 
Observation of $\ha,\hh$ peaks in the $b\anti b$ mass distribution
$m_{b\anti b}$ created by this bremsstrahlung tail may be possible.
The region of the $(\mha,\tanb)$ parameter space plane for which
a peak is observable depends strongly on the $b\anti b$
invariant mass resolution. For an excellent $m_{b\anti b}$
mass resolution of order $\pm 5\gev$ and integrated luminosity
of $L=50\fbi$ ($200\fbi$) at $\rts=500\gev$, 
the $\ha,\hh$ peak(s) are observable for $\tanb\geq 5-6$ ($3.5-4.5$)
if $250\gev\leq\mha\leq 500\gev$ --- \ie\ the LHC/FMC gap 
for $\tanb\gsim 3$ is essentially closed at the
higher luminosity.

Finally, even if
a $\sqrt{s}\sim 500\gev$ muon collider does not have sufficient energy
to discover heavy supersymmetric Higgs bosons in the $s$-channel,
construction of a higher energy machine would be possible;
a popular reference design is one for $\sqrt{s}=4\tev$. 
Such an energy would allow discovery of $\mupmum\to\ha\hh$ 
and $\hp\hm$ pair production,
via the $b\anti b$ or $t\anti t$ decay channels of the $\hh,\ha$
and $t\anti b,\anti t b$ decay channels of the $\hp,\hm$,
up to masses very close to $\mha\sim \mhh\sim \mhpm \sim 2\tev$.~\cite{gk}

\subsection{The NMSSM at the NLC}

Consider a CP-conserving Higgs sector for the NMSSM, and the three CP-even
Higgs bosons $\h_{1,2,3}$, labelled in order of increasing mass.
The first question is whether or not a $\sqrt s=500\gev$ $\epem$
collider would still be guaranteed to discover at least one
of the NMSSM Higgs bosons.  We have seen that in the MSSM
there is such a guarantee
because the $\hl$ has an upper mass bound, {\it and} because 
it has near maximal $\hl ZZ$ coupling when $\mhl$ approaches its upper limit.
In the NMSSM model, it is in principle possible to choose
parameters such that $\h_{1,2}$ 
have such suppressed $ZZ$ coupling strength that their $Z\h_{1,2}$
and $WW\rta \h_{1,2}$ production rates are too low for observation,
while the heavier $\h_3$
Higgs is too heavy to be produced in the $Z \h_3$ or $WW\rta \h_3$ modes.
This issue has been studied recently.~\cite{kimoh,kot,KW,ETS,eurostudy}
If the model is placed within the
normal unification context, with simple boundary conditions at $M_X$,
{\it and if all couplings are required to remain perturbative
in evolving up to scale $M_X$} (as is conventional), then 
it is found that the above situation does not arise. At least
one of the three neutral scalars will have $\sigma(\epem\rta Z \h)\gsim 0.04\pb$
for any $\epem$ collider with $\sqrt s\geq 300\gev$.
For $L=10\fbi$, this corresponds to roughly 30 events in the clean
$Z\h$ with $Z\rta \lplm$ recoil-mass discovery mode.  
However, these same studies all make it clear that
there is no guarantee that LEP2 will detect a Higgs boson of the NMSSM.
This is because the Higgs boson with significant $ZZ$-Higgs coupling
can easily have mass beyond the kinematical reach of LEP2.
Of course, once a neutral Higgs bosons is discovered,
it will be crucial to measure all its couplings and to 
determine its CP character (using techniques discussed
earlier), not only to 
try to rule out the possibility that it is the SM $\hsm$, but
also to try to determine whether or not the supersymmetric
model is the MSSM or the NMSSM (or still further extension).

\subsection {The NMSSM in $s$-channel collisions at the FMC}

This interesting topic is currently under investigation.
Unlike the standard $\rts=500\gev$ modes for discovery,
the sensitivity in the $s$-channel depends significantly on
the $\mupmum$ couplings of the various Higgs bosons.

\section{Conclusions}\label{sconcl}

Models in which electroweak symmetry breaking occurs via
a Higgs sector, leaving behind physical spin-0 Higgs bosons,
continue to be very attractive and imply a rich phenomenology.
Our focus has been on supersymmetric theories, in which context
Higgs bosons are completely natural and the
Higgs sector is highly constrained.
Not only must the Higgs sector consist
of just two doublets (and no more, and no triplets, unless intermediate
scale matter is introduced to fix up coupling unification)
plus possible singlets, but also the Higgs sector couplings
are closely related to gauge couplings. As a result,
there is a strong upper bound on the mass of the lightest CP-even
Higgs boson ($\hl$) in a supersymmetric model. Further, in the decoupling
limit (\eg\ for large $\mha$ in the MSSM), the $\hl$ becomes SM-like.
Thus, Higgs phenomenology for the simpler supersymmetric models is relatively
well defined. This review has outlined the more important highlights regarding
experimentally probing a supersymmetric Higgs sector
at the LHC, NLC and FMC colliders. 
Some principle points and conclusions include the following.
\begin{itemize}
\item 
In the minimal two-doublet/no-singlet MSSM model, discovery of at least one
Higgs boson is guaranteed both at the LHC and in Higgstrahlung or Higgs pair 
production at a $\rts\geq 300\gev$ NLC and/or FMC, regardless of $\mha$. 
For large $\mha$, it is the SM-like $\hl$ that would be observed.
Since $\mhl<2\mw$, the SM-like $\hl$ could also be
produced at a high rate in $s$-channel collisions at the FMC.
\item
In the two-doublet/one-singlet NMSSM model, 
discovery of at least one Higgs boson is certain at the NLC and/or FMC,
but can no longer be absolutely guaranteed at the LHC (although
it is highly probable).
\item
Precision determination of all the couplings, the total width
and the mass of a light SM-like Higgs boson will be possible. The LHC and
NLC often provide highly complementary information.  For
example, in the MSSM-preferred $\leq 130\gev$ mass range, data
from the LHC, from $\epem$ collisions at the NLC (or $\mupmum$ collisions
at the FMC) and from $\gam\gam$ collisions at the NLC are {\it all} necessary
in order to extract the absolute coupling magnitudes and the total width
with good precision in a model-independent way.
Production in the $s$-channel at 
the FMC would provide additional precision measurements of relative
branching ratios and a direct measurement of the total width.
Such precision measurements could make it possible to distinguish
a SM-like $\hl$ of the MSSM from the SM $\hsm$.
\item
Thus, if the light $\hl$ of the supersymmetric models is SM-like,
it would be enormously advantageous to have, in addition to the LHC
and an NLC operating at $\rts\sim 500\gev$,
an FMC concentrating on $s$-channel Higgs studies. 
All three facilities are needed
in order to maximize the precision with which the properties
of a light SM-like Higgs boson can be determined. The value
of the resulting precision is great.  If $L=200\fbi$ of data 
is accumulated both in NLC running at $\rts=500\gev$ 
and in $s$-channel production at the FMC,
then $\hl$ vs. $\hsm$ discrimination is possible at the $\geq 3\sigma$ level
for $\mha\leq 600\gev$, 
and $\mha$ can be determined to within $\sim \pm50\gev$.
\item
Prospects for discovery of the heavier, non-SM-like 
$\hh,\ha$ of the MSSM are excellent.
They will be found at the LHC if $\mha\leq 200\gev$ or
at higher $\mha$ if either $\tanb\leq 3$ or $\tanb$ is large. 
If they are not found at the LHC, then the resulting
limits on $\mha$ and $\tanb$ 
tell us how to distribute the luminosity of a $L=200\fbi$ 
$s$-channel scan
at the FMC so as to guarantee their discovery (if $\mha\leq 500\gev$).
Even if the FMC is run at maximal $\rts$,
discovery of the $\hh,\ha$ is still possible
for $\mha$ below $\rts$ if $\tanb$ is large; peaks in the $b\anti b$
mass distribution, deriving from $s$-channel production via 
the bremstrahlung tail in the $\mupmum$ energy spectrum, would be visible.
Discovery of the $\hh,\ha$ in the $\gam\gam$ collider mode of operation 
at the NLC is possible if $\mha\leq
0.8\rts$ and integrated luminosity of $L\geq 100-200\fbi$ is accumulated.
At the NLC or FMC, $\hh\ha$
and $\hp\hm$ pair production will be observable provided $\rts> 2\mha$.
Once discovered, detailed studies of the decays of the $\hh$, $\ha$
and $\hpm$ would be possible and would strongly constrain
the GUT-scale soft-supersymmetry-breaking parameters.
\end{itemize}
To summarize, if Higgs bosons exist, they are very likely
to be part of a supersymmetric theory, and experimental efforts directed towards
fully studying the Higgs bosons will provide one of the most
exciting programs at the next generation of colliders.
Substantial progress has been made in detailing
the strategies required at the different accelerators for the
discovery and study of the Higgs bosons of a supersymmetric model
Higgs sector. 

\section*{Acknowledgments}

This work was supported in part by the U.S. Department of Energy
under Grant No.~DE-FG03-91ER40674.
Further support was provided by the Davis Institute for High Energy Physics.

\end{document}